\documentclass[pdflatex,sn-mathphys-num]{sn-jnl}
\usepackage{graphicx}%
\usepackage{amsmath,amssymb,amsfonts}%
\usepackage{amsthm}%
\usepackage{lmodern}
\usepackage{derivative}
\usepackage{physics}
\usepackage{multicol}
\usepackage{multirow}
\usepackage{array}
\usepackage{bm}
\newcolumntype{P}[1]{>{\centering\arraybackslash}p{#1}}
\usepackage{ragged2e}

\usepackage[labelsep=none]{caption}
\DeclareCaptionLabelFormat{bar}{#1 \textbf{#2} \textbar}
\usepackage{pdfpages}

\usepackage{xr}
\usepackage{xcolor}
\usepackage{soul}

\renewcommand{\figurename}{}
\renewcommand{\thefigure}{\textbf{Fig. \arabic{figure}}}

\begin{document}

\title[Article Title]{Raman fingerprint of high-temperature superconductivity in compressed hydrides}


\author[1]{ \fnm{Philip} \sur{Dalladay-Simpson}}
\equalcont{ These authors contributed equally to this work.}

\author[2]{\fnm{Guglielmo} \sur{Marchese}}
\equalcont{ These authors contributed equally to this work.}

\author[3]{\fnm{Zi-Yu} \sur{Cao}}
\equalcont{ These authors contributed equally to this work.}

\author[2,4]{\fnm{Paolo} \sur{Barone}}

\author[2]{\fnm{Lara} \sur{Benfatto}}

\author[5]{\fnm{Gaston} \sur{Garbarino}}

\author*[2]{\fnm{Francesco} \sur{Mauri}}
\email{ francesco.mauri@uniroma1.it}

\author*[1,6,7]{\fnm{Federico Aiace} \sur{Gorelli}}
\email{ federico.gorelli@hpstar.ac.cn}

\affil[1]{\small HPSTAR, Center for High Pressure Science and Technology Advanced Research, 1690 Cailun Road, Shanghai, 201203, China}

\affil[2]{\small Dipartimento di Fisica, Universit\`a di Roma La Sapienza, Piazzale Aldo Moro 5, I-00185 Roma, Italy}

\affil[3]{\small School of Physics Science and Information Technology, Liaocheng< University, Liaocheng 252059, China}

\affil[4]{\small CNR-SPIN, Area della Ricerca di Tor Vergata, Via del Fosso del Cavaliere 100, I-00133 Rome, Italy}

\affil[5]{\small ESRF,European Synchrotron Radiation Facility, 71 Avenue des Martyrs, Grenoble 38000, France}

\affil[6]{\small CNR-INO, Instituto Nazionale di Ottica, Consiglio Nazionale delle Ricerche, 50019 Sesto Fiorentino, Italy}

\affil[7]{\small SHARPS, Shanghai Advanced Research in Physical Sciences, Shanghai, 201203, China}



\abstract{The discovery of high-temperature superconductivity in hydrogen-rich compounds under extreme pressures \cite{drozdov_conventional_2015, drozdov_superconductivity_2019} has prompted great excitement, intense research, but also debate over the past decade. Electrical transport has been the primary diagnostic tool for identifying superconductivity in these systems, whereas complementary probes, including magnetic, spectroscopic, tunnelling  and ultrafast methods, remain mostly qualitative due to experimental constraints and sample heterogeneity \cite{minkov_magnetic_2022, capitani_spectroscopic_2017,  cao_pointcontact_2023,wu_ultrafast_2024,du_tunneling_2024}. Recent concerns over their reliability have fuelled controversy, leading to scepticism \cite{Hirsch_PRB2021,Eremets_commentPRB2024} and pointing out the need for alternative, quantitative approaches \cite{bhattacharyya_imaging_2024, du_superconducting_2025}. In this study, we acquired unprecedented high-quality Raman spectra of hexagonal LaH$_{10}$ at approximately 145 GPa and low temperatures, in conjunction with electrical transport measurements. Upon cooling, we observe a drop of resistivity and simultaneous remarkable variations of phonon frequencies and linewidths. These effects are interpreted and perfectly reproduced by the Migdal–Eliashberg theory, providing a definitive proof of phonon-mediated superconductivity and enabling a quantitative determination of the superconducting energy gap. Our results establish Raman spectroscopy as a robust, contact-free probe with micrometric resolution for studying high temperature superconductivity, opening a powerful route to its discovery and characterization.}
\keywords{Raman, high-pressure, superconductivity, electron-phonon}

\maketitle

\newpage


\section*{Introduction}

Over the past decade, intense research has focused on hydrogen-rich compounds under pressure to realize superconductivity at ambient temperatures \cite{boebinger_hydride_2025}. This pursuit stems from Ashcroft's seminal proposal \cite{ashcroft_metallic_1968} that crystals containing light elements, with intrinsically high lattice vibration energies, could sustain phonon-mediated superconductivity at elevated critical temperatures.  
The first experimental evidence supporting this idea came from the observation of an abrupt resistance drop in sulphur hydride \cite{drozdov_conventional_2015}, later reproduced by several groups in different compounds \cite{drozdov_conventional_2015,drozdov_superconductivity_2019,kong_superconductivity_2021,ma_high-temperature_2022}. However, concerns have also been raised regarding the reliability and reproducibility of these findings \cite{hirsch_electrical_2023}. To address the resulting persistent scepticism in the field, it is now crucial to extend the experimental characterization of the superconducting (SC) state beyond electrical transport techniques and to support such studies with quantitative theoretical analysis.

\begin{figure}[ht]
  \centering
  \includegraphics[width = \textwidth]{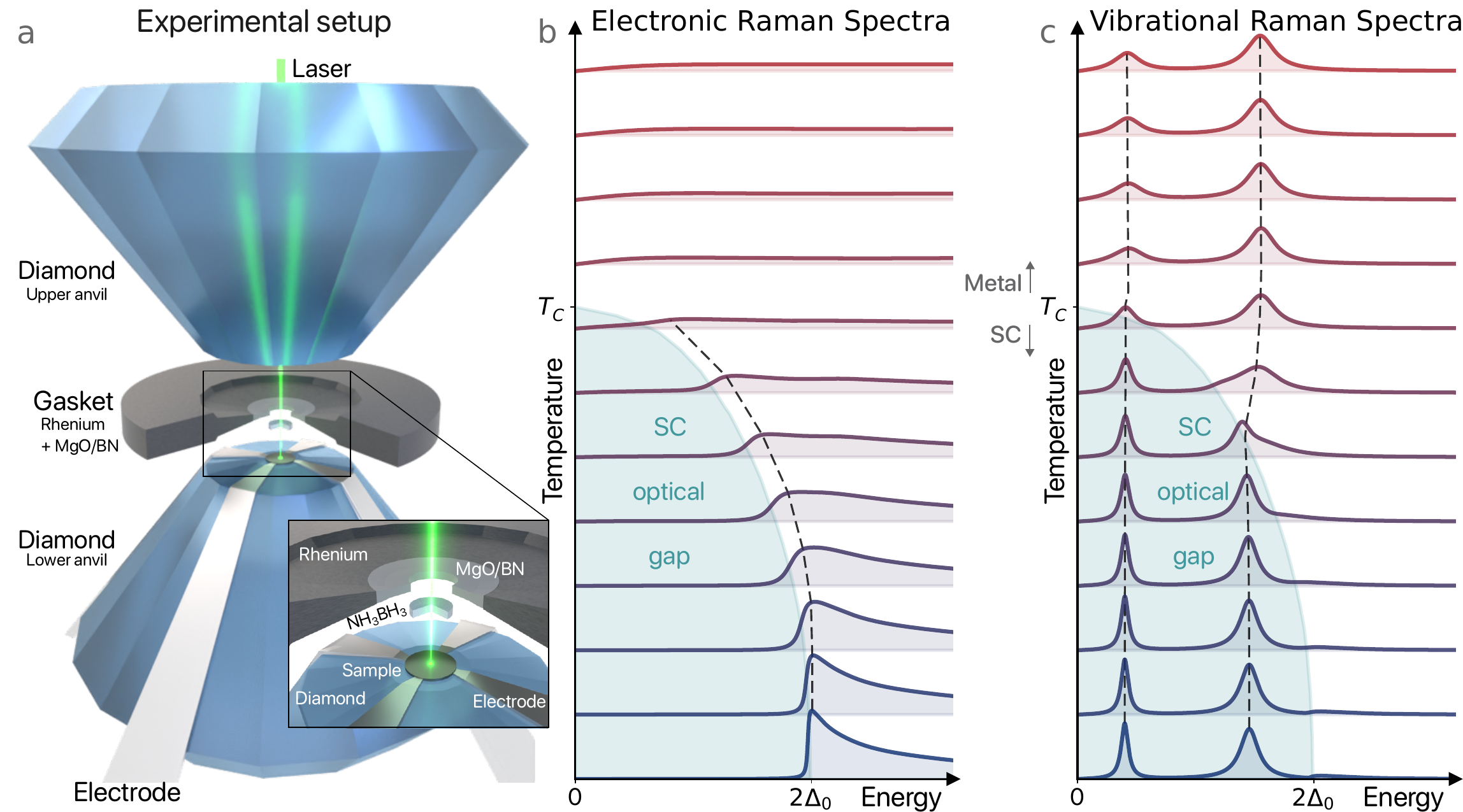}
  \caption{\textbf{High-pressure setup and schematic evolution of Raman spectra across the SC transition.} 
\textbf{a}, Schematic illustration (not to scale) of diamond anvil cell (DAC) for combined Raman and transport measurements. The hydride sample is compressed between two opposing diamond anvils, with four electrodes for resistance measurements and ammonia borane (NH$_{3}$BH$_{3}$) as pressure medium. The sample is confined by concentric gaskets: an insulating inner ring (MgO or BN), and an outer Rhenium gasket.
\textbf{b}, \textbf{c} Theoretical temperature evolution of the electronic (\textbf{b}) and vibrational (\textbf{c}) Raman spectra across the SC transition at $T_{\rm c}$ predicted by the ME approach, see Methods.
 Upon cooling, the SC gap energy $2\Delta(T)$ opens and saturates at $2\Delta_0 \sim 4k_BT_c$. The electronic Raman spectrum develops a pair-breaking peak that blue-shifts and gains intensity below $T_c$ \cite{devereaux_inelastic_2007}. Phonon modes with strong electron–phonon coupling and energies near the gap edge ($\omega_{\text{ph}} \lesssim 2\Delta_0$) exhibit, upon cooling, a redshift and linewidth narrowing when $\omega_{\text{ph}} < 2\Delta(T)$, preceded by a transient broadening around $T_c $ \cite{zeyher_zpb90,zeyher_phonon_1988, thomsen_phonon_1990, FriedlCardona_PhysRevLett_1990,cardona_raman_1991, hardy_prb93,tajima_prb00, hardy_prb04,kim_prl21}. The redshift magnitude decreases for lower $\omega_{\text{ph}} / (2\Delta_0)$ ratios, while linewidth narrowing persists. These behaviours reflect the modification of electron–phonon scattering induced by the opening of the SC gap in the electronic density of states.
}
\label{fig1:abstract}
\end{figure}

The main challenge arises from the experimental constraints imposed by high pressure measurements. The micrometer sized sample must be encapsulated within a diamond anvil cell (DAC) (\ref{fig1:abstract}\textbf{a}), which severely limits the use of conventional probes used to assess superconductivity at ambient conditions. The hallmark of superconductivity, the perfect diamagnetism, remains difficult to detect because of strong background signals\cite{minkov_magnetic_2022,bhattacharyya_imaging_2024}. 
As an alternative, spectroscopic evidence of gap formation has been explored via infrared reflectivity \cite{capitani_spectroscopic_2017},  ultrafast \cite{wu_ultrafast_2024} and tunneling spectroscopy \cite{du_tunneling_2024,du_superconducting_2025}. While these methods reveal signatures of electron pairing, their quantitative analysis is hindered by limited spatial resolution and sample inhomogeneity. Hydride samples are often multiphasic, comprising micrometric domains formed during the synthesis process\cite{laniel_high-pressure_2022}. Raman spectroscopy, by contrast, offers micrometric spatial resolution and is widely used to probe vibrational and electronic properties. 

Signatures of the SC transition can be detected in Raman spectra. 
Below the SC critical temperature ($T_{\rm c}$), the formation of Cooper pairs causes a redistribution of the spectral weight of electron-hole excitations, that can give rise to the so called pair-breaking peak in the electronic Raman spectra at twice the SC gap energy, the \textit{optical} gap $2\Delta(T)$ \cite{devereaux_inelastic_2007}.
As shown in  \ref{fig1:abstract}\textbf{b}, this feature 
closely follows the evolution of the SC optical gap, exhibiting a pronounced blueshift and intensity enhancement upon cooling. 
The temperature-dependent modification of the electronic density of states across the SC transition can also affect phonons, producing a distinctive temperature evolution of phonon frequencies and linewidths (\ref{fig1:abstract}\textbf{c}) which differs markedly from that of the electronic pair-breaking peak (\ref{fig1:abstract}\textbf{b}). This occurs
when the phonon energy ($\hbar\omega_{\mathrm{ph}}$) is comparable to or smaller than the zero-temperature SC optical gap, $2\Delta(0) = 2\Delta_0 \sim 4k_{\mathrm{B}}T_{\mathrm{c}}$, so that the phonon is resonant with  electron–hole excitations modified by the formation of the SC state. 
In particular, the opening of the SC gap, irrespective of the pairing mechanism, suppresses decay channels for phonons with $\hbar\omega_{\mathrm{ph}} < 2\Delta(T)$, leading to a pronounced peak narrowing and a redshift relative to the normal-state phase. 
Moreover,  at the temperature for which $\hbar\omega_{\mathrm{ph}}\sim 2\Delta(T)$, the accumulation of electronic states just above the gap enhances the electron-phonon scattering, causing a transient broadening. 
This behaviour, predicted by strong-coupling theory \cite{zeyher_zpb90} and shown in~\ref{fig1:abstract}\textbf{c}, constitutes a clear fingerprint of the phonon coupling to the SC electronic excitations.

In low-$T_{\mathrm{c}}$ superconductors, only acoustic phonons at small momenta have energies comparable to $4k_{\mathrm{B}}T_{\mathrm{c}}$,  requiring   techniques alternative to Raman, which is a zero-momentum probe, to detect their SC-induced temperature evolution. Indeed, the phenomenology described in~\ref{fig1:abstract}\textbf{c} has been  observed for acoustic phonons of elemental superconductors Nb and Pb measured by inelastic neutron scattering~\cite{shapiro_measurements_1975,Keller_NRSE_PhysRevLett.96.225501}.
In contrast, in high-$T_{\mathrm{c}}$ cuprates, the energy scale $4k_{\mathrm{B}}T_{\mathrm{c}}$ is sufficiently large to match that of optical phonons, enabling these effects to be observed directly by vibrational Raman spectroscopy ~\cite{zeyher_phonon_1988, thomsen_phonon_1990, FriedlCardona_PhysRevLett_1990,cardona_raman_1991, hardy_prb93,tajima_prb00, hardy_prb04,kim_prl21}.
The even higher $T_{\rm c}$ and stronger electron-phonon coupling of hydrides suggests that Raman spectroscopy should be ideally suited to probe the SC gap formation in these systems. Yet, despite its potential, the weak signal caused by the limited light penetration in metals has so far restricted the use of Raman spectroscopy in hydrides under extreme pressures mostly to the detection of structural and compositional transitions of insulating phases \cite{drozdov_conventional_2015,drozdov_superconductivity_2019,kong_superconductivity_2021,ma_high-temperature_2022}. Achieving the sensitivity required for accurate phonon measurements in high-pressure SC hydrides demands effective suppression of the parasitic background from the DAC environment, which typically overwhelms the weak Raman signal of the metallic state.

In this study, we report high-quality Raman measurements of hexagonal LaH$_{10}$ (h-LaH$_{10}$) at 145 GPa, enabled by a custom setup designed to detect the intrinsically weak metallic hydride signal even at low temperatures. The resulting spectra are exceptionally clean, permitting direct quantitative analysis without any background subtraction. The temperature evolution of the Raman spectra reveals multiple phonon modes displaying clear anomalies at the onset of the SC transition, independently identified by simultaneous electric transport measurements.  
By combining \textit{ab initio} density functional theory (DFT) calculations of the electron-phonon interaction with a many-body Migdal-Eliashberg (ME) framework for electron-hole excitations in the SC state~\cite{zeyher_zpb90, marsiglio_phonon_1992}, we obtain an excellent agreement between theory and experiment. Together with the observation of a zero-resistance state quantitatively reproduced by a percolative transport model, these results provide a coherent picture of a strong-coupling, phonon-mediated SC transition in h-LaH$_{10}$.

\clearpage

\begin{figure}[ht]
  \centering
  \includegraphics[width = \textwidth]{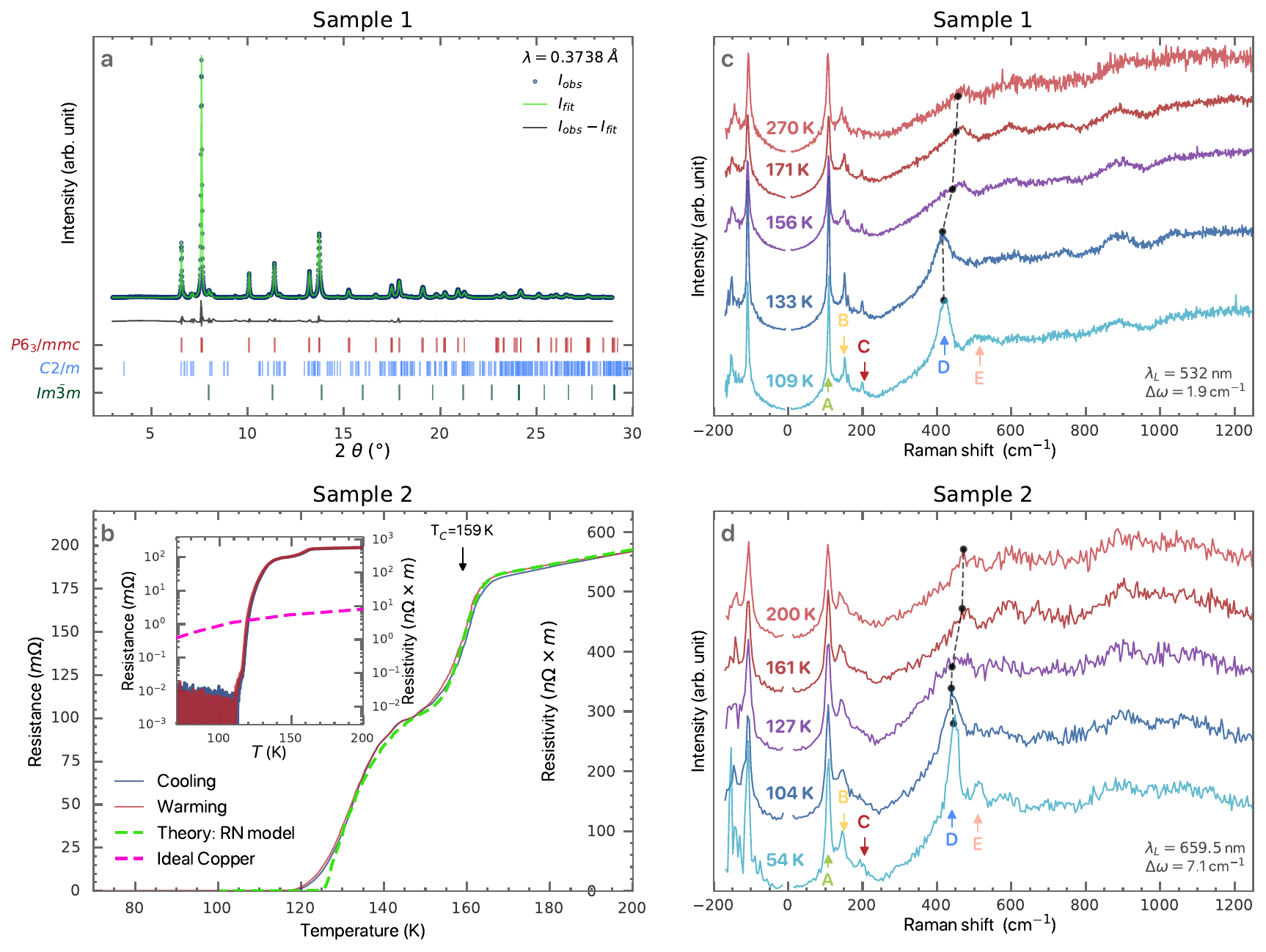}
  \caption{
  \textbf{Experimental investigation of La hydride.} 
\textbf{a}, Synchrotron X-ray diffraction of sample 1 collected at 140~GPa on beamline ID27 (ESRF) using 0.3738~\AA\ radiation focused to a 0.5~$\mu$m spot. The Le Bail fit (solid line) reveals a dominant hexagonal phase (P6$_3$/mmc) with minor cubic and monoclinic contributions, consistent with diffraction mapping. The refined volume of 34.45~\AA$^3$ per La atom agrees with previous reports \cite{laniel_high-pressure_2022, chen_high-temperature_2024}, confirming LaH$_{10}$ stoichiometry.
\textbf{b},  Electrical transport of sample 2 measured in a four-lead van der Pauw configuration \cite{Pauw_1958}. The resistance drops sharply by nearly five orders of magnitude below the SC transition, reaching values indistinguishable from zero ($2 \pm 2~\mu\Omega$). The absence of hysteresis confirms a second-order phase transition, as expected for superconductivity, while a shoulder near 140~K indicates phase coexistence. The green line shows a two-$T_{\rm c}$ percolation model fit with components  at \(159 \pm 4\,\mathrm{K}\) and \(115 \pm 15\,\mathrm{K}\) (Methods), the higher $T_{\rm c}$ matching previous reports for the hexagonal phase \cite{chen_high-temperature_2024}.
The temperature evolution of the resistivity of ideal Copper (magenta curve from \cite{matula_electrical_1979}) is presented to underline the incompatibility of the abrupt discontinuity measured in LaH$_{10}$ with a normal metal.
\textbf{c}, \textbf{d}, Raman spectra of samples 1 without leads and 2 with leads at selected temperatures. Spectra were corrected for the Bose population factors (Methods). The low-frequency sharp peaks (A-C), attributed to La vibrations, redshift and narrow on cooling. Higher frequency modes (D-E), assigned to H vibrations, also redshift but show a non-monotonic linewidth evolution, broadening just below $T_{\rm c}$ before narrowing at lower temperatures, with mode D evolving into a prominent peak below $T_{\rm c}$. All features are reproducible across thermal cycles, with pressure rising from 140 to 150 GPa (sample 1) and from 135 to 145 GPa (sample 2) during cooling.
Spectra presented in \textbf{c}  and \textbf{d} where obtained probing the system with laser of different wavelengths $\lambda_L$ and have different frequency resolution $\Delta\omega$. 
Frequency regions within $\pm$ 7~cm$^{-1}$ (\textbf{c}) and $\pm$ 12~cm$^{-1}$ (\textbf{d}) are omitted due to laser and filters artifacts.
}
  \label{fig2:exp}
\end{figure}

\section*{Experimental characterization}

h-LaH$_{10}$ samples were synthesized near 140 GPa by laser heating a layered LaH$_3$ and ammonia borane (NH$_{3}$BH$_{3}$) mixture  inside a DAC, prepared without (sample 1)  and with (sample 2) integrated leads for electrical transport measurements. Synchrotron X-ray diffraction (XRD) with sub-micron resolution identified the dominant phase as hexagonal (space group $P$6$_{3}$/$mmc$), with two La atoms per unit cell in a hexagonal close-packed arrangement. Le Bail refinement of the hexagonal phase (\ref{fig2:exp}\textbf{a}) yielded a unit cell volume of 34.45 \AA$^3$ per La atom at 140 GPa, consistent with stoichiometric LaH$_{10}$ \cite{laniel_high-pressure_2022, geballe_synthesis_2018, chen_high-temperature_2024}. Minority cubic and monoclinic domains were also detected, revealing structural heterogeneity. Full synthesis and characterization details are given in Methods.

 
Four-probe resistance measurements on sample 2 reveal a sharp and reproducible SC transition at about 165 K (\ref{fig2:exp}\textbf{b}) with negligible hysteresis, indicative of a second-order phase transition~\cite{eremets_high-temperature_2022}. The resistance decreases by nearly five orders of magnitude on cooling across the transition, from 200\,m\(\Omega\) to below the measurement threshold \( (2 \pm 2)\; \mu\Omega\), confirming a zero-resistance state (inset of~\ref{fig2:exp}\textbf{b}). The corresponding upper-limit resistivity, estimated using the van der Pauw method \cite{Pauw_1958}, is approximately $\sim10^{-11}\,\Omega\,\mathrm{m}$ at at 100\,$\mathrm{K}$ (see Methods), about two orders of magnitude lower than that of copper~\cite{matula_electrical_1979}. At 200 K, the resistivity is conversely about 2 orders of magnitude higher than that of copper, a temperature dependence inconsistent with conventional metallic behaviour, thereby supporting the identification of a genuine SC transition and ruling out alternative explanations~\cite{hirsch_electrical_2023}.
A closer inspection of the temperature dependence of the resistivity reveals two distinct steps, consistent with inhomogeneous superconductivity and percolative transport. This behaviour is well captured by a random resistor network (RRN) model~\cite{caprara_effective_2011}, in which local resistors \( R_i(T) \) become superconducting at temperatures \( T_{\rm c}^i \) and are distributed with probability \( P_i \) (see Methods). Effective-medium analysis yields two Gaussian components centred at \(159 \pm 4\,\mathrm{K}\) and \(115 \pm 15\,\mathrm{K}\) (\ref{fig2:exp}\textbf{b}).
The higher $T_{\rm c}$ value agrees with previous reports for h-LaH$_{10}$~\cite{chen_high-temperature_2024}, and is lower than that of the cubic phase (c-LaH$_{10}$), which is stable above 160\,GPa~\cite{sun_high-temperature_2021}.

The Raman spectra display spatial variations reflecting local crystallinity and phase purity (see \ref{fig:Sample_1_after}), 
in line with the multi-phasic character inferred from the diffraction mapping and transport measurement. In the following, we focus on the spectra from micro-metric regions showing a well-resolved low-frequency multiplet below 250 cm$^{-1}$, which we identify as a structural fingerprint for a well-crystallized h-LaH$_{10}$ phase.  
Spectra, collected with different laser wavelengths, across multiple thermal cycles, on different samples, exhibit reproducible features, confirming the same dominant phase. The raw intensities were corrected for the Bose population factors of both Stokes and anti-Stokes sides. The resulting spectra are symmetric about the Rayleigh line, indicating negligible background contributions from diamond or sample fluorescence. The correction also enables a direct determination of the sample temperature, thereby accounting for potential laser-induced heating, which is estimated to remain below $<20$ K for 5 mW excitation power.

Three sharp and well-defined low-frequency peaks are observed below 250\,cm\(^{-1}\) (A, B, C in~\ref{fig2:exp} \textbf{c}, \textbf{d}). Their frequencies fall within the range of the acoustic branches of the cubic phase~\cite{errea_quantum_2020}, suggesting vibrations of the La sublattice. The presence of at least three distinct peaks, however, contradicts the single Raman-active La mode allowed in $P$6$_3$/$mmc$, implying a lower-symmetry structure. First-principles calculations, guided by symmetry analysis, identify a dynamically stable rearrangement of the hydrogen sublattice forming a $P$6$_3$/$m$, 2$\times$2$\times$1 superstructure. Although indistinguishable from $P$6$_3$/$mmc$ in powder X-ray diffraction—due to minimal La displacement and weak H scattering—the $P$6$_3$/$m$ model accurately reproduces the observed low-frequency Raman features, making it the most plausible structural candidate (see Methods and Supplementary Information).
Upon cooling below $T_c$, peaks A–C (\ref{fig2:exp}) undergo a substantial linewidth narrowing, consistent with a suppression of scattering channels in the SC state and confirming their phononic nature. At higher frequencies, broader peaks D and E, (\ref{fig2:exp}) associated with hydrogen vibrations \cite{errea_quantum_2020} display a pronounced temperature dependence (\ref{fig:raman_el_waterfall}). Peaks D and E are already visible above $T_c$ but become considerably sharper and more intense below it, peak D evolving into a particularly prominent feature upon cooling. Both exhibit a marked redshift and a non-monotonic linewidth evolution across the SC transition. This behaviour contrasts with the blueshift expected for the electronic pair-breaking peak at $2\Delta(T)$ (\ref{fig1:abstract}\textbf{b}) and instead matches the phonon temperature evolution predicted for modes strongly coupled to the electronic continuum (\ref{fig1:abstract}\textbf{c}).

\clearpage

\section*{Data analysis and theoretical interpretation}
\begin{figure}[ht]
  \centering
  \includegraphics[width = \textwidth]{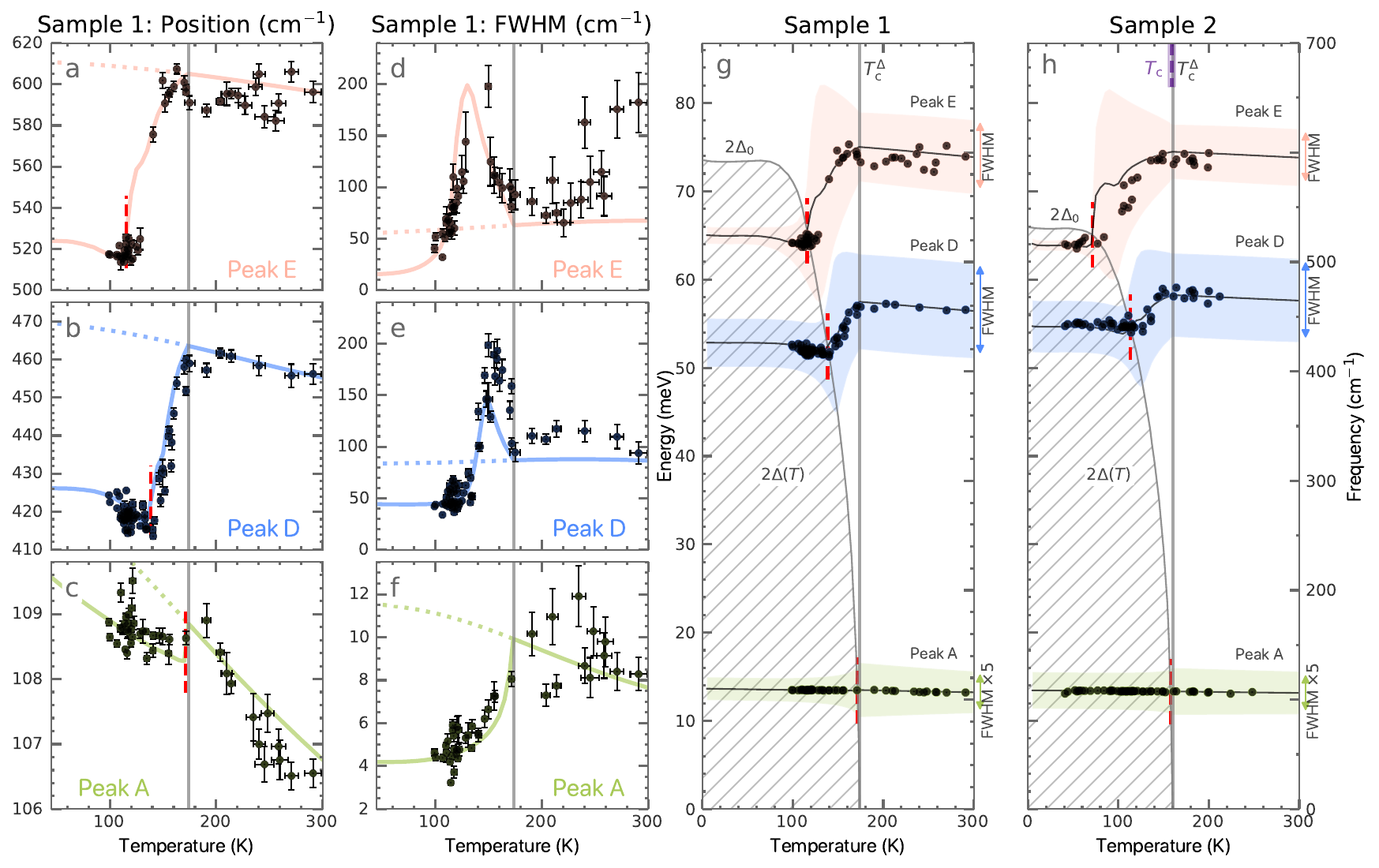}
  \caption{\textbf{Raman signatures of superconductivity.}
Positions (\textbf{a-c}) and linewidths (\textbf{d-f}) of the most pronounced peaks A, D, E. 
Scatter points illustrate the experimental data as extracted by the multipeaks procedure detailed in the Methods, whose error bars report the fitting uncertainties.
The grey vertical line at 174 K marks the critical gap temperature $T^{\Delta}_{\rm c}$  where all phonons exhibit an abrupt change in their temperature evolution. Solid lines represent the theoretical model (with $r_{\omega}=1$), illustrating its ability to capture the characteristic temperature evolution observed in the experimental spectra. Dashed lines indicate the expected behaviour of the normal metallic state, evaluated within the same theoretical framework.
\textbf{g}, \textbf{h} depict the interplay between the SC condensate (hatched region up to the pair-breaking energy $2\Delta(T)$) and key spectral features for sample 1 and 2, respectively. Experimental peak positions (circles) overlay theoretical frequencies (solid lines). Theoretical linewidths are represented by coloured shaded areas. 
In \textbf{h}, the dashed purple tick labelled $T_{\rm c}$ marks the critical temperature obtained from the transport measurement in sample 2.
Red dashed vertical lines in \textbf{a-c, g-h} indicate the temperatures at which the calculated phonon energy intersects the optical gap energy $2\Delta(T)$, nearly coinciding with the temperature where the phonon position is most strongly renormalized by electron–phonon interactions. Consequently, for each phonon mode, the superconducting gap magnitude at the crossing temperature can be experimentally inferred from the phonon peak position at its maximum redshift. 
}\label{fig3:raman_analysis}
\end{figure}

The temperature evolution of the Raman resonances was quantified by a multi-peak fit. 
As shown in  \ref{fig3:raman_analysis}~\textbf{a-f}, the observed positions and linewidths of  all three peaks A, D, E present a remarkable non-monotonic $T$-dependence with a slope-discontinuity (a kink) at the same temperature, $T^{\Delta}_{\rm c}=174$ K for sample 1 and 160 K for sample 2.  Such a behaviour is captured by 
the ME description of the phonon spectral function \cite{marsiglio_phonon_1992,zeyher_zpb90}, which predicts, for all modes coupled with Fermi-surface electrons, a kink in the phonon properties at the critical temperature where superconductivity forms. This interpretation is also supported by the fact that, in sample 2, the value of $T^{\Delta}_{\rm c}$ obtained from Raman measurements is in excellent agreement with the $T_{\rm c}=159$ K obtained from the RRN analysis of the resistivity of the same sample.

To use the ME approach \cite{zeyher_zpb90,marsiglio_phonon_1992},
we need to first model the SC gap formation at the electron Fermi-surface. This requires the knowledge of the Eliashberg spectral function for h-LaH$_{10}$, $\alpha^2F(\omega)$, that describes the electron-phonon interaction mediating the SC Cooper pairing. The large unit cell of the $P6_3/m$ superstructure makes unfeasible a direct DFT calculation of this quantity.
However, we can take advantage from the observation that the SC properties are mostly sensitive to the total coupling $\lambda=2\int_0^{\infty} d\omega \alpha^2F(\omega)/\omega$ and to the  phonon-mediator energy-scale $\omega_{\rm log}$,  being other finer details of the Eliashberg function less important~\cite{allen_theory_1983}. 
We thus adapt the function previously computed for the much smaller unit-cell of c-LaH$_{10}$, $\alpha^2F_{\rm cub}(\omega)$ \cite{errea_quantum_2020}, to construct the one of the h-LaH$_{10}$, by rescaling $\lambda$ and assuming that the phonon energies of the two phases are similar, so that $\alpha^2F(\omega)=(\lambda/\lambda_{\rm cub})\alpha^2F_{\rm cub}(\omega)$. We then obtain the value of $\lambda$ by imposing that the ME $T^{\Delta}_{\rm c}$ coincides with the one extracted from the Raman spectra. 

To simulate the vibrational peaks, for each phonon mode $\mu$ we need also to determine 3 
parameters: the bare phonon frequency $\omega^{\rm dyn}_{\mu}$, the static phonon frequency $\omega^{\rm stat}_{\mu}$, whose departure from $\omega^{\rm dyn}_{\mu}$ quantifies the coupling of the mode to the Fermi-surface electrons, and $\Gamma^{\rm intr}_{\mu}$, which accounts for the linewidth contribution arising from residual scattering process. 
For the D and E mode we obtain good results with $T$-independent parameters. Instead, for the low-energy mode A, we have also to add to both $\omega^{\rm dyn}_{\mu}$ and  $\omega^{\rm stat}_{\mu}$ a frequency shift varying as $-c\times T$. Such extra dependence, not associated with electron-phonon interaction, is tiny and visible only in the A peak, which has $T$-line-shifts that are one order of magnitude smaller than those of the D and E peaks.
We identify optimal values of the mode parameters, achieving a remarkable agreement at all temperatures  with the experimental data, as shown in~\ref{fig3:raman_analysis}~\textbf{a-f} for both positions and linewidths.  Finally, for the A mode, because of the small anharmonic zero-point fluctuation of the heavy La atoms, it is also possible to obtain  $\omega^{\rm stat}_{\mu}$ and $\omega^{\rm dyn}_{\mu}$ from first principles in the $P6_3/m$ h-LaH$_{10}$ superstructure. The computed DFT values nicely agree with those fitted on experiments, see Table~\ref{tab:dfpt_phononfreq_mtds}.

The quality of the ME predictions proves the validity of the theoretical framework. However, this result alone does not yet exclude an unconventional nature of the SC pairing interaction.
Indeed, the ME applicability extends well beyond the original picture of lattice-mediated Cooper pairing for which it was developed, and has inspired numerous efforts to describe SC arising from alternative interactions, as e.g. the exchange of collective electronic excitations, such as antiferromagnetic fluctuations in cuprates \cite{scalapino_d-wave_1986,monthoux_toward_1991,monthoux_spin-fluctuation-induced_1992,monthoux_weak-coupling_1992}). 
Nonetheless, different mediating mechanisms act over distinct energy scales. 
Thus we test the influence of the mediator energy $\omega_{\rm log}$ in the ME simulation of the Raman spectra by modifying its value with respect to that computed by DFT for phonon of c-LaH$_{10}$, $\omega^{\rm cub}_{\rm log}=597$ cm$^{-1}$. Namely we assume in the ME approach that $\omega_{\rm log}=r_{\omega}\omega^{\rm cub}_{\rm log}$ and $\alpha^2F(\omega)=(\lambda/\lambda_{\rm cub})\alpha^2F_{\rm cub}(\omega/r_{\omega})$. For each value of the scaling factor $r_{\omega}$,  $\lambda$ is chosen to match the experimental $T^{\Delta}_{\rm c}$. A reduced (enhanced) $\omega_{\rm log}$ requires larger (smaller) coupling amplitude $\lambda$ to obtain the same $T^{\Delta}_{\rm c}$. This reflects in stronger (weaker) bonding of Cooper pairs, as exhibited by the larger (smaller) zero-$T$ gap energy, $\Delta_0$, see \ref{fig:rescale_vs_sc}. Since the D and E phonon peaks have energy close to $2\Delta_0$,
their simulated $T$-evolution is very sensitive to the mediator energy scale, see \ref{fig:energy_scale}. 
The best agreement is obtained with values of  $r_{\omega}$ between $0.85$  and $1$, namely with an energy scale which is very close to that computed for phonons in the c-LaH$_{10}$ structure. Correspondingly, we obtain that  $2.2<\lambda<2.6$, and  $73~\mathrm{meV}<2\Delta_{0}<78~\mathrm{meV}$.
This result clearly indicates that superconductivity in h-LaH$_{10}$ originates from a conventional phonon-mediated pairing in a strong coupling regime.

\subsection*{Outlook}

The combined Raman, X-ray diffraction, and transport measurements presented here provide a comprehensive characterization of hexagonal LaH$_{10}$ as a phonon-mediated superconductor, and reveal the remarkable potential of Raman spectroscopy for addressing key challenges in the quest for room-temperature superconductors. 
Indeed, a very high SC transition temperature, $T_{\rm c}$, is particularly favourable for Raman investigations. The strong pairing interaction responsible for the high $T_{\rm c}$ suppresses the optical conductivity in the visible range, increasing the laser penetration depth, an otherwise limiting factor for Raman spectroscopy in weakly-correlated metals.  A large electron-phonon interaction assures a remarkable dependence of the vibrational peak on electronic phase-transitions. The presence of multiple vibrational modes with energies comparable with a large SC gap,  which is proportional to $T_{\rm c}$, offer an ideal platform to scan spectroscopically the presence of the gap and its temperature evolution.  Finally, the Raman micrometric spatial resolution is invaluable not only in extreme-condition environments, but also for the detection of very high-$T_{\rm c}$ superconductivity in small or heterogeneous materials at ambient-pressure.

\clearpage
\section*{Methods}

\subsection*{Sample Preparation and Synthesis}
Symmetric diamond anvil cells (DACs) were utilized to generate the high pressures required for this study. The DACs were equipped with type-IIas Boehler-Almax (BA) diamonds with 100 $\mu$m culets and bevelled to 300 $\mu$m at 8.5$^{\circ}$, the design allowed for an opening of approximately 60$^{\circ}$ (4$\theta$). 
 The force was applied to the DAC by screws or by a metallic membrane. The sample pressure was determined by the calibrated shift of stressed edge of the Raman diamond phonon\cite{akahama_pressure_2006}. The composite gasket was obtained by indenting a 250 $\mu$m thick Re foil to about 25 $\mu$m thickness and drilling a hole at the center of the indent of about 75 $\mu$m of diameter, subsequently filled with a mixture of ceramic (MgO or BN) and epoxy, finally drilled in the center with a $\sim$50 $\mu$m diameter hole, as illustrated in \ref{fig:Sample_1_after}. The ceramic mixture ring provides an electrical insulating gasket around the sample. Four platinum electrical leads, with tips tapered to about 1 $\mu$m thickness have been placed on one diamond in a Van der Pauw arrangement for resistance measurements. 

 Commercial precursors lanthanum trihydride (LaH$_3$) and ammonia borane (AB) were loaded into the sample chamber under an inert argon atmosphere. A LaH$_3$ flake was first pre-pressed between the diamond anvils to form a layer approximately 30 × 30 $\mu$m in size and about 1 $\mu$m thick. This LaH$_3$ layer was placed on the diamond culet, and on the electrical leads when present, and then pressed by a flake of AB, with both materials confined inside the gasket hole that defines the sample chamber. This configuration ensures that LaH$_3$ remains in direct contact with one diamond, allowing clean Raman measurements from that side, while AB, in contact with the opposite diamond, enables efficient indirect laser heating during synthesis.

LaH$_3$ and AB were compressed at room temperature up to pressures ranging from 120 to 140 GPa. The release of molecular hydrogen from AB and the synthesis of high-stoichiometry La-hydrides were achieved through laser heating. Laser heating was conducted using 10-15 W of a 1064 nm Nd:YAG laser, with the beam focused onto the sample via a near-infrared (NIR) 15X microscope objective, producing a heating spot approximately $5\mu$m in diameter. The NIR laser coupled directly to the metallic LaH$_3$ through the AB, and was systematically rasterised across the entire sample. Although temperatures were not measured using optical pyrometry, the visible emissions suggest that temperatures exceeded 1000 K. 
 The molecular hydrogen released by the heated AB was readily absorbed by the  LaH$_3$ layer increasing its hydrogen content, which could be appreciated visually by its observed volume expansion.
A motorized translation stage enabled automated switching between Raman measurement and laser heating, facilitating rapid assessments of hydrogen release from ammonia borane and the transformation of the sample.

\subsection*{Synchrotron X-ray diffraction}
X-ray Diffraction was carried out at the ID27 Beamline at the European Synchrotron Radiation Facility (ESRF) and ID15U at the Shanghai Synchrotron Radiation Facility (SSRF). The incident X-ray spot was $0.3738$\,\AA$ $ ($0.6199$\,\AA) in wavelength and around $0.8$\,$ \mu$m ($2$\,$ \mu$m) in diameter in ESRF (SSRF). The resultant diffraction was imaged in ESRF using the Eiger2 CdTe 9M detector (MAR detector in SSRF) and  the pattern was then integrated into a 1D dataset using Dioptas software \cite{prescher_dioptas_2015}. Subsequent Le Bail refinements \cite{le_bail_ab-initio_1988} were then carried out using the Jana package using well structurally characterized La-hydrides \cite{laniel_high-pressure_2022}.

\subsection*{Low-temperature Raman spectroscopy}
Raman experiments were conducted using a custom confocal optical system integrated with an optical cryostat. Raman excitation was achieved using diode-pumped solid-state (DPSS) lasers with wavelengths of 532 nm and 660 nm. The laser beam was focused to a spot size of approximately   $\sim$2\,$\mu$m  using a long working distance microscope objective with 20$\times$ magnification. To minimize sample heating during measurements, the incident laser power was maintained below 5 mW for both excitation wavelengths. Due to the inherently weak Raman signal characteristic of these materials, spectra were acquired with exposure times of up to 5 minutes. Backscattered light was filtered using three ultra-narrow notch filters and dispersed with an aberration-corrected spectrometer equipped with a back-illuminated, deep-depletion, liquid nitrogen-cooled CCD detector. Measurements were performed on the side of the sample in direct contact with the diamond, avoiding spurious signals originating from AB.

Low-temperature experiments were conducted using a custom optical cryostat, with the cold finger internally cooled by liquid nitrogen (L-N$_2$) or liquid helium (L-He). The cryogens were transported from the dewar through a low-loss transfer line and extracted using a diaphragm pump acting on the exhaust vapours. Temperature monitoring and control were achieved using two silicon diode sensors providing feedback to a temperature controller. One diode was mounted on the cryostat's cold finger, while the second was mechanically fixed to the diamond anvil cell (DAC) body, positioned near the sample chamber. To mitigate errors in transport measurements caused by temperature gradients, very slow cooling/warming rates were employed typically $<$1 K/min. For Raman measurements, the system was allowed to stabilize at the target temperature before data collection.
 The sample pressure has been estimated by the Raman shift of the stressed edge of the diamond peak using, as illustrated in \ref{fig:pressure}.

Raman measurements were performed using a 320 mm focal length spectrograph and a laser with energy $\hbar \omega_{\rm L}$ of 2.329 eV (532 nm) and either a 300 gr/mm or a 1200 gr/mm grating on sample 1 and with $\hbar \omega_{\rm L}=1.88$ eV (660 nm) and  300 gr/mm grating in sample 2. The intensities measured using the 1200 gr/mm grating are calibrated by dividing for a correction factor $[1+5.639(\omega/\omega_{\rm L})]$, where $\omega$ is the Raman shift and $\omega_{\rm L}$ is the laser frequency for the 532 nm laser.  No correction was required for the other grating.  The Raman raw intensity $S^{\rm raw}(\omega)$ is modulated by the Bose population factor $n(\omega,T)=(e^{\hbar\omega/k_{\rm B} T} -1)^{-1}$. Stokes ($\omega>0$) and anti-Stokes ($\omega<0$) intensities scale with $n(\omega, T) + 1$ and $n(|\omega|, T)$, respectively, enhancing the signal and distorting lineshape profiles especially at low frequencies. The Raman intensity has also a dependence on the scattered photon energy, scaling as $\omega_{\rm s}^3=(\omega_L-\omega)^3$ when using photon counting detectors \cite{kauffmann_use_2019}. 
The population corrected Stokes and anti-Stokes intensities then result:
\begin{equation}
S^{\rm corr}_{\delta,\beta}(\omega,T) = \frac{S^{\rm raw}(\omega+\delta) - \beta}{[n(|\omega|,T) + \theta(\omega)]\omega_{\rm s}^3}
 \label{eq:raw_spectrum_corr},
\end{equation}
where $\theta(\omega)$ is the theta function, equal to 1 and 0 in the Stokes and anti-Stokes side respectively, $\delta$ and $\beta$ are small corrections taking into account a calibration offset and the presence of small parasitic-light background.
The resulting Stokes and anti-Stokes sides of $S^{\rm corr}_{\delta,\beta}(\omega,T)$ are perfectly symmetric around the Rayleigh line for the exact sample temperature, allowing in this way to take into account the potential laser-induced heating during Raman acquisition. 
For each spectra the three parameters $\{\delta,\beta,T\}$ were determined through a fitting procedure that minimizes the quadratic difference between the Stokes and anti-Stokes components in the low-energy region (15–250 cm$^{-1}$).
We confirmed a posteriori that the parasitic background $\beta$ results in small corrections, being at maximum 50 photon counts over several thousands at the peak maximum. 
Subsequently, a multi-peak fitting analysis on the Stokes side ($75<\omega <1000 $ cm$^{-1}$), for sample 1, and $75<\omega <1100 $ cm$^{-1}$, for sample 2), \ref{fig:raman_ind_sample1}, was conducted to obtain from $S^{\rm corr}(\omega)$ a quantitative description of the peak positions and linewidths as a function of temperature with the following target function:
\begin{align}
 f_{\rm fit}(\omega) &= a + b \frac{2\gamma\omega}{\omega^2+4\gamma^2 } + C \frac{\mathcal{G}_{\Omega
 \Gamma}(\omega) -\mathcal{G}_{\Omega 
 \Gamma}(-\omega) }{\omega}
 + \sum_{i=1}^{N_p}c_i \left[ \mathcal{V}_{\omega_{i}\Gamma_{i}}^{\Gamma_{\rm res}}(\omega) - \mathcal{V}_{\omega_{i}\Gamma_{i}}^{\Gamma_{\rm res}}(-\omega) \right]
 \label{eq:fit_fx}
\end{align}
where $\mathcal{G}_{\Omega \Gamma}(\omega)$ denotes a broad Gaussian distribution centred at $\Omega$ with a FWHM of $\Gamma$ and
$\mathcal{V}_{\omega_{i}\Gamma_{i}}^{\Gamma_{\rm res}}(\omega)$
 is a narrow Voigt profile centred at $\omega_i$, with a Lorentzian FWHM $\Gamma_i$ and a Gaussian FWHM $\Gamma_{\rm res}$ that accounts for the   instrumental resolution.
All parameters in Eq.~\eqref{eq:fit_fx} are fitted simultaneously, except for $\Gamma_{\rm res}$, which is independently determined from the Neon Raman lines. We estimate $\Gamma_{\rm res}$ as 7.1~cm$^{-1}$ (or 2.0~cm$^{-1}$) when using a 660~nm laser with a 300~gr/mm (or 1200~gr/mm) grating, and as 11.8~cm$^{-1}$ when employing a 532~nm laser with a 300~gr/mm grating.
The fitting parameters $\{a,b,\gamma\}$ in Eq.~\eqref{eq:fit_fx} describe a smooth baseline that accounts for the contribution of H-dominated vibrations that can not be resolved into narrow peaks, because of a too strong interaction/disordered broadening, as well as for any residual electronic background.
The parameters $\{C,\Omega,\Gamma\}$ model the broad low-frequency (0–250 cm$^{-1}$) component associated with La vibrations of aperiodic or  poorly crystallized phases.
The number of fitted peaks, $N_p$, is determined for each spectrum according to the individual peak visibility and is chosen to maximize the number of identifiable spectral features included in the analysis.

\subsection*{Electrical transport measurements}
Four platinum probes were placed in a Van der Pauw configuration \cite{Pauw_1958}, enabling the estimation of sheet resistance. A Keithley 6221 current source and a Keithley 2182A nano-voltmeter were employed in delta mode, which automatically alternates the current polarity and measures the voltage at each polarity. This technique effectively cancels constant thermoelectric offsets, ensuring accurate voltage readings and reducing substantially the noise compared to standard methods, a crucial improvement when measuring the very low resistance values characteristic of superconducting materials. 
The voltage measured in response of a 1 mA current in sample 2 reached almost the minimum value detectable by the Keithley 2182A in delta measurement mode.
The resulting positive-defined resistance below 110 K observed in the warming (cooling) hysteresis cycle fluctuates around 2 (5) $\mu\Omega$ with an error margin around 2 (4) $\mu\Omega$. Furthermore, an estimate of the resistivity of the sample using the Van Der Pauw formula \cite{Pauw_1958} ($\rho=2/ln(2)\cdot d\cdot R$) assuming a two dimensional homogeneous sample of 1 $\mu$m thickness provides an upper limit value of about 10$^{-11}$ $\Omega$m. This value is about two orders of magnitude smaller than copper at the same temperature (at 110 K $\rho = 3 \times 10^{-9}$ $\Omega$m \cite{matula_electrical_1979} ), ruling out a possible objection \cite{hirsch_electrical_2023} that the observed resistance drop is not an indication of truly zero resistance state. 
 
\subsection*{Structural analysis} 

The structural landscape of lanthanum hydrides at mega-bar pressures has been investigated both experimentally~\cite{laniel_high-pressure_2022,geballe_synthesis_2018, chen_high-temperature_2024} and theoretically~\cite{peng_hydrogen_2017,shipley_stability_2020,errea_quantum_2020}. The superconducting phase most consistently identified as stable adopts the high-symmetry cubic $Fm\bar{3}m$ space group~\cite{drozdov_superconductivity_2019,liu_dynamics_2018}. It features a face-centered cubic (fcc) closed-packed lattice of La atoms arranged in ABC-stacked triangular layers along the [111] direction (\ref{fig:structure}\textbf{a}). Hydrogen atoms occupy tetrahedral and octahedral interstitial positions realizing a cage around each La atom that can be seen as a chamfered cube, a highly symmetric polyhedron with six squares and twelve flattened hexagons having internal angles of $109.47^\circ$ and $125.26^\circ$ (\ref{fig:structure}\textbf{e}). Several lower-symmetry structures, including rhombohedral ($R\bar{3}m$), orthorhombic ($Immm$), and monoclinic ($C_2$, $C_2/m$) phases, have also been proposed~\cite{errea_quantum_2020,sun_high-temperature_2021,eremets_high-temperature_2022}, all preserving the ABC stacking of La layers. These phases can be interpreted as a symmetry-lowering deformation of the cubic LaH$_{10}$ (c-LaH$_{10}$) primarily due to a rearrangement of hydrogen atoms.

A fundamentally different and competing phase with hexagonal $P6_3/mmc$ symmetry has also been proposed as thermodynamically stable and hosting a high-$T_{\rm c}$ superconducting phase \cite{shipley_stability_2020}. Signatures of this structural phase have been reported as impurities in otherwise cubic structures \cite{drozdov_superconductivity_2019,sun_high-temperature_2021}, whereas Al doping has recently been shown to stabilize the metastable h-LaH$_{{10}}$ \cite{chen_high-temperature_2024}. 
In this structure La atoms form an hexagonal closed-packed (hcp) lattice where La triangular layers display a different ABAB stacking along the hexagonal $c$ axis, that can be related to the fcc structure by a 3-fold rotation of the C layer about an axis orthogonal to the triangular layers and passing through a La atom (see \ref{fig:structure}\textbf{a,b}). 
Although the $P6_3/mmc$ phase displays the largest number of symmetries within the hexagonal crystal systems, the hydrogen cage around each La atom appears to be substantially distorted. It can be interpreted as a deformation of the chamfered cube where the lower half of the polyhedron with respect to the triangular-layer plane is rotated by 120$^\circ$ about a perpendicular axis passing through the La atom at the centre of the hydrogen cage (\ref{fig:structure}\textbf{f}).
In close analogy with the several proposed modifications of the fcc phase, one can anticipate symmetry-lowering rearrangements of H atoms for the hcp phase also, that have not been explored yet to the best of our knowledge. This expectation is confirmed by the structured Raman features observed between 100–200 $cm^{-1}$, that are incompatible with the single Raman-active La mode allowed in the $P6_3/mmc$ phase. 

To identify the possible symmetry-lowering structural distortion, we first optimized the $P6_3/mmc$ structure reported in \cite{shipley_stability_2020} at a classical pressure of 135 GPa, as quantum effects have been shown to add around 10-15$\%$ of extra pressure \cite{errea_quantum_2020,errea_quantum_2016,belli_2022} (see Supplementary Information for details). The calculated lattice parameters, reported in \ref{tab:dft_structural_parameters}, are within 1$\%$ error with respect to the refined XRD lattice constants, whereas the enthalpy is 4.1 meV/f.u. above fcc phase at same pressure. The phonon calculation reveals several dynamically unstable phonon modes (\ref{fig:lah10_194_phdispersion}). Projection onto atomic character clearly relates them to light H atoms, while La atoms mostly contribute to phonon modes in the 0--250 cm$^{-1}$ range. The only Raman-active La mode in this range is the $E_{2g}$ ($\Gamma_5^{+}$) one at approximatively 90 $cm^{-1}$, with the other optically-inactive $B_{1g}$  ($\Gamma_4^{+}$) mode at $\sim 250$ cm$^{-1}$. 
The most unstable hydrogen phonon is a $M_2^{+}$ mode at $M$ point of the $P6_3/mmc$ Brillouin zone, that could lead to a 2$\times$1$\times$1 reconstruction with orthorhombic $Pnma$ structure or to a 2$\times$2$\times$1 hexagonal $P6_3/m$ phase comprising eight formula units. We discard the 2$\times$1$\times$1 orthorhombic reconstruction as it will turn on the Raman activity of the zone-centre $B_{1g}$ mode, falling outside the experimental frequency range where La phonons are observed. Following the symmetry-lowering phonon mode $M_2^{+}$ we therefore optimized the hexagonal $P6_3/m$ structure and found an enthalpy gain of -2.8 meV/f.u. over c-LaH$_{10}$. In the reconstructed cell La atoms undergo very tiny displacements (at most of 4$\times 10^{-3}$ \AA) from the ideal positions of the $P6_3/mmc$ phase, while hydrogen displacements are one order of magnitude larger. The zone-centre $E_{2g}$ mode of undistorted h-LaH$_{10}$ is blue-shifted to $\sim$102 cm$^{-1}$ in the lower-symmetry structure, while other zone-folded phonon modes appear to become Raman-active in the frequency range between 100 and 200 cm$^{-1}$ in clear agreement with the experimental spectra, as shown in  \ref{fig:main_reconstruction}. More details can be found in the supplementary material.

\subsection*{Random-resistor network analysis of the resistance}
To analyse the temperature dependence of the resistance shown in \ref{fig2:exp} we implemented an effective-medium solution \cite{kirkpatrick73} of a random-resistor-network model. As discussed e.g. in \cite{caprara_effective_2011}, this approach successfully accounts for the experimental observation reported in several classes of materials, where the rounding of the resistive transition from the metallic to the SC state cannot be simply attributed to paraconductivity effects. We can then model the system as a network of local resistors $\rho_i$ that undergo a SC transition at a local transition temperature $T_{\rm c}^i$, such that $\rho_i(T)=\rho_0\theta(T-T_{\rm c}^i)$. If we denote with $P_i$ the probability occurrence of the local $\rho_i$ the global network resistance $\rho$ at each temperature is solution of the self-consistent equation \cite{caprara_effective_2011}:
\begin{equation}
    \sum_i P_i \frac{\rho-\rho_i}{\rho+(d-1)\rho_i}=0
\end{equation}
where $d$ is the sample dimensionality. As the temperature starts to decrease below the highest $T_{\rm c}^i$ of the network, some resistors undergo a SC transition (so that $\rho_i=0$) and $\rho(T)$ starts to decrease with respect to the normal-state value $\rho_0$. One can easily show \cite{caprara_effective_2011} that the global transition temperature $T_{\rm c}$ of the network occurs when a fraction $1/d$ of the network undergoes a SC transition, corresponding to a percolation threshold of $1/3$ in $d=3$. To account for the two-step transition observed in our sample (see \ref{fig2:exp}\textbf{b}), we modelled the probability distribution of the local $T_{\rm c}$ with two Gaussian components:
\begin{equation}
    P_i=w_h \frac{e^{-(T_{\rm c}^i-T_h)^2/2\sigma_h^2}}{\sqrt{2\pi}\sigma_h}+w_l \frac{e^{-(T_{\rm c}^i-T_l)^2/2\sigma_l^2}}{\sqrt{2\pi}\sigma_l}
\end{equation}
where $T_h=159$ K, $\sigma_h=3.5$ K are the parameters of the high-temperature phase, with relative weight $w_h=0.13$, while $T_l=115$ K, $\sigma_l=15$ K and $w_l=0.87$ are the parameters of the low-temperature phase. Finally, we phenomenologically reproduced the temperature dependence of the resistance in the normal state near to the transition temperature as $\rho_0(T)=a+T/b$, where $a=0.11$ Ohm and $b=2300$ K/Ohm. The result of the fit is shown as a dashed line in  \ref{fig2:exp}\textbf{b}.

\subsection*{Theoretical analysis of the Raman spectra}

The Raman spectra in metals exhibit, for values of Stokes shift $\omega$ in the  far IR region, both a vibrational  $S^{\rm vib}(\omega)$ and an  electronic $S^{\rm el}(\omega)$ contribution. 
The vibrational contribution within the Placzek approximation \cite{cardona_raman_1999}, in the Stoke side, is given by
\begin{equation}
S^{\rm vib}(\omega) =\frac{n(\omega)+1}{\omega}
  \sum_{\mu}^{\rm modes} K^{\rm vib}_{\mu}A_{\mu}(\omega),
  \label{eq:vib_raman}
\end{equation}
where  $n(\omega)$ is the Bose occupation factor, $\mu$ indicates the optical phonon branch, $K^{\rm vib}_{\mu}$ is the phonon Raman activity, which weakly depends on $\omega$, and $A_{\mu}(\omega)$ is the $\mu$ phonon spectral function at zone centre, that can be defined 
via the retarded (mass-reduced) displacement-displacement phonon  Greens' function $D_{\mu}(z)$ as (see Supplementary Section \ref{si:dyn_mat_bubble}):
\begin{align}
  A_{\mu}(\omega) & = -\lim_{z\rightarrow \omega+i0^+}\Im \frac{2z}{\pi } D_{\mu}(z).
  \label{eq:main_spectral_fx}
\end{align}
Here we assume that
the phonon normal-mode eigenvectors are independent on $z$ and on temperature, namely that they diagonalize both the ``bare'' dynamical matrix with eigenvalues $(\omega^{dyn}_{\mu})^2$ and the phonon self-energy with eigenvalues $\Sigma_{\mu}(z)$, so that:
\begin{equation}
  D_{\mu}(z) = \frac{1}{z^2 - (\omega^{\rm dyn}_{\mu})^2 - \Sigma_{\mu}(z)}.
  \label{eq:dphonon}
\end{equation}
The ``bare" phonon frequencies $\omega^{\rm dyn}_{\mu}$ are obtained by screening the nuclear displacements only by electron-hole interband excitations (i.e. those away from the Fermi level). We adopt   a ``dyn'' superscript following the notation of \cite{marchese_born_2024}. In insulators, there are not additional low-energy electronic contributions. On the other hand, in metals also the intra-band electron-hole excitations at the Fermi level dress the phonon frequency and their contribution is described by the self-energy $\Sigma_{\mu}(z)$ of Eq.\ \eqref{eq:dphonon}. Its real part shifts the position of the maximum of the phonon spectral function $A_\mu(\omega)$ as compared to the insulating case, while its imaginary part gives rise to a finite linewidth broadening.

We describe the self-energy in both the normal and superconducting state by using the ``Dressed Bubble" approach developed in \cite{zeyher_phonon_1988,marsiglio_phonon_1992}, where the Fermi- surface electrons are described within the Midgal-Eliashberg framework, which accounts for both impurity and electron-phonon scattering. We can then define:
\begin{equation}
\label{eq:defs}
  \Sigma_{\mu}(z) =  [(\omega^{\rm stat}_{\mu})^2-(\omega^{\rm dyn}_{\mu})^2]I(z).
\end{equation}
Here the strength of the coupling of the $\mu$-mode with the electrons at the Fermi-surface is parametrized  by the difference:
\begin{equation}
(\omega^{\rm dyn}_{\mu})^2-(\omega^{\rm stat}_{\mu})^2=\frac{2}{N_k}\sum_{{\bf k}i}\left|\left\langle {\bf k}i\left|\frac{\partial V}{\partial{u}_{\mu}}\right|{\bf k}i \right\rangle\right|^2
\delta(\varepsilon_{{\bf k}i}-\varepsilon_{\rm F})>0,\label{defomegastatdyn}
\end{equation}
 where $|{\bf k}i\rangle$ and $\varepsilon_{{\bf k}i}$ are the (non-interacting) electronic Bloch  wave-function and energy of band $i$, $\varepsilon_{\rm F}$ is the Fermi energy, $N_k$ is the number of the k-points, and ${\partial V}/{\partial{u}_{\mu}}$ is the derivative of the electron Hamiltonian with respect to the $\mu$ phonon displacement.
The dimensionless complex function $I(z)$ is the Fermi-surface averaged polarization bubble of the  intra-band  electron-hole excitations:
\begin{align}
    I(z) & = -\frac{k_{\rm B} T}{2N_{\rm F} N_k} \sum_{{\bf k}i, \Omega_n}
 {\rm Tr} \left[ \hat{\tau_3} \hat G_{\mathbf{k}i}(i\hbar\Omega_n-\varepsilon_{\rm F}) \hat{\tau_3} \hat G_{\mathbf{k}i}(i\hbar\Omega_n+i\hbar\omega_m-\varepsilon_{\rm F}) \right]_{i\omega_m \Rightarrow z}.
  \label{eq:Ifx_dress_main}
\end{align}
Here $\hat G_{\mathbf{k}i}$ denotes the dressed electronic Green's function for band $i$ in Nambu notation, $i\Omega_n$ ($i\omega_m$) are the fermionic (bosonic) Matsubara frequencies, $N_{\rm F}=\sum_{{\bf k}i}\delta(\varepsilon_{{\bf k}i}-\varepsilon_{\rm F})/N_k $ is the density of state per spin at $\varepsilon_{\rm F}$, $\Rightarrow$ indicates the analytical continuation to $z$ and $\tau_3$ is the Pauli matrix, encoding the structure of the density operator in Nambu notation. 
Since $I(0)=1$ at any $T$ \cite{marsiglio_phonon_1992}, if we approximate  $\Sigma_{\mu}(z)$ with its {\it static} value $\Sigma_{\mu}(0)$,  we see from Eq.s \eqref{eq:defs} and \eqref{eq:main_spectral_fx} that $A_{\mu}(\omega)$ becomes a $\delta$ function centred at $\omega^{\rm stat}_{\mu}$. If we retain the frequency dependence of the self-energy,  the behaviour of $\Sigma_\mu(\omega)$ is set by the comparison between the electron quasiparticle broadening $\Gamma^{\rm el}_{\mathbf{k}i}$, encoded in the interacting Green's functions $\hat G_{\mathbf{k}i}$,  and the phonon scale, which is of the order of  $\omega^{\rm dyn}_{\mu}$. For fast relaxing electrons, namely when $\Gamma^{\rm el}_{\mathbf{k}i}\gg  \omega^{\rm dyn}_{\mu}$, $I(\omega)\simeq 1$ and $A_{\mu}(\omega)$ exhibits a narrow peak at $\omega^{\rm stat}_{\mu}$ \cite{saitta_giant_2008,marchese_born_2024}, as in the purely static limit. Conversely, for non-relaxing electrons,   $\Gamma^{\rm el}_{\mathbf{k}i}\ll \ \omega^{\rm dyn}_{\mu}$, $I(\omega)\simeq 0$ and $A_{\mu}(\omega)$ has a narrow peak at $\omega^{\rm dyn}_{\mu}$. These observations justify the parametrization of the self-energy as in Eq.\ \eqref{eq:defs}
above. 

We  compute $\hat G_{\mathbf{k}i}$ and $I(\omega)$  with the isotropic ME approach in the real-axis formulation of \cite{marsiglio_phonon_1992}.
Finally, in order to describe the broadening of the phonon spectral function not related to interactions with the Fermi-surface electrons, as due e.g. to anharmonic and disorder scattering, we include an additional imaginary contribution to the phonon propagator with the following phenomenological substitution of the $z^2$ term in Eq. \eqref{eq:dphonon}  with $(z+i\Gamma^{intr}_{\mu}/2)^2$, where $\Gamma^{intr}_{\mu}$ is a real positive quantity which determines the FWHM of the phonon peak when $\Sigma_{\mu}=0$.

In a DFT \textit{ab initio} framework, as pointed out in \cite{lazzeri_nonadiabatic_2006,saitta_giant_2008},
$\omega^{\rm stat}_{\mu}$ (named as $\omega^{\rm A}$ in \cite{saitta_giant_2008})  can be computed  from the second derivative of the total energy with respect to a {\it static} phonon displacement or, equivalently, from {\it static} density-functional perturbation theory \cite{baroni_phonons_2001}. Whereas, $\omega^{\rm dyn}_{\mu}$ (named as $\omega^{\rm NA}$ in \cite{saitta_giant_2008}) can be obtained by removing from  $\omega^{\rm stat}_{\mu}$  the dressing by Fermi-surface electrons with Eq. \eqref{defomegastatdyn}.
For h-LaH$_{10}$, the size of the unit cell and the requirement to consider the strong quantum anharmonic Hydrogen-ions fluctuations \cite{errea_quantum_2020}  make unfeasible to evaluate from first-principles $\omega^{\rm stat}_{\mu}$ and $\omega^{\rm dyn}_{\mu}$ of H modes. As described in the next Section, we instead obtain $\omega^{\rm stat}_{\mu}$ and $\omega^{\rm dyn}_{\mu}$ from a fit on the experimental data.   Nonetheless, for the low-energy  La phonon modes, weakly sensitive to H fluctuations, we also computed the DFT values finding a good agreements with the fitted ones, see  \ref{tab:dfpt_phononfreq_mtds}. 

The electronic Raman contribution within the ``effective mass'' approximation \cite{devereaux_inelastic_2007} is also linked to the dimensionless function $I(\omega)$. Indeed, by denoting with $K^{\rm el}$ an overall prefactor linked to the electronic Raman activity, one can approximate the electronic part as:
\begin{equation}
  S^{\rm el}(\omega) =[n(\omega)+1]
  K^{\rm el}\; \Im [I(\omega+i0^+)]
  \label{eq:el_raman}.
\end{equation}
As discussed in~\cite{devereaux_inelastic_2007}, the prefactor $K^{\rm el}$ accounts for the effects of the light polarization, that can be particularly relevant when the gap has nodes in momentum space, as it happens e.g. in high-temperature cuprate superconductors. However, current spectroscopic determination of the SC gap of hydrides are consistent with a nodeless gap, justifying the approximate description of polarization effects in an overall prefactor in Eq.~\eqref{eq:el_raman}. 
 
\subsection*{Theoretical model parameters optimization procedure}
Based on the ME method of \cite{zeyher_phonon_1988}, we build a theoretical model exploiting a reduced number of free parameters, divided in two subsets: those defining the SC electronic state and those describing the coupling of the lattice vibrations to the electronic state. 

The SC properties are defined by the Eliashberg function $\alpha^2F(\omega)$, the strength of elastic impurity $\eta_{\rm{imp}}$ and the Morel-Anderson pseudo-potential\cite{morel_calculation_1962} $\mu^*$. 
 We set $\mu^*$ to the 0.15 value found for the cubic phase in \cite{elatresh_optical_2020} and $\eta_{\rm{imp}}$ to 887 cm$^{-1}$, a value commonly accepted in literature for this class of materials  \cite{capitani_spectroscopic_2017}.
The SC properties depend mostly on  $\lambda$ and on the energy scale $\omega_{\rm{log}}=\exp[2\int_0^{\infty}d\omega \ln(\omega)\alpha^2F(\omega)/(\omega\lambda)]$\cite{allen_theory_1983}.
Thus, for the Eliashberg function in the hexagonal phase we choose to exploit the $\alpha^2F_{\rm cub}(\omega)$ calculated in \cite{errea_quantum_2020} for the cubic $Fm\bar{3}m$ phase of LaH$_{10}$ at 129 GPa properly rescaled, as described in the main text. Fixed the SC subset of parameters ($\mu^*,\eta_{\rm{imp}}, \alpha^2F(\omega) $), we solve the ME systems and obtain the dimensionless function $I(\omega)$ of Eq.~\eqref{eq:Ifx_dress_main} in real frequencies \cite{marsiglio_phonon_1992} using 1024 Matsubara frequencies and calculating $\Sigma^{el}(\omega)$ on 1000 real frequencies up to 4032 cm$^{-1}$, with the small imaginary part of the frequency ($z=\omega+i\eta_{\lambda}$) equal to $\hbar\eta_{\lambda}=0.8$ cm$^{-1}$.

To determine the vibrational parameters for peaks D and E,
we first find $(\omega_{\mu}^{\mathrm{stat}}, \omega_{\mu}^{\mathrm{dyn}})$ by matching the theoretical position of the peaks with the values measured at the highest and lowest temperature. A $\chi^2$ optimization between the theoretical linewidth and the experimental one fixes the value of $\Gamma^{\rm{intr}}_{\mu}$. 
For peak A, we find the optimal values of the four parameters ($\omega_{\mu}^{\mathrm{stat}},\omega_{\mu}^{\mathrm{dyn}},\Gamma^{\rm{intr}}_{\mu}, c$ ) with a least-squares minimization applied contextually on both linewidths and positions. 
Owing to the non-Lorentzian profile of the calculated peaks, the position is defined as the midpoint between the half-maximum crossings, while the linewidth corresponds to the full width at half maximum (FWHM), see \ref{fig:raman_ind_sample1}. The results of the optimisation for sample 2 are shown in \ref{fig:inpurities}. The optimal parameters are listed in \ref{fig:energy_scale} and \ref{fig:inpurities}. The agreement between theory and measurements is robust against variations of the impurity scattering rate $\eta_{\rm imp}$ used in the ME calculation, as shown in \ref{fig:inpurities}.

\section*{Acknowledgements}

We acknowledge financial support from the European Union ERC-SYN MORE-TEM no. 951215 (F.M., L.B., P.B. and G.M.), and CINECA award under ISCRA Initiative grant no. HP10C8TW7J (G.M.) and HP10CMZFMM (P.B.). L.B. acknowledges financial support by the Italian MIUR under project PRIN2022-CoInEx (2022WS9MS4). P. D-S. acknowledges the support of the National Natural Science Foundation of China (NSFC) reference code W2532012. ZYC acknowledges funding from the NSFC, No. 12504160, the Natural Science Foundation of Shandong Province, No. ZR2024QA058, and the Special Construction Project Fund for Shandong Province Taishan Scholars. 

\section*{Author contributions}

F.M. and F.A.G. conceived the study and designed the research;
F.A.G. and P.D.-S. designed the DACs, the Raman setup and the optical cryostat.
P.D.-S. and Z.-Y.C. prepared sample 1 and sample 2 respectively.  F.A.G. P.D.-S. and Z.-Y.C. synthesized the hydride by lasers heating and performed Raman and transport measurements at low temperature; F.A.G. and G.M. analysed the Raman spectra; G.G., F.A.G. P.D.-S. performed and analysed XRD experiments; F.M. and P.B. achieved the \textit{ab initio} structural characterization of the samples; F.M., L.B., P.B. and G.M. carried out theoretical analyses and data interpretation; F.A.G. and G.M. wrote the paper with inputs from all other authors; F.M. and F.A.G. supervised the project.

\section*{Competing interests}
The authors declare no competing interests.

\section*{Data availability}
Transport measurements, X-ray data and Raman spectra (with and without the Bose-Einstein correction) are available on the Zenodo public database.


\section*{Other declarations}
Supplementary Information is available for this paper.
Correspondence and requests for materials should be addressed to Federico A. Gorelli and Francesco Mauri.

\bibliography{Bibliography}

\setcounter{figure}{0}
\setcounter{table}{0}
\renewcommand{\figurename}{}
\renewcommand{\thefigure}{\textbf{Extended Data Fig. \arabic{figure}}}
\renewcommand{\tablename}{\textbf{}}
\renewcommand{\thetable}{\textbf{Extended Data Table \arabic{table}}}

\clearpage
\section*{Extended Data}




\begin{figure}[ht]
  \centering
  \includegraphics[width=0.8\textwidth]{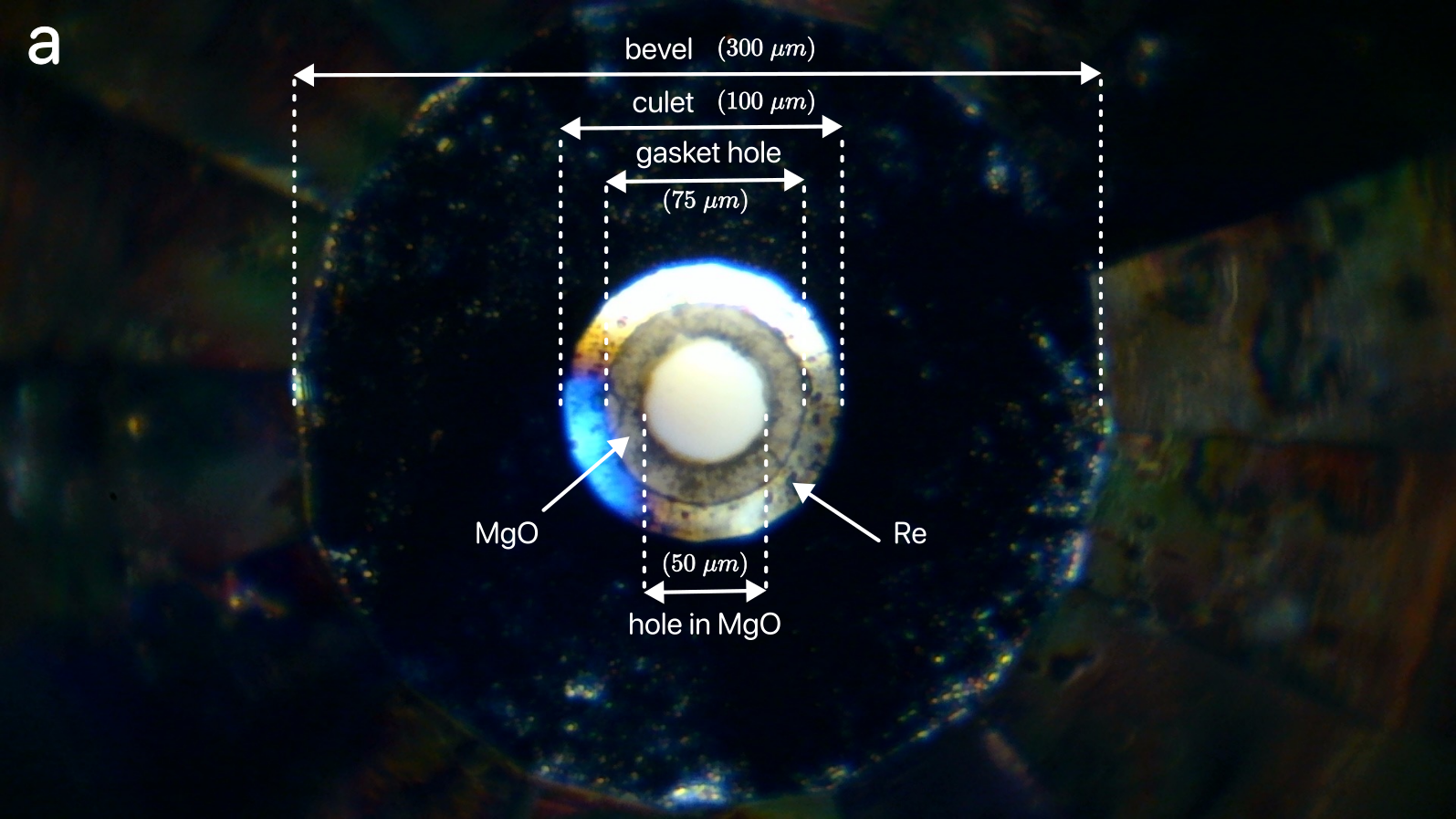}
  \includegraphics[width=0.8\textwidth]{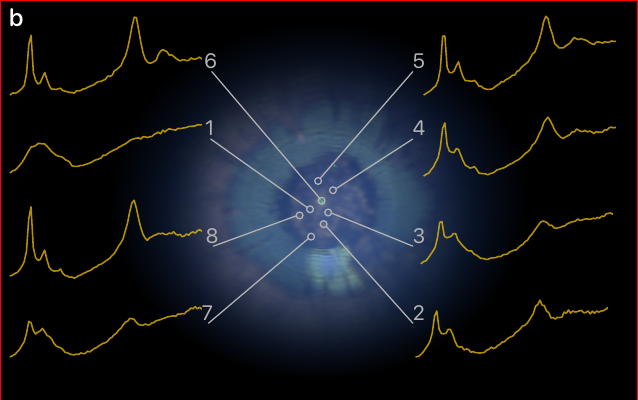}
  \caption{ \textbf{
      Photograph of the gasket and sample heterogeneity.} \textbf{a}, optical top view of the DAC before sample loading. The outer ring (barely visible) marks the edge of the diamond bevel. A composite gasket, consisting of an external rhenium (Re) support and an internal MgO ring, is positioned on the bevel. The gasket core is filled with ammonia borane (NH$_3$BH$_3$), which serves as the pressure-transmitting medium and is in direct contact with the sample (see inset of Fig.~\ref{fig1:abstract}\textbf{a}). \textbf{b}, top-view image of sample~1 inside the DAC at 150~GPa. Small circles indicate the laser spots corresponding to selected measurement positions used for Raman mapping. The iridescent background arises from the MgO layer in the gasket. Each position is labeled by a number, and the continuous yellow curve represents the corresponding Raman spectrum acquired at 75~K. Variations in crystallinity and local purity are reflected in the sharpness of the low-energy Raman peaks: positions~1, 2, 3, and~7 exhibit broader, less defined features indicative of lower crystalline quality, whereas positions~4, 5, 6, and~8 display sharper phonon modes consistent with higher crystallinity. Spectra from positions~6 and~8 were primarily used for the analysis presented in ~\ref{fig3:raman_analysis}.
    }
  \label{fig:Sample_1_after}
\end{figure}

\clearpage
\begin{figure}[ht]
  \centering
  \includegraphics[width=1.\textwidth]{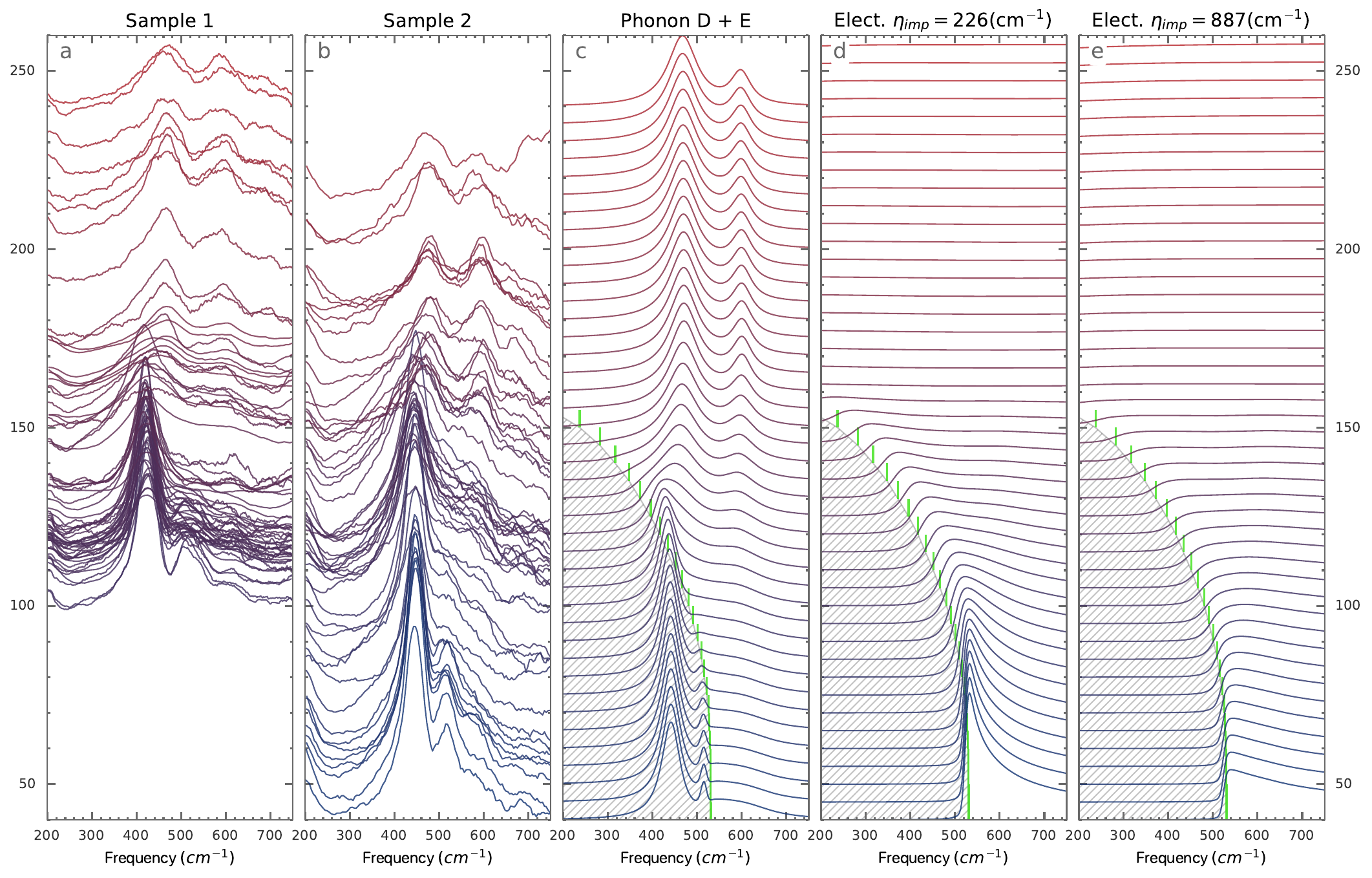}~
  \caption{ \textbf{ Temperature evolution waterfall of Raman spectra.}
  \textbf{a},\textbf{b}, experimental Raman spectra for samples~1 and~2, respectively. For graphical clarity only (and not for any quantitative analysis), the spectra have been processed to reduce noise and highlight the temperature evolution of the most prominent features. Specifically, the baseline was subtracted, and the data from sample~1 (sample~2) were smoothed using an 8-point (4-point) moving average. 
  \textbf{c-e}, theoretical calculations performed within the Migdal–Eliashberg (ME) framework and the parameters used for modelling sample 2 (see Methods). The temperature evolution of the vibrational (c) and electronic (d,e) contributions to the Raman signal exhibits distinct signatures governed by the evolution of the pair-breaking energy $2\Delta(T)$ (indicated by green bars and the shaded grey region).
The calculated vibrational Raman spectra (\textbf{c}) show that peaks D and E remain visible across all temperatures. Their positions evolve discontinuously around the resonant condition $2\Delta(T)=\omega_{\rm ph}$, marking the transition between high- and low-temperature limits. The corresponding linewidths display a similar trend: at low temperature, the peaks are sharpest, while at higher temperatures they broaden as $2\Delta(T)<\omega_{\rm ph}$, reflecting enhanced electron–phonon scattering as spectral weight is released from the condensate.
    The electronic contribution (\textbf{d,e} for two impurity levels $\eta_{imp}$) features a distinct peak emerging at the pair-breaking energy $2\Delta(T)$.
    As temperature increases, the peak shifts to lower energy and vanishes smoothly once the superconducting gap closes above $T^{\Delta}_{\rm c}$, merging into the smooth background.
    This contribution displays a broad temperature dispersion near the gap-closing regime $T\sim T^{\Delta}_{\rm c}$, similar to that observed in cuprates \cite{bohm_balancing_2014}, and becomes sharper at lower temperatures. The impurity level $\eta_{imp}$ acts primarily as a spectral broadening factor, as illustrated by comparing \textbf{d} ($\eta_{imp}=226$) and \textbf{e}  ($\eta_{imp}=887$ cm$^{-1}$).
  }
  \label{fig:raman_el_waterfall}
\end{figure}

\clearpage
\begin{table}[h!]
  \resizebox{1\textwidth}{!}{%
    \begin{tabular}{|
      P{0.4cm}P{1cm}P{1cm}|
      P{0.4cm}P{1cm}|
      P{1.3cm}P{1cm}P{1cm}||
      P{1cm}P{1cm}|P{1cm}P{1cm}|P{1cm}|
      }
      \multicolumn{5}{c|}{$P6_3/mmc$ (\# 194)}  & \multicolumn{3}{c||}{$P6_3/m$ (\# 176)} & \multicolumn{4}{c}{Parameters from measurements}  & Peak                                                                                                                                                                                                                                                                                                                                                      \\
      \hline
      \multicolumn{3}{|c|}{$\Gamma$ $(D_{6h})$} & \multicolumn{2}{c|}{$M$ ($D_{2h}$)}   & \multicolumn{3}{c||}{$\Gamma$ $(C_{6h})$} & \multicolumn{2}{c|}{Sample 1} & \multicolumn{2}{c|}{Sample 2} &   label                                                                                                                                                                                                                                                                           \\
      \hline
                                                  & $\omega^{\rm stat}_{\mu}$                                   & $\omega^{\rm dyn}_{\mu}$                       &                               & $\omega^{\rm stat}_{\mu}$                 &                                & $\omega^{\rm stat}_{\mu}$                   & $\omega^{\rm dyn}_{\mu}$                     & $\omega^{\rm stat}_{\mu}$                 & $\omega^{\rm dyn}_{\mu}$                & $\omega^{\rm stat}_{\mu}$                 & $\omega^{\rm dyn}_{\mu}$                & \\[0.2cm]
      $\Gamma_5^+$                                & 89.5                                     & 125.8                         &                               &                        & $\Gamma_3^+\oplus\Gamma_5^+$   & 102.2                    & \textcolor{black}{154.4 }   & $110$                  & $143$                  & \multicolumn{2}{c|}{$\sim 108$}            & A     \\[0.2cm]
                                                  &                                          &                               & \multirow{2}*{$M_3^+$}        & \multirow{2}*{139.8 }  & $\Gamma_4^+\oplus\Gamma_6^+$   & 150.6                    & 163.6                       & \multicolumn{2}{c|}{ \multirow{2}{*}{$\sim 151$}} & \multicolumn{2}{c|}{ \multirow{2}{*}{$\sim 145$}} & \multirow{2}{*}{B}\\[-0.01cm]
                                                  &                                          &                               &                               &                        & \textcolor{gray}{$\Gamma_2^+$} & \textcolor{gray}{178.0}  & \textcolor{gray}{179.6}     &                        &                        &                        &                        & \\[0.2cm]
                                                  &                                          &                               & \multirow{2}*{$M_1^+$}        & \multirow{2}*{142.7  } & $\Gamma_3^+\oplus\Gamma_5^+$   & 164.6                    & \textcolor{black}{182.4 }   &                        &                        &                        &                        &   \\[-0.01cm]
                                                  &                                          &                               &                               &                        & $\Gamma_1^+$                   & 178.1                    & 179.0                     &                        &                        &                        &                          &  \\[0.2cm]
                                                  &                                          &                               & \multirow{2}*{$M_2^+$}        & \multirow{2}*{182.6 }  & $\Gamma_1^+$                   & 185.1                    & 192.8                     & \multicolumn{2}{c|}{ \multirow{2}{*}{$\sim 199$}}   &  \multicolumn{2}{c|}{ \multirow{2}{*}{$\sim 195$}}  & \multirow{2}{*}{C}\\[-0.01cm]
                                                  &                                          &                               &                               &                        & $\Gamma_3^+\oplus\Gamma_5^+$   & 197.0                    & 210.9                     &                        &                        &                        &                          &  \\[0.2cm]
      \textcolor{gray}{$\Gamma_4^+$}              & \textcolor{gray}{254.1 }                 & \textcolor{gray}{256.1}       &                               &                        &                                &                          &                           &                        &                        &                        &                          &      \\[0.1cm]
      \hline
    \end{tabular}
  }
  \caption{\justifying\textbf{Low-energy phonon resonances: comparison between \textit{ab initio} and experimental fit.}
  A vertical divider separates the \textit{ab initio} (left) and experimental (right) characterizations of the La-derived phonon modes in h-LaH$_{10}$. The \textit{ab initio} calculations were performed for both the high-symmetry P6$_3$/mmc phase and its 2×2×1 reconstructed P6$_3$/m derivative. Phonon modes are labeled according to the irreducible representations of the corresponding point group and phonon momentum, following the Miller–Love notation. Optically inactive modes are shown in grey.
  For each mode, the two frequencies ($\omega^{\rm stat}_{\mu}$ and $\omega^{\rm dyn}_{\mu}$) are given in cm$^{-1}$ as computed in the limit of infinitely light hydrogen mass to remove spurious coupling with higher-energy modes introduced by the harmonic approximation \cite{errea_quantum_2020}. For the higher-symmetry h-LaH$_{10}$ phase (P6$_3$/mmc), the two La-derived phonons at the $\Gamma$ point are reported, only one of which is Raman active.
  The second column lists the phonon frequencies of inversion-even modes at the $M$ point, whose folding into the lower-symmetry 2$\times$2$\times$1 reconstruction gives rise to additional Raman-active phonon modes at the $\Gamma$ point of the P6$_3$/m phase.
    The three rightmost columns report the static and dynamic frequencies and the peak labels obtained fitting the low-energy resonance triplet observed experimentally in sample 1 and 2 with the ME model using $r_{\omega}=1$ and $\eta_{\rm imp}=887$ (cm$^{-1}$). The small lineshifts of these modes, comparable to the experimental resolution, allow unambiguous determination of $\omega^{\rm stat}_{\mu}$  and  $\omega^{\rm dyn}_{\mu}$ only for peak~A in sample~1. For the remaining peaks, only averaged positions could be extracted; these are denoted in the table by a preceding tilde symbol.
  }
  \label{tab:dfpt_phononfreq_mtds}.
\end{table}

\clearpage
\begin{figure}[ht]
  \centering
  \includegraphics[width=0.7\textwidth]{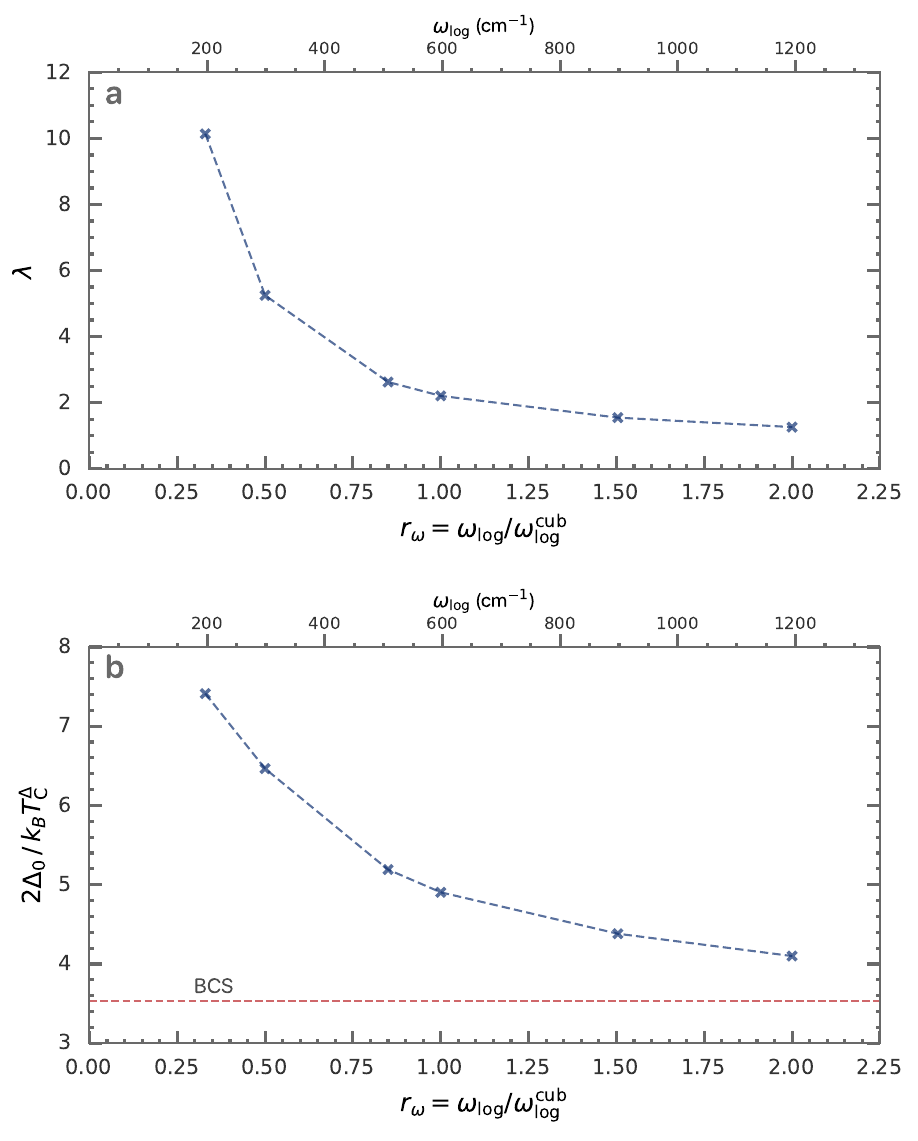}
\caption{\textbf{Study SC properties varying the mediators energy scale at fixed critical temperature}. 
The ME calculation is mainly determined by the Eliashberg function $\alpha^2F(\omega)$, characterized by the average frequency $\omega_{\rm log}$ and the integrated coupling strength $\lambda$. In this analysis, we explore the impact of the mediator energy scale by fixing the sc critical temperature to the experimental value $T^{\Delta}_{\rm c}=174$ K and rescaling the characteristic frequency $\omega^{\rm cub}_{\rm log}$ of the cubic phase $\alpha^2F_{\rm cub}(\omega)$ \cite{errea_quantum_2020} by a factor $r_{\omega}=\omega_{\rm log}/\omega^{\rm cub}_{\rm log}$.
\textbf{a} shows how the coupling strength $\lambda$ must vary as a function of the rescaling factor $r_{\omega}$ in order to preserve $T^{\Delta}_{\rm c}$. 
\textbf{b} displays the ratio $2\Delta_0/ k_{\mathrm{B}}T_{\mathrm{c}}$ obtained by solving the self-consistent ME equations. The horizontal dashed line marks the value predicted by the weak-coupling BCS theory. 
}
  \label{fig:rescale_vs_sc}
\end{figure}

\clearpage
\begin{figure}[ht]
  \centering
  \includegraphics[width=0.95\textwidth]{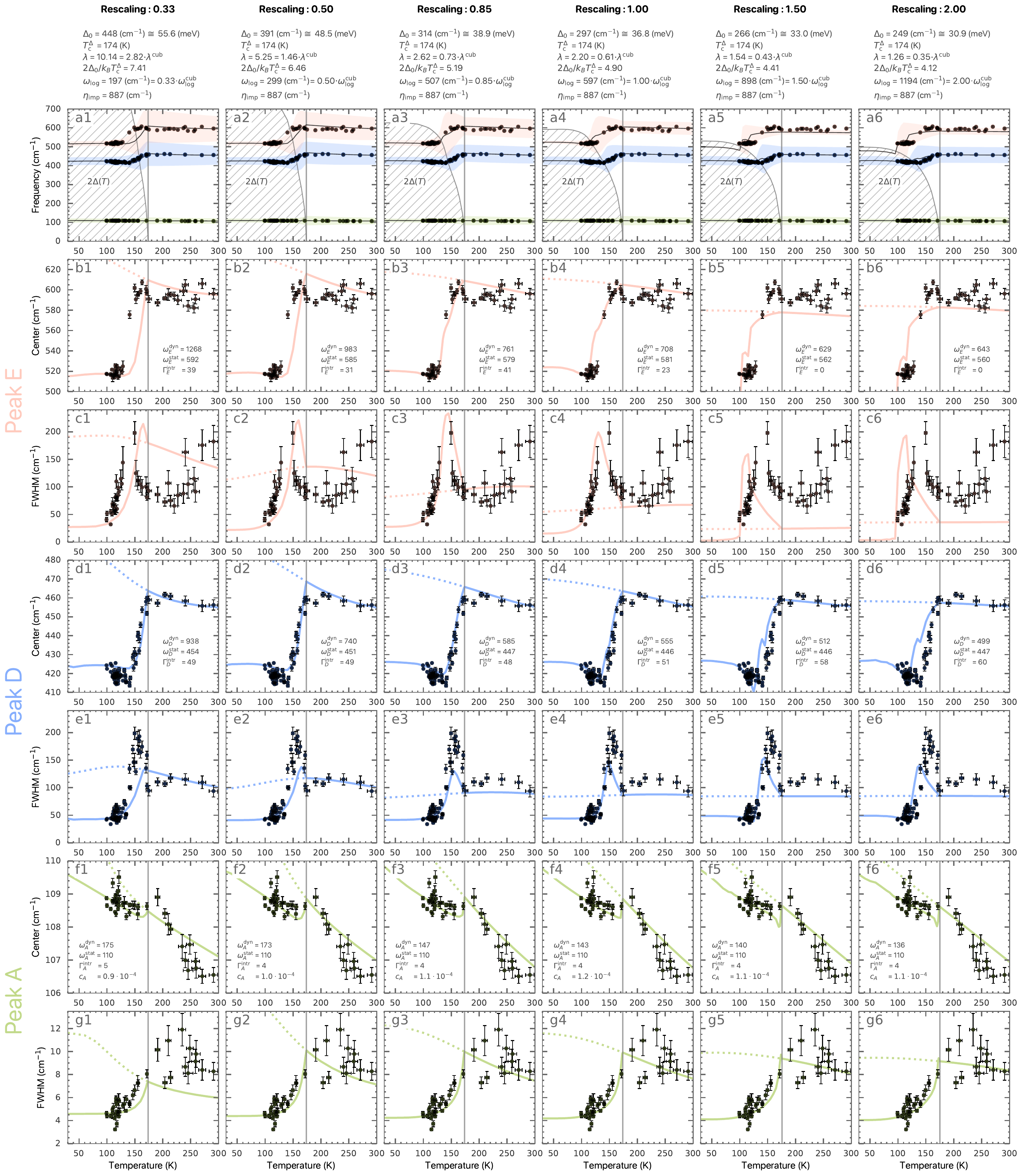}
\caption{\textbf{Boson mediators energy scale}. 
The ME method describes the formation of a SC condensate given the density of states and coupling of Boson mediators. In the case of electron-phonon SC, this role is played by the Eliashberg function $\alpha^2 F(\omega)$. To assess the agreement with different possibile pairing mechanisms, we modified the energy scale of the cubic phase $\alpha^2 F_{\rm cub}(\omega)$ by rescaling its average frequency by a factor $r_{\omega}=\omega_{\rm log}/\omega^{\rm cub}_{\rm log}$, while adjusting the coupling amplitude to reproduce the experimental critical temperature $T^{\Delta}_{\rm c}=174$ K.
The four columns compare theoretical scenarios corresponding to energy rescaling factors $0.33$, $0.5$, $1$ and $2$ for the main spectroscopic peaks A, D, E. For each value of  $r_{\omega}$ (i.e., for each column), the upper \textbf{a1-a5} show the comparison between theoretical predictions (solid lines) and experimental peak positions (dots), with shaded regions indicating the theoretical linewidth.
The lower \textbf{b-f} provide a more detailed comparison of the temperature dependence of the peak position and linewidth, as in \ref{fig3:raman_analysis}.
As clearly evident for peaks D and E, only a scaling factors $r_{\omega}$ between 0.85 and 1, corresponding to minor modification of the \textit{ab initio}  energy range for c-LaH$_{10}$, yields a satisfactory agreement with the experimental findings. }
  \label{fig:energy_scale}
\end{figure}

\clearpage
\begin{figure}[ht]
  \centering
  \includegraphics[width=0.95\textwidth]{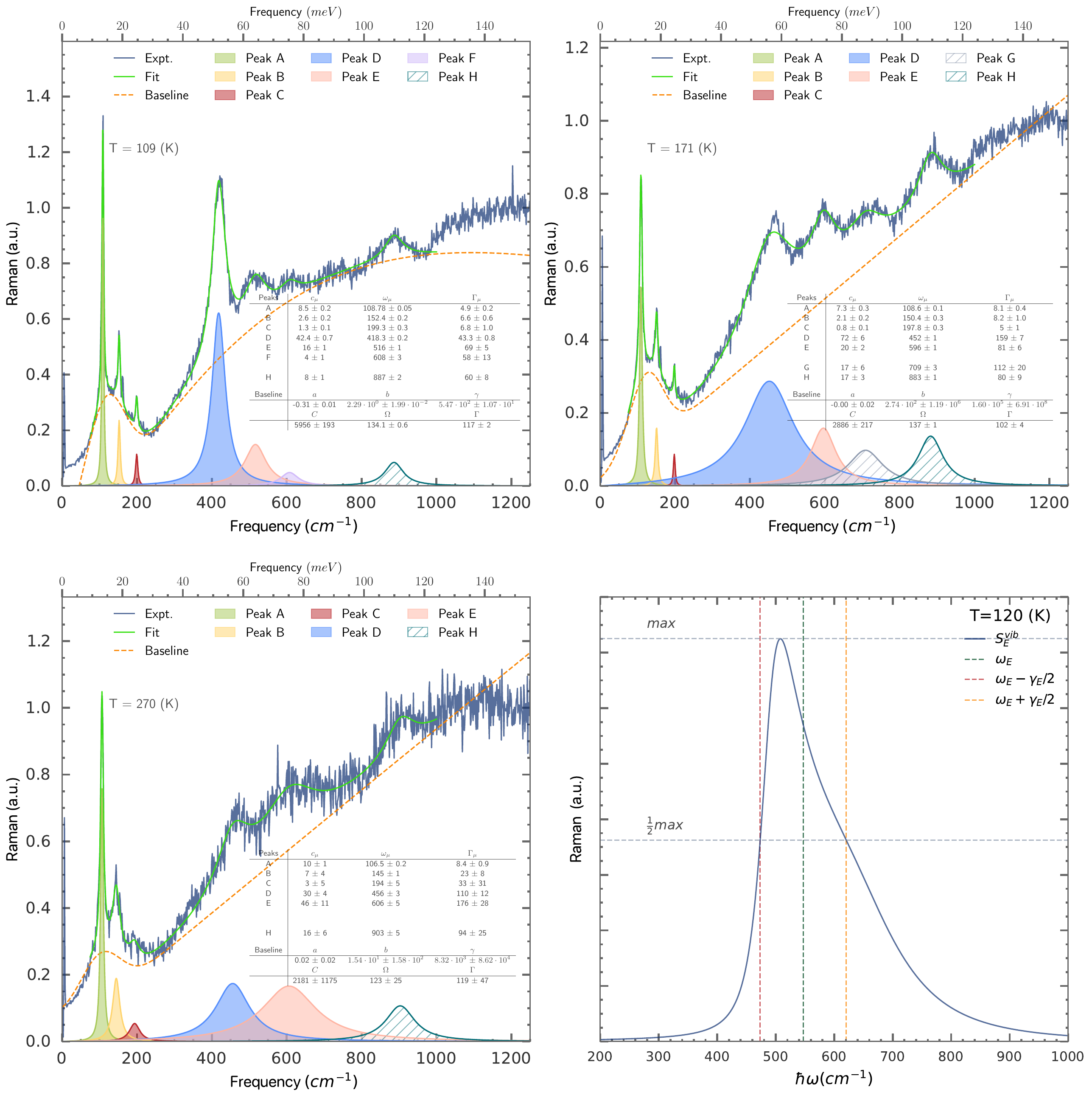}
  \caption{ \textbf{Extraction of peak parameters from experimental and theoretical spectra.}
    \textbf{a-c} illustrate the fitting procedure applied to the experimental Raman data (see Methods and Supplementary Information for details).
    The green curve represents the optimized fit, comprising comprising contributions from individual phonon modes (coloured regions) and a smooth background (orange dashed line).
    The overlaid table reports the resulting fitting parameters for each spectrum. We note that the integrated area of the three La low-energy peaks (A, B, C) accounts for  $\sim 20$ \% of the total spectral weight within the 0–225~cm$^{-1}$ range and this ratio is temperature-independent.
    \textbf{d} illustrates how the vibrational Raman contribution calculated with the ME approach is characterized by a significant asymmetry. 
    Consequently, the peak linewidth is estimated as the full width at half maximum ($\Gamma_{\mu}$), while the peak position $\omega_{\mu}$ (green line) is defined as the midpoint between the half-maximum crossings (red and yellow vertical lines).    
  }
  \label{fig:raman_ind_sample1}
\end{figure}

\clearpage
\begin{figure}[ht]
  \centering
  \includegraphics[width=0.9\textwidth]{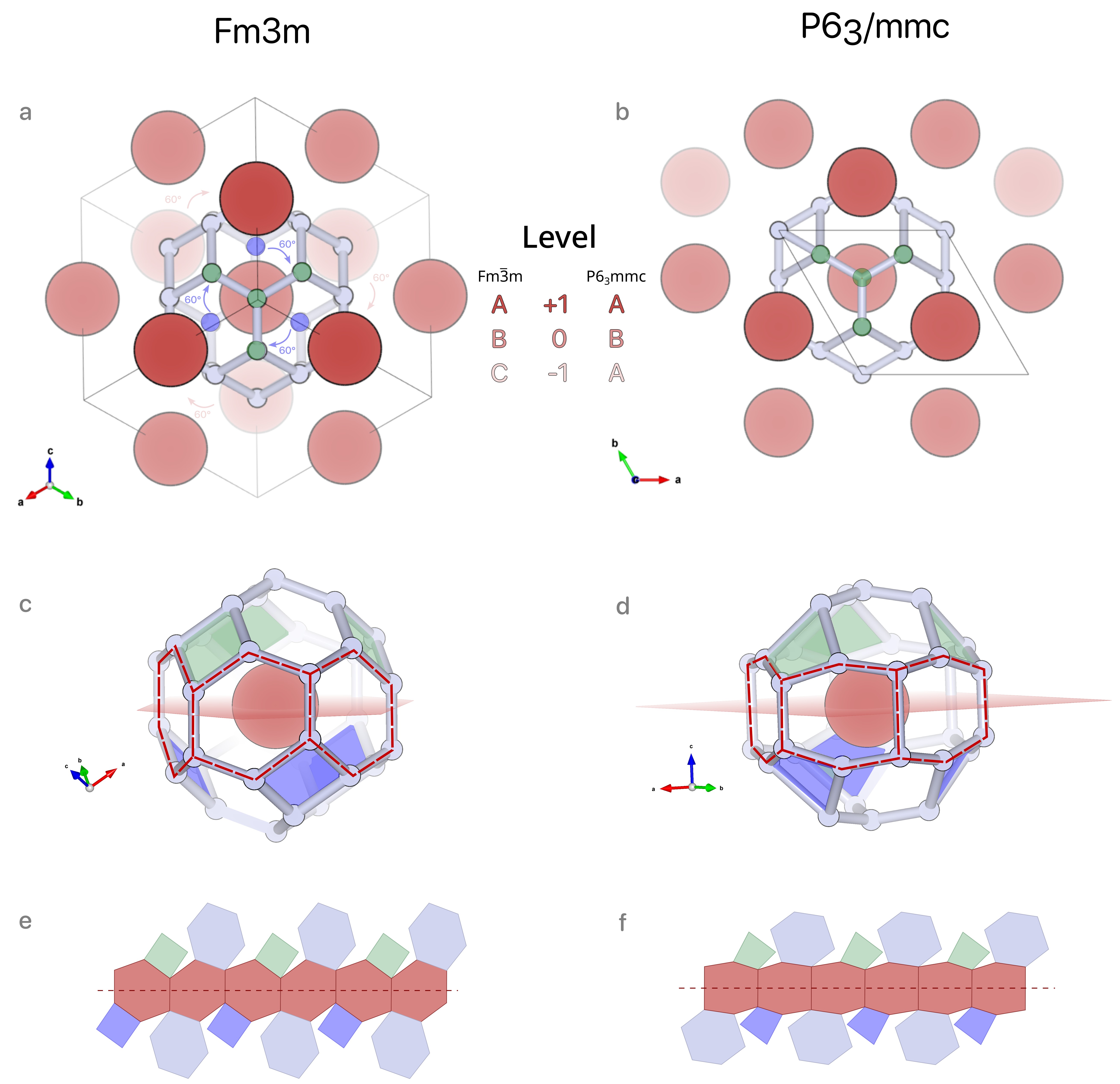}
  \caption{ \textbf{Relation between cubic and hexagonal closed-packed structures of LaH$_{10}$.}
    The cubic $Fm\bar{3}m$ and hexagonal $P6_3/mmc$ phases represent the highest-symmetry realization of two different close-packing of La atoms, the $fcc$ and hcp, respectively. In both structures the closed-packed lattices formed by La atoms comprises triangular layers orthogonal to the [111] axis of the cubic cell or the $c$ axis of the hexagonal one: these layers are $ABC$-stacked in the $fcc$ phase (\textbf{a}), while they display an $AB$ stacking in the hcp one (\textbf{b}), that can be obtained by a 3-fold rotation of the $C$ plane about the [111] axis of the cubic cell.     
    Hydrogen atoms occupy the tetrahedral and octahedral interstitial voids of both closed-packed lattices. The higher symmetry of the $fcc$ phase causes the H atoms to form a regular chamfered-cubic cage around each La atom, comprising square and flattened hexagonal faces whose internal angles are fixed by symmetry. The chamfered cube is displayed in \textbf{c} and \textbf{d}, highlighting the triangular-layer plane and the squares that correspond to one face of the regular H cubes sitting at octahedral voids of the $fcc$ lattice. For clarity we use different colours for square faces above or below the triangular-lattice plane. \textbf{e} shows a polyhedral-net representation of the chamfered cube where the chain of edge-joined flattened hexagons is bisected by the triangular-layer plane, corresponding to the hexagonal faces highlighted with a dashed red line in the central \textbf{c}. As a consequence of the 3-fold rotation relating the fcc and hcp lattices, the hydrogen atoms form distorted tetrahedra and cubes in tetrahedral and octahedral voids, resulting in a distorted cage around each La atom (\textbf{d}) that can be visualized as the result of rotating the lower half of the chamfered cube by 120$^\circ$ about the hexagonal $c$ axis, as shown in \textbf{f}.
  }
  \label{fig:structure}
\end{figure}

\clearpage
\begin{figure}[ht]
  \centering
  \includegraphics[width = \textwidth]{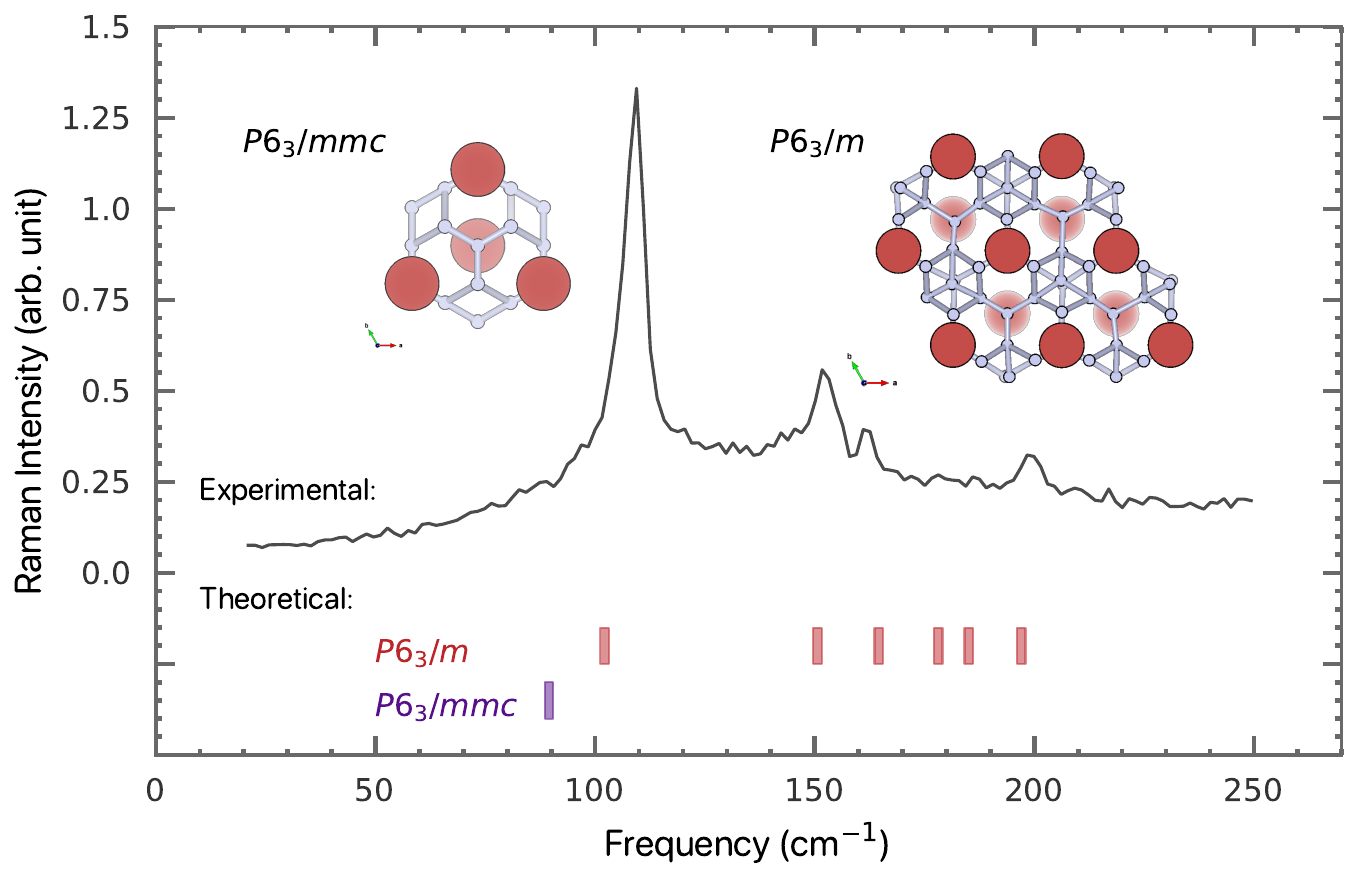}
  \caption{ \textbf{Agreement between Raman structural characterization and \textit{ab initio} calculation.}
  The low-energy range of the Raman spectra collected at low-temperature (solid curve, representative spectrum of sample 1 at 109~K) exhibit a spectroscopic fingerprint in excellent agreement with the first-principles predictions for the P6$_3$/m phase of h-LaH$_{10}$ (red squares).
  The violet square indicates the calculated position of the sole low-frequency Raman-active mode of the P6$_3$/mmc phase.
  The insets illustrate the unit-cell structures of these two phases, highlighting that the lower-symmetry P6$_3$/m arises from a minor rearrangement of the hydrogen positions within a 2x2x1 supercell of the P6$_3$/mmc phase. The reduced intensity of the experimental peaks at 150 and 200~cm$^{-1}$ relative to the 110~cm$^{-1}$ peak supports their origin as folded mode of the parent P6$_3$/mmc phase.
  }
  \label{fig:main_reconstruction}
\end{figure}

\clearpage
\begin{figure}[ht]
  \centering
  \includegraphics[width=1\textwidth]{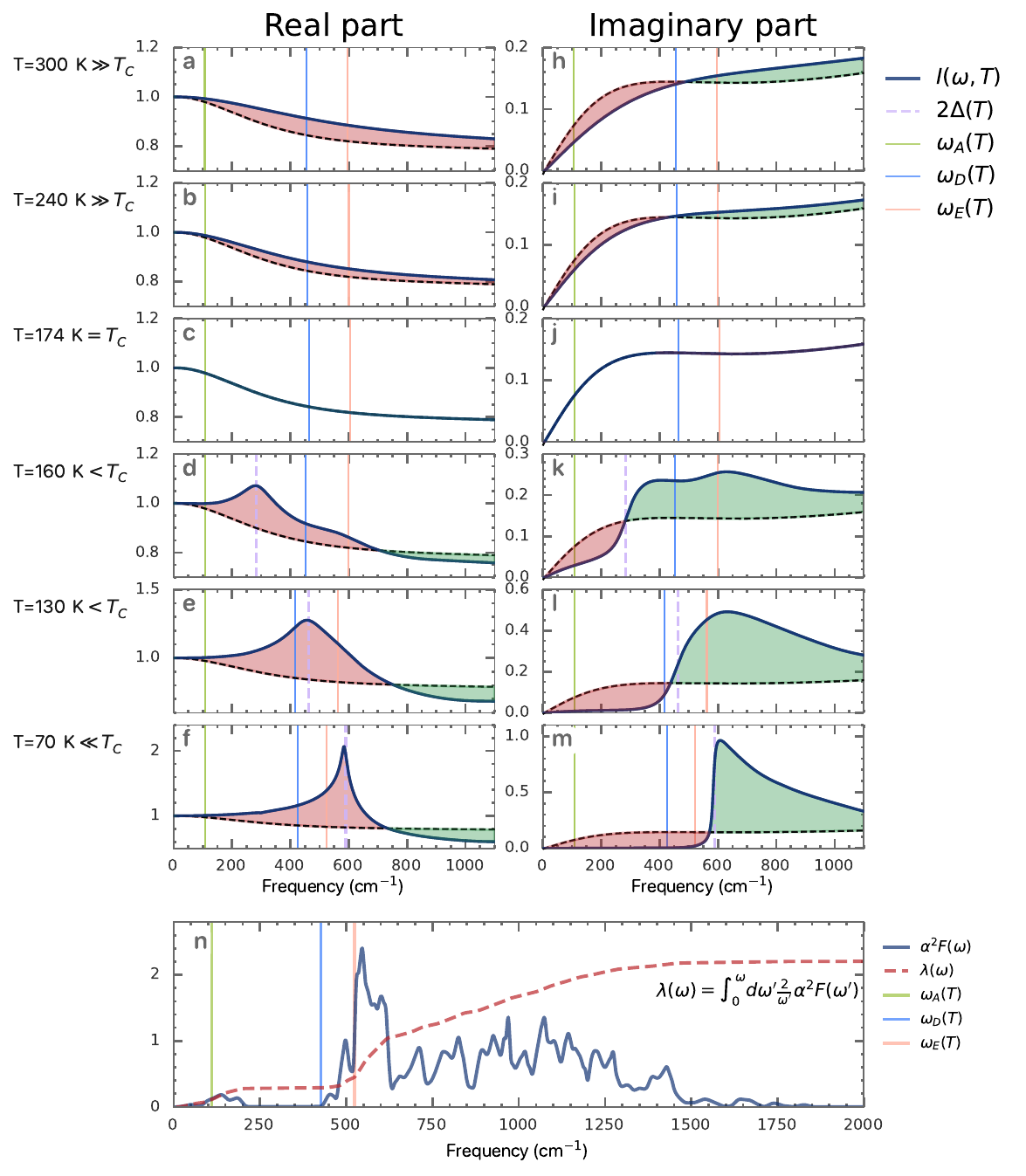}
\caption{\textbf{Temperature dependence of the dimensionless retarded function $I(T,\omega)$}. 
\textbf{a–m} display, in two adjacent columns, the real (left) and imaginary (right) components of the dimensionless retarded function $I(T,\omega+i0^+)$ defined in Eq.~\eqref{eq:Ifx_dress_main}, shown for a sequence of temperatures. Each row corresponds to a distinct temperature and is arranged in a vertical waterfall format, , from lowest (bottom) to highest (top).
Vertical solid lines indicate the position of the main spectroscopic features A, D, E identified in \ref{fig3:raman_analysis}, while violet dashed lines denote the temperature-dependent pair-breaking energy $2\Delta(T)$. 
Coloured regions highlight the relative variation of $I(T,\omega)$ with respect to its value at $T_{\rm c}=174$ K--green indicating positive and red indicating negative changes.
A positive (negative) variation real part  of $I(T,\omega)$ corresponds to a redshift (blueshift) of phonons falling in that red  (green) region. 
A negative (positive) variation in the imaginary part of $I(T,\omega)$ reflects a suppression (enhancement) of electron-phonon scattering due to electronic excitations close to the Fermi level.
\textbf{n} shows the Eliashberg function $\alpha^2F(\omega)$ used for the calculation of $I(T,\omega)$ in the other panels and for the theoretical modelling of sample~1 in Fig.~\ref{fig3:raman_analysis}. It was obtained by applying a global multiplicative scaling to the $\alpha^2F_{\rm cub}(\omega)$ computed for cubic LaH$_{10}$ at 129~GPa~\cite{errea_quantum_2020}, so that $r_{\omega}=1$ and $\lambda=2.2$.
}
  \label{fig:I_waterfall}
\end{figure}

\clearpage
\begin{figure}[ht]
  \centering
  \includegraphics[width=1\textwidth]{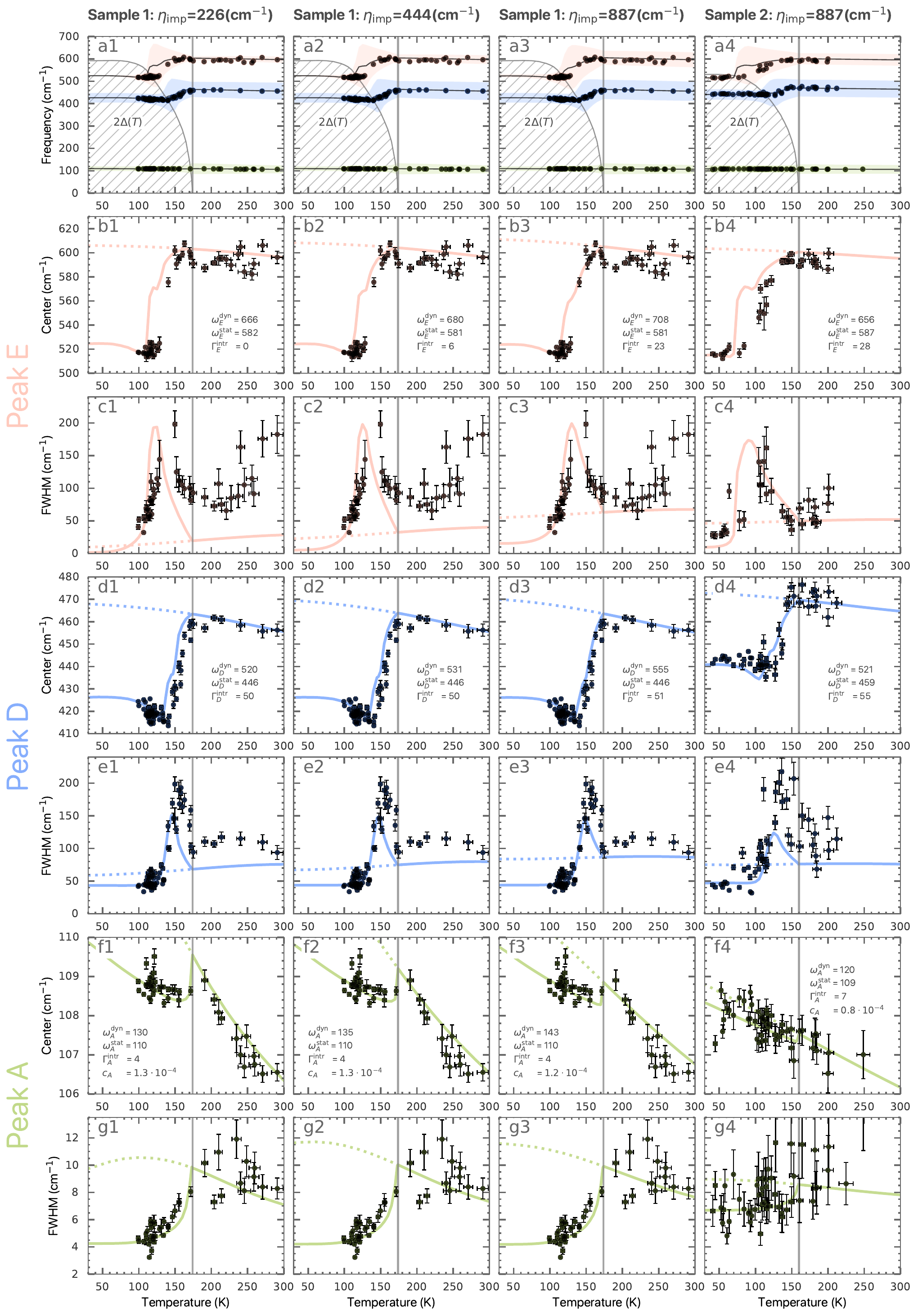}
\caption{\textbf{Effect of impurity level}. The first three columns show the analysis of the main spectroscopic features in sample 1 calculated with $r_{\omega}=1$ for varying impurity level $\eta_{\rm imp}$ in the theoretical model. 
While the impurity primarily affects the sharpness of the dimensionless function $I(\omega)$, its impact on the calculated peak positions and linewidths is largely mitigated, being effectively absorbed into the other fitted parameters ($\omega^{\rm dyn}_{\mu}, \omega^{\rm stat}_{\mu}$).
For comparison, the last column presents the corresponding analysis for sample~2, using the optimal impurity value of 110~meV ($\cong 887~\mathrm{cm}^{-1}$).
}
\label{fig:inpurities}
\end{figure}

\clearpage



\section*{Supplementary material}

# SUPPLEMENTARY INFORMATION

**Table of contents**

# I  Structural characterization

When tackling such a difficult sample synthesis procedure, samples characterization is pivotal. The possible structures of Lanthanum hydrides at Mbar pressures have been studied in experimental [Laniel et al. 2022] as well as purely theoretical structural search works [Peng et al. 2017; Shipley et al. 2020; Errea, Belli, et al. 2020]. In this wide literature, the phase most commonly recognized to be stable and that show strong SC behavior displays a $Fm\bar{3}m$ space-group symmetry [Drozdov et al. 2019; Liu et al. 2018]. This comprises a cubic close-packed $fcc$ structure of La atoms, that are arranged in triangular layers with an ABC stacking along the [111] direction of the conventional cubic cell (as shown in **Extended Data Fig. 6**, top left panel). The much smaller H atoms occupy the interstitial positions of such highly symmetric close-packed lattices. Hydrogen atoms can be symmetry-wise grouped in two sets of Wyckoff positions, $8c$ and $32f$: the former are located at tetrahedral interstitial positions, the latter instead arrange in regular cubes at octahedral interstitial sites or in regular tetrahedra around the $8c$ Hydrogen at tetrahedral sites. As a result, they form a cage around each La atom that can be seen as a chamfered cube, a highly symmetric convex polyhedron with 6 congruent squares and 12 congruent flattened hexagons having two internal angles of $109.47°$ and the remaining four of $125.26°$ (**Extended Data Fig. 6**, bottom left). We will refer to this highly symmetric phase as $fcc$-LaH$_{10}$. Several other structural phases have been proposed, including rhombohedral $R\bar{3}m$, orthorhombic $Immm$ and monoclinic $C_2$ or $C_2/m$ ones [Errea, Belli, et al. 2020; Sun et al. 2021] [Eremets et al. 2022]. All such phases retain the ABC stacking of La triangular layers and can be seen as a symmetry-lowering deformation of the cubic close-packed $fcc$ structure mostly involving a rearrangement of the hydrogen atoms and a consequent deformation of the chamfered-cubic cage.

A fundamentally different and competing hexagonal phase with $P6_3/mmc$ space-group symmetry and two formula units has been also proposed as thermodynamically stable and hosting a high-$T_c$ superconducting phase [Shipley et al. 2020]: signatures of this structural phase have been reported as impurities in otherwise cubic structures [Drozdov et al. 2019; Sun et al. 2021]. In this structure La atoms form an hexagonal closed-packed ($hcp$) lattice where La triangular layers display a different ABAB stacking along the hexagonal $c$ axis, that can be related to the $fcc$ structure by a 3-fold rotation of the C layer about an axis orthogonal to the triangular layers and passing through a La atom (see **Extended Data Fig. 6** top panels). Hydrogen atoms can be grouped in three sets of Wyckoff positions of the highest-symmetric hexagonal $P6_3/mmc$ space group, namely $4e$, $4f$ and $12k$. Here the $4f$ H atoms sit at the center of tetrahedral interstitial sites while $4e$ and $12k$ H atoms form both distorted tetrahedra around $4f$ positions and distorted cubes filling the octahedral voids of the $hcp$ lattice. The resulting substantial deformation of the hydrogen cage around each La atom can also be interpreted as a deformation of the chamfered cube where the lower half of the polyhedron with respect to the triangular-layer plane is rotated by $120°$ about a perpendicular axis passing through a La atom at the center of the H cage (**Extended Data Fig. 6**, bottom right panel). Nonetheless, the $P6_3/mmc$ phase displays the largest number of symmetries within the hexagonal crystal systems; henceforth we will refer to it as the $hpc$-LaH$_{10}$ phase. Symmetry-lowering rearrangements of H atoms can be foreseen also for this $hpc$ lattice, in close analogy with the several proposed modification of the $fcc$ phase, that to the best of our knowledge have not been explored so far.

The different arrangement of La atoms in $fcc$ and $hcp$ lattices results in a clear identification of a hexagonal structure of La atoms from X-ray diffraction measurements, that instead cannot provide information on the Hydrogen positions due to its small atomic mass. Nonetheless, useful structural information can be extracted from Raman spectra, especially in the low-frequency range where vibrational signatures are mostly contributed by La phonons. La atoms do not contribute to any Raman-active phonon mode in the highly symmetric $fcc$ $Fm\bar{3}m$ phase. On the other hand, the highly symmetric $P6_3/mmc$ phase allows for a single Raman-active La phonon, contrasting the structured spectra we observe in the 100-200 $cm^{-1}$ frequency range. The latter observation, alongside the XRD characterization of the La hexagonal lattice, suggests a substantial rearrangement of Hydrogen atoms that may cause an overall symmetry reduction with a consequent activation of other La phonon modes. We explored this possibility by performing first-principles calculations of the candidate structures within density functional theory (DFT) using the generalized gradient approximation (GGA) as parametrized by Perdew, Burke and Ernzerhof (PBE) [Perdew et al. 1996].

Harmonic phonon frequencies were calculated within density functional perturbation theory (DFPT) [Baroni et al. 2001] making use of the Quantum ESPRESSO code [Paolo Giannozzi et al. 2009; Giannozzi et al. 2017]. We used optimized norm-conserving Vanderbilt pseudopotentials [Hamann 2013] from the Pseudo dojo repository [van Setten et al. 2018], including 11 electrons for La, with a plane-wave cut-off energy of 110 Ry for the kinetic energy. We adopted $\Gamma$-centered Monkhorst–Pack grid of $k$-points with a spacing of $\sim 0.13$ Å$^{-1}$ and a Gaussian smearing of 0.03 Ry for Brillouin-zone integration. For structural optimization we adopted convergence criteria such that forces on the atoms were less than 2.5 meV Å$^{-1}$ and stresses were less than 0.01 GPa. The static and dynamic phonon frequencies were obtained diagonalizing the dynamical matrix in these two limits as defined in the method and we tested a posteriori the negligible difference between static and dynamic eigen-vectors.

We first optimized the $P6_3/mmc$ structure previously reported [Shipley et al. 2020] at a classical pressure of 135 GPa, as quantum effects have been shown to add around 10-15% of extra pressure [Errea, Belli, et al. 2020; Errea, Calandra, et al. 2016; Belli and Errea 2022]. The calculated lattice parameters are $a = 3.72$ Å and $c = 5.68$ Å, roughly within 1% error with respect to refined XRD lattice constants (structural details are given in **Supplementary Table 1**). The enthalpy of such structure is 4.1 meV/f.u. above that of the $fcc$ phase evaluated at the same pressure. We calculated the phonon dispersion by Fourier-interpolating the harmonic force constants on a $2\times2\times2$ grid comprising the high-symmetry points $\Gamma$, $A$, $L$ and $M$ of the hexagonal Brillouin zone. Several phonon modes display imaginary frequencies, shown as negative values in Supplementary Figure **Supplementary Figure 1**: projection onto the La atomic character clearly shows that unstable modes mostly involves light H atoms, while La atoms mostly contribute to phonon modes in the 0–250 cm$^{-1}$ range. As materials rich in light elements are sensible to quantum anharmonic effects, typically reducing the tendency to dynamical instabilities, we recalculate La-dominated phonon frequencies by artificially reducing the hydrogen atomic mass, thus disentangling the H contribution to those phonon modes**Supplementary Figure 2**. At the $\Gamma$ point of the $P6_3/mmc$ (displaying $D_{6h}$ point-group symmetries) there are only two optical phonon modes due to the La vibrations, among which only the $E_{2g}$ ($\Gamma_5^+$ in the Miller-Love notation) mode at approximatively 90 cm$^{-1}$ is Raman-active, as shown in Fig. **Supplementary Figure 2a**. The most unstable hydrogen phonon is a $M_2^+$ mode at $M$ point of the hexagonal Brillouin zone ($B_{1g}$ using the Mulliken notation for the $D_{2h}$ point group of point $M$), that could lead to a $2\times1\times1$ reconstruction with orthorhombic $Pnma$ structure or to a $2\times2\times1$ hexagonal $P6_3/m$ phase comprising eight formula units. Interestingly, La-derived phonon modes appear in the frequency range 120-200 cm$^{-1}$ at the $M$ point, that might induce additional Raman-active modes when folded at the center of the Brillouin zone after the symmetry-lowering reconstruction (see coloured area in Fig. **Supplementary Figure 2a**). We discard the $2\times1\times1$ orthorhombic reconstruction as it will turn on the Raman activity of the zone-center La phonon $B_{1g}$ at $\sim 250$ cm$^{-1}$, falling outside the targeted frequency range. Following the symmetry-lowering phonon mode $M^{2+}$ we therefore optimized the hexagonal $P6_3/m$ structure and found an enthalpy gain of -2.8 meV/f.u. over $fcc$-LaH$_{10}$. In the reconstructed structure La atoms undergo very tiny displacements (at most of $4\times10^{-3}$ Å) from the ideal positions of the $P6_3/mmc$ phase, while Hydrogen displacements are one order of magnitude larger. We computed phonon frequencies of the reconstructed cell limiting the calculations to inversion-even modes, i.e., those that may display Raman activity, in order to reduce the computational burden in such a large supercell. The zone-center $\Gamma_5^+$ mode of undistorted $hpc$-LaH$_{10}$ is blue-shifted to $\sim 102$ cm$^{-1}$, while other zone-folded phonon modes appear to become Raman-active in the frequency range between 100 and 200 cm$^{-1}$ as shown in Fig. **Supplementary Figure 2b**. Among these, we identify those contributing to satellite peaks at $\sim 150$ and $\sim 197$ cm$^{-1}$ as $E_{1g}$ and $E_{2g}$ modes arising from the folding of La $M^{3+}$ and $M^{2+}$ modes of the $P6_3/mmc$ phase, respectively.

| Space group | Lattice parameters | Wyckoff positions | |
|---|---|---|---|
| $Fm\bar{3}m$ | $a = 5.13709$ Å | La **4b** | $\left[\frac{1}{2}, \frac{1}{2}, \frac{1}{2}\right]$ |
| | | H **8c** | $\left[\frac{1}{4}, \frac{1}{4}, \frac{1}{4}\right]$ |
| | Volume/f.u. = 33.89 Å³ | H **32f** | $[x_f, x_f, x_f]$    $x_f = 0.87908$ |
| $P6_3/mmc$ | $a = 3.72120$ Å | La **2c** | $\left[\frac{1}{3}, \frac{2}{3}, \frac{1}{4}\right]$ |
| | $c = 5.68038$ Å | H **4e** | $[0, 0, z_e]$    $z_e = 0.35384$ |
| | | H **4f** | $\left[\frac{1}{3}, \frac{2}{3}, z_f\right]$    $z_f = 0.61510$ |
| | Volume/f.u. = 34.06 Å³ | H **12k** | $[x_k, 2x_k, z_k]$    $x_k = 0.83410, z_k = 0.08601$ |
| $P6_3/m$ | $a = 7.45130$ Å | La **2c** | $\left[\frac{1}{3}, \frac{2}{3}, \frac{1}{4}\right]$ |
| | $c = 5.67802$ Å | La **6h** | $[x_h, y_h, \frac{1}{4}]$    $x_h = 0.83259, y_h = 0.16563$ |
| | | H **4e** | $[0, 0, z_e]$    $z_e = 0.34603$ |
| | | H **4f** | $\left[\frac{1}{3}, \frac{2}{3}, z_f\right]$    $z_f = 0.61619$ |
| | | H **12i** | $[0.16664, 0.08117, 0.91759]$ |
| | | H **12i** | $[0.16724, 0.58398, 0.91395]$ |
| | | H **12i** | $[0.67170, 0.08756, 0.91236]$ |
| | | H **12i** | $[0.65834, 0.57907, 0.91780]$ |
| | | H **12i** | $[0.99935, 0.49028, 0.64746]$ |
| | Volume/f.u. = 34.13 Å³ | H **12i** | $[0.33147, 0.17916, 0.88389]$ |

**Supplementary Table 1**: **Computed structural parameters** Structural parameters of $fcc$, $hcp$ and distorted $hcp$ phases evaluated from DFT calculations at a classical pressure of 135 GPa. Wyckoff positions are given in crystal coordinates and only for a representative site for each Wyckoff orbit, as the other positions can be obtained by applying the corresponding space-group symmetry operations.

| | $P6_3/mmc$ (# 194) | | | | | | $P6_3/m$ (# 176) | | | |
|---|---|---|---|---|---|---|---|---|---|---|
| | $\Gamma$ ($D_{6h}$) | | | $M$ ($D_{2h}$) | | | $\Gamma$ ($C_{6h}$) | | | |
| | | Static | Dynamic | | | Static | | | Static | Dynamic |
| $\Gamma_5^+$ | $(E_{2g})$ | 89.5 [83.3] | 125.8 [120.7] | | | | $\Gamma_3^+ \oplus \Gamma_5^+$ | $(E_{2g})$ | 102.2 [90.3] | 154.4 [144.8] |
| | | | | $M_3^+$ | $(B_{2g})$ | 139.8 [135.4] | $\Gamma_4^+ \oplus \Gamma_6^+$ | $(E_{1g})$ | 150.6 [131.9] | 163.6 [159.8] |
| | | | | | | | $\Gamma_2^-$ | $(B_g)$ | 178.0 [172.9] | 179.6 [173.9] |
| | | | | $M_1^+$ | $(A_g)$ | 142.7 [134.2] | $\Gamma_3^+ \oplus \Gamma_5^+$ | $(E_{2g})$ | 164.6 [160.4] | 182.4 [178.4] |
| | | | | | | | $\Gamma_1^+$ | $(A_g)$ | 178.1 [176.8] | 179.0 [177.8] |
| | | | | $M_2^+$ | $(B_{1g})$ | 182.6 [181.3] | $\Gamma_1^+$ | $(A_g)$ | 185.1 [180.9] | 192.8 [189.5] |
| | | | | | | | $\Gamma_3^+ \oplus \Gamma_5^+$ | $(E_{2g})$ | 197.0 [192.5] | 210.9 [198.3] |
| $\Gamma_4^+$ | $(B_{1g})$ | 254.1 [253.3] | 256.1 [255.4] | | | | | | | |

**Supplementary Table 2**: **Calculated phonon frequencies in the static harmonic approximation.** La-derived phonon modes that can display Raman optical activity. Phonon modes are labeled according to the irreducible representations of the point group relevant for the crystal structure and phonon momentum, using the Miller-Love notation (Mulliken notation in round brackets). Phonon (static) frequencies are given in cm$^{-1}$ in the limit of infinitely small Hydrogen mass, the corresponding value obtained for the physical $m_H$ being given in square brackets. For the higher-symmetry $hcp$-LaH$_{10}$ phase we list the two La phonons at $\Gamma$, only one of which is Raman active; the optically inactive phonon is written in gray. We further provide the phonon frequencies of inversion-even modes at the $M$ point whose folding in the lower-symmetric $2{\times}2{\times}1$ reconstruction gives rise to additional Raman-active phonon modes. The folding of the $M_3^+$ mode in the zone center of the $P6_3/m$ produces a Raman-active $E_{1g}$ mode and an optical inactive $B_g$ mode (gray-shaded) that appears to hybridize with the zone-center phonon mode labeled as $\Gamma_4^+$ in the high-symmetry structure. **Extended Data Table 1** compares the ab initio frequencies of the $P6_3/m$ reconstruction with the values obtained from the analysis of the experimental spectra.

.

3

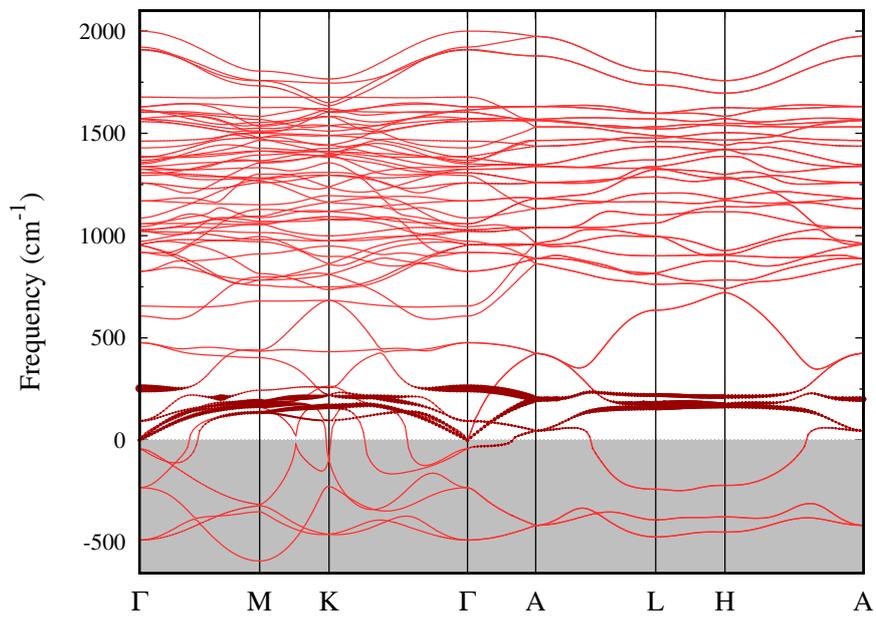

**Supplementary Figure 1**: **Phonon dispersion of $hpc$-LaH$_{10}$ from DFT.** Harmonic phonon dispersion of hexagonal LaH$_{10}$ in the $P6_3/mmc$ phase, interpolated from a 2×2×2 grid comprising the high-symmetry points $\Gamma$, $A$, $L$ and $M$. The phonon modes with dominant La character in the frequency range 0-250 cm$^{-1}$ is highlighted by dark-red points whose size is proportional to the La contribution. The grey shading marks the region with unstable (imaginary) phonon frequencies, here depicted as negative frequencies. The strongest dynamical instability occurs for a $B_{1g}$ ($M_2^+$ in Miller-Love notation) phonon mode contributed only by Hydrogens at the $M$ point.

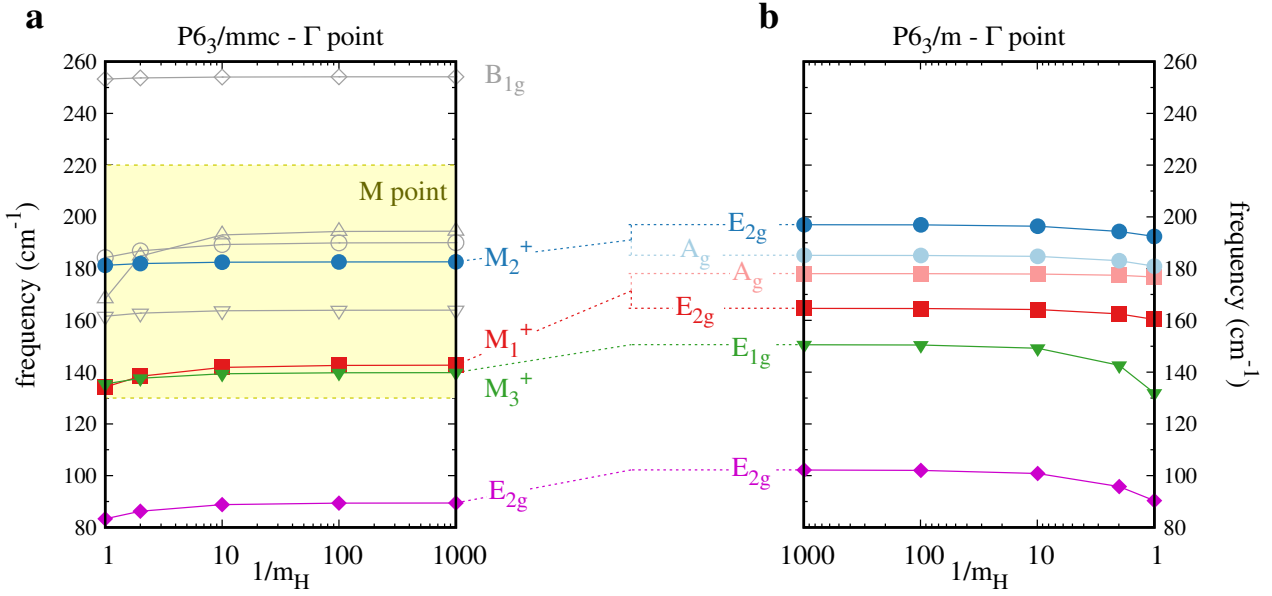

**Supplementary Figure 2**: **Lanthanum-derived Raman-active modes in ideal** $hpc$-**LaH$_{10}$ and in reconstructed cell.** In order to identify the interplay of La and H contribution in low-energy phonon modes with dominant Lanthanum character, we show the evolution of phonon frequencies as a function of inverse Hydrogen mass, where the Hydrogen contributions can be fully suppressed in the limit of vanishing $m_H$. **a** At $\Gamma$ point of $hpc$-LaH$_{10}$ only two modes are significantly contributed by La atoms, a Raman-active $E_{2g}$ and an optically inactive $B_{1g}$ mode, displayed as grey empty symbols. Phonons are labeled using the Mulliken notation for the irreducible representations of the $D_{6h}$ point group describing the symmetry properties at the $\Gamma$ point. The coloured region highlights six La-derived phonon modes found at $M$ point of LaH$_{10}$ in the $P6_3/mmc$ phase, labeled using the Miller-love notation. Among these only three are even under inversion and expected to give rise to Raman-active folded phonons. They are displayed with solid symbols with red, blue and green colours, while the inversion-odd partners are shown as empty grey symbols. Using Mulliken notation for the $M$ point with $D_{2h}$ point-group symmetries, the inversion-even modes correspond respectively to $A_g$, $B_{1g}$ and $B_{2g}$ irreducible representations. **b** Raman-active La-derived phonon modes for the $2 \times 2 \times 1$ reconstructed cell with $P6_3/m$ symmetry (crystallographic point grop $C_{6h}$). The lowest $E_{2g}$ mode corresponds to the zone-center Raman-active mode of the higher-symmetry $P6_3/mmc$ phase, that is however blue-shifted at slightly higher frequencies most probably as a consequence of the different H-sublattice. The folding of $M_1^+$ and $M_2^+$ phonons of the higher-symmetry structure split into $E_{2g} + A_g$ modes at the $\Gamma$ point of the lower-symmetry phase, with a splitting of approximatively 12-14 cm$^{-1}$. Finally, $M_3^+$ phonons are zone-folded and split in a Raman-active $E_{1g}$ mode (green solid triangles) strongly sensitive to the H-sublattice, and in a silent $B_g$ mode (not shown) that is blue-shifted by $\sim$ 30 cm$^{-1}$.

| Convention | "ladder" operator $\phi_\mu$ | "displacement" operator $u_{s\alpha}$ |
|---|---|---|
| Phonon propagator | $\tilde{D}_{\mu\nu}(t-t') = -i\left\langle \hat{T}\phi_\mu(t)\phi_\nu^\dagger(t')\right\rangle$ | $D_{\mu\nu}(t-t') = -i\sqrt{M_s M_{s'}}\left\langle \hat{T}u_{s\alpha}(t)u_{s'\beta}^\dagger(t')\right\rangle \xi_{s\alpha,\mu}\xi_{s'\beta,\nu}$ |
| Bare Phonon propagator | $\tilde{D}_{\mu\mu}^{(0)}(z) = \frac{2\omega_\mu^0}{z^2-(\omega_\mu^0)^2}$ | $D_{\mu\mu}^{(0)}(z) = \frac{1}{z^2-(\omega_\mu^{\mathrm{dyn}})^2}$ |
| Dyson Equation | $\tilde{D}_{\mu\mu}(z) = 2\omega_\mu^0\left[z^2 - (\omega_\mu^0)^2 - 2\omega_\mu^0\tilde{\Sigma}_\mu(z)\right]^{-1}$ | $D_{\mu\mu}(z) = \left[z^2 - (\omega_\mu^{\mathrm{dyn}})^2 - \Sigma_\mu(z)\right]^{-1}$ |
| Phonon self energy | $\tilde{\Sigma}_\mu(z) = -\hbar 2 N_F \left\langle |g_\mu|^2\right\rangle_{S_F} I(z)$ | $\Sigma_\mu(z) = (\omega_\mu^{\mathrm{stat}} - \omega_\mu^{\mathrm{dyn}})^2 I(z)$ |
| Phonon spectral func. | $\tilde{A}_\mu(\omega) = \lim_{z\to\omega+i0^+} \mathrm{Im}\,\frac{-2}{\pi}\tilde{D}_{\mu\mu}(z)$ | $A_\mu(\omega) = \lim_{z\to\omega+i0^+} \mathrm{Im}\,\frac{-2z}{\pi}D_{\mu\mu}(z)$ |
| Phonon spectral func. norm | $\int_{-\infty}^{\infty} \tilde{A}_\mu(\omega) = 0$ | $\int_{-\infty}^{\infty} A_\mu(\omega) = 2$ |
| Raman profile | $\tilde{\mathcal{P}}_\mu(\omega) = \frac{1}{4\omega_\mu^0}\tilde{A}_\mu(\omega)$ | $\mathcal{P}_\mu(\omega) = \frac{1}{2\omega}A_\mu(\omega)$ |
| Parameters | $\left\{\omega_\mu^0\,;\, -2N_F\left\langle |g_\mu|^2\right\rangle_{S_F}\,;\, \eta_{imp}\right\}$ | $\left\{\omega_\mu^{\mathrm{dyn}}\,;\, \omega_\mu^{\mathrm{stat}}\,;\, \eta_{imp}\right\}$ |

**Supplementary Table 3**: Definition of pivotal quantities in the "Dressed bubble" approximation for phonon propagator according to different conventions.

## II  Ab initio dressed bubble approximation for phonon spectral function

### A  Ladder operators convention

The central quantity in the many-body description of crystal vibrations is the phonon propagator $D_{\mu\nu}(z)$ defined as the retarded correlation of the phonon field operator $\phi_\mu$ [Mahan 2000, p. 118]

$$\tilde{D}_{\mu\nu}(t-t') = -i\theta(t-t')\left\langle \phi_\mu(t)\phi_\nu^\dagger(t')\right\rangle,\tag{S1}$$

where $\theta(t)$ is the Heaviside function.

In order to describe the low-energy interaction between the electronic and ionic degrees of freedom, the electron-phonon coupling $g_{klm,\mathbf{q}\mu}$, relating two electronic states $|kl\rangle$ and $|\mathbf{k}+\mathbf{q}m\rangle$ and a phonon mode $\mu$, is introduced. Furthermore, in many cases, it is reasonable to neglect isolated degeneracy of the Fermi surface and thus it is reasonable to identify the Fermi contribution with the electronic intraband ($l=m$) excitation and higher-energy contributions with interband ($l\neq m$) transitions. In second quantization lattice vibrations are described by the creation (annihilation) operators $a_{\mathbf{q}\mu}^\dagger$ ($a_{\mathbf{q}\mu}$), so that the phonon field operators are $\phi_{\mathbf{q}\mu}(t) = \left(a_{\mathbf{q}\mu}e^{i\omega_{\mathbf{q}\mu}t} + a_{\mathbf{q}\mu}^\dagger e^{-i\omega_{\mathbf{q}\mu}t}\right)$. The physical displacement $u_{s\alpha}^L(t)$ of the $s$-th ion in the $L$ unit-cell, along the $\alpha$ direction can be retrieved as [Mahan 2000, p. 10 eq. 1.85]

$$u_{s\alpha}^L(t) = \sum_{\mathbf{q}} \frac{e^{i\mathbf{q}\cdot\mathbf{R}_L}}{\sqrt{N}} u_{s\alpha}^{\mathbf{q}}(t) = \sum_{\mathbf{q}\mu} \frac{e^{i\mathbf{q}\cdot\mathbf{R}_L}}{\sqrt{N}}\sqrt{\frac{\hbar}{M_s 2\omega_{\mathbf{q}\mu}^0}}\xi_{\mathbf{q},\mu}^{s\alpha}\phi_{\mathbf{q}\mu}(t),\tag{S2}$$

where $N$ is the number of unit cells considered in the Born-Von Karman boundary conditions, $\xi_{\mathbf{q},\mu}^{s\alpha}$ is the eigenvector of the phonon mode $\mu$ at the $\mathbf{q}$ point and ionic mass $M_s$. The system is then described by the following low-energy Hamiltonian [Maksimov and Shulga 1996, Eq. (2)]:

$$H = \sum_{\mathbf{k}l}\varepsilon_{\mathbf{k}l}c_{\mathbf{k}l}^\dagger c_{\mathbf{k}l} + \sum_{\mathbf{q}\mu}\hbar\omega_{\mathbf{q}\mu}^0\left(a_{\mathbf{q}\mu}^\dagger a_{\mathbf{q}\mu} + \frac{1}{2}\right) + \sum_{\mathbf{k}l,\mathbf{q}\mu}\hbar g_{\mathbf{k}l,\mathbf{q}\mu}c_{\mathbf{k}l}^\dagger c_{\mathbf{k}l}(a_{\mathbf{q}\mu} + a_{-\mathbf{q}\mu}^\dagger)\tag{S3}$$

where electronic states of energy $\varepsilon_{\mathbf{k}l}$ are populated by $c_{\mathbf{k}l}^\dagger$ operators, with "l" labelling both the band index and the electronic spin for a lighter notation.

From now on we consider only the $q=0$ case, being interested in the spectroscopic response of the system. The first building block for an evaluable expression of the phonon propagator, is the "bare" phonon propagator defined as the correlation in the ground state at zero temperature $\langle\cdot\rangle_0$

$$\tilde{D}_{\mu\nu}^{(0)}(t-t') = -i\theta(t-t')\left\langle \phi_\mu(t)\phi_\nu^\dagger(t')\right\rangle_0.\tag{S4}$$

Since at zero temperature the Fermi surface does not contribute, the bare phonon propagator coincides with the insulator-like contribution and thus it is completely described by the bare phonon frequency

$$\tilde{D}_{\mu\mu}^{(0)}(z) = \frac{2\omega_\mu^0}{z^2-(\omega_\mu^0)^2}\tag{S5}$$

where we express the time-dependent correlation in the complex frequency space $z = \omega + i\eta$.

The complete phonon propagator Eq. (S1) includes the effects of intraband electron-phonon interaction accounted for by the phonon self-energy $\tilde{\Sigma}_{\mu\mu'}(\omega)$ as defined by the following Dyson equation [Mahan 2000, Eq. (2.123)]:

$$\tilde{D}_{\mu\nu}(z) = \frac{\tilde{D}_{\mu\nu}^{(0)}(z)}{1 - \tilde{D}_{\mu\nu}^{(0)}(z)\tilde{\Sigma}_{\mu\nu}(z)} = \frac{2\omega_\mu^0}{z^2 - (\omega_\mu^0)^2 - 2\omega_\mu^0\tilde{\Sigma}_{\mu\nu}(z)}. \tag{S6}$$

For the sake of simplicity, we neglect the mode mixing that due to $\tilde{\Sigma}_{\mu\nu}(z)$ and focus on just the diagonal terms. Besides being a valid approximation whenever the phonon frequencies are well-separated, this approximation can be easily lifted. Secondly, we assume that the contributions at the Fermi level are isotropic. Neglecting anisotropy greatly simplified the derivation since it allows reducing the 3D integral over momenta $\mathbf{k}$ to the energy integral over $|\mathbf{k}|$. Furthermore, it implies that all quantities computed at the Fermi level coincide with their average over the Fermi surface, starting from the electron-phonon coupling $g_{\mathbf{k}l,\mu}$ whose Fermi average is:

$$\langle |g_{\mathbf{k}l,\mu}|^2 \rangle_{S_F} = -\frac{1}{2N_{\mathrm{F}}} \frac{2}{N_{\mathbf{k}}} \sum_{\mathbf{k},l} f'(\varepsilon_{\mathbf{k}l})|g_{\mathbf{k}l\mu}|^2 \tag{S7}$$

where $N_{\mathrm{F}}$ stands for the electronic density of state per spin at Fermi. The validity of the isotropic approximation is system-specific and has to be verified. While it was considered commonly valid [Allen and Mitrović 1983, p. 45] [Shepelev 1969], the recent study [Lucrezi et al. 2024] questions its applicability to super-Hydrides. The agreement between the theoretical isotropic model and experimental observations supports the validity of the isotropic approximation in the case of h-LaH$_{10}$. The general anisotropic theory is discussed in [Allen and Mitrović 1983] and [Marchese et al. 2024], we leave the generalization of the present implementation to the anisotropic case to subsequent studies.

The isotropic phonon self-energy reads as in [Zeyher and Zwicknagl 1988, Eq. (13)] and [Maksimov and Shulga 1996, Eq. (13)]:

$$\tilde{\Sigma}_\mu(z) = -\hbar 2N_{\mathrm{F}} \langle |g_{\mathbf{k}l,\mu}|^2 \rangle_{S_F} I(z) \tag{S8}$$

where the dimensionless function $I(z)$ defined as [Zeyher and Zwicknagl 1988, Eq. (8)]:

$$I(z) = -\frac{k_{\mathrm{B}}T}{2N_{\mathrm{F}}N_k} \sum_{\mathbf{k}i,\Omega_n} \mathrm{Tr} \left[ \hat{\tau}_3 \hat{G}_{\mathbf{k}i}(i\Omega_n)\hat{\tau}_3 \hat{G}_{\mathbf{k}i}(i\Omega_n + i\omega_m) \right]_{i\omega_m \Rightarrow z}, \tag{S9}$$

encodes completely the temperature and frequency dependence of the phonon self-energy. The electronic propagator appearing in (S9) is written in the Nambu spinor representation and accounts for the electron-phonon and impurity scattering and the Cooper pairs formation across the SC transition. Its explicit form is then found solving the Migdal-Eliashberg systems of equations for a given Eliashberg function $\alpha^2 F(\omega)$ determining the electron-phonon coupling. Once the dimensionless function and thus the phonon propagator is found, the peak profile entering the vibrational Raman contribution can be written as:

$$\tilde{\mathcal{P}}_\mu^{vib}(\omega) = \frac{\tilde{A}_\mu(\omega)}{4\omega_\mu^0}, \tag{S10}$$

where the phonon spectral function is commonly defined as [Mahan 2000, Eq. (3.114)]

$$\tilde{A}_\mu(\omega) = -\lim_{z\to\omega+i0^+} \mathrm{Im} \frac{2}{\pi}\tilde{D}_{\mu\mu}(z), \tag{S11}$$

with the following normalization properties:

$$\int_{-\infty}^{\infty} d\omega \tilde{A}_\mu(\omega) = 0 \qquad \int_0^\infty d\omega \tilde{A}_\mu(\omega) = 2\omega_\mu^0. \tag{S12}$$

In conclusion, this many-body approach requires estimating only three parameters entering the (low-energy) effective model: the bare phonon frequency $\omega_\mu^0$, the electron-phonon coupling contribution of the Fermi surface $-2N_{\mathrm{F}} \langle |g_{\mathbf{k}l,\mu}|^2 \rangle_{S_F}$ and the level of impurities $\eta_{imp}$. An example of obtaining the electron-phonon coupling contribution from the experimental data can be found in [Cappelluti 2006], where the value of $-2N_{\mathrm{F}} \langle |g_{\mathbf{k}l,\mu}|^2 \rangle_{S_F}$ is estimated by the zero-point motion while $\omega_\mu^0$ is treated as a fitting parameter of the model. In the following section, we rewrite these quantities in a slightly different convention required for expressing these phenomenological parameters in terms of ab initio quantities evaluable in DFT. **Supplementary Table 3** reports the main quantities for the "dressed bubble" approximation in the two conventions.

## B The displacement operator convention

To draw a connection between the many-body formalism and the ab initio framework based on DFPT, it is convenient to adopt a different definition of the phonon propagator based on displacement operators $u_{s\alpha}$ rather than field operators $A_\mu$ [Abrikosov et al. 1975]:

$$D_{ss'\alpha\beta}(t-t') = -i\sqrt{M_s M_{s'}}\theta(t-t')\left\langle u_{s\alpha}(t)u^\dagger_{s'\beta}(t')\right\rangle. \tag{S13}$$

The equivalence of the two conventions is insured by the linearity of Eq. (S2) relating field and displacement operator. Choosing the displacement operator as the descriptor of the lattice vibrations implies that the coupling between the electronic and the ionic degrees of freedom should be replaced accordingly. Thus, the "electron-displacement" coupling $V_{\mathbf{k}l,\mathbf{q}s\alpha}$, also known as "deformation potential", assumes the role of the electron-phonon coupling $g_{\mathbf{k}l,\mu}$. If we write the deformation potential in the normal mode basis as:

$$V_{\mathbf{k}l,\mu} = \sum_{s\alpha}\sqrt{\frac{1}{\hbar M_s}}\xi_{\mu,s\alpha}V_{\mathbf{k}l,s\alpha}, \tag{S14}$$

we find that the two coupling are related by a linear relation [Giustino 2017]:

$$V_{\mathbf{k}l,\mu} = \sqrt{2\omega_\mu^0}g_{\mathbf{k}l,\mu}. \tag{S15}$$

Following the same steps of the previous section we find that the bare phonon propagator in this convention reads

$$D_\mu^{(0)}(z) = \xi_{\mu,s\alpha}\xi_{\mu,s'\beta}D_{ss'\alpha\beta}^{(0)}(z) = \frac{1}{z^2 - (\omega_\mu^0)^2}; \tag{S16}$$

the complete phonon propagator is then

$$D_\mu(z) = \frac{1}{z^2 - (\omega_\mu^0)^2 - \Sigma_\mu(z)} \tag{S17}$$

and the self-energy at leading order is

$$\Sigma_\mu(z) = -2N_F\hbar\left\langle|V_{\mathbf{k}l\mu}|^2\right\rangle_{S_F} I(z). \tag{S18}$$

The spectral function in this convention has the following definition for the phonon spectral function (as [Berges et al. 2023] equivalent to [Abrikosov et al. 1975, Eq. (7.29)])

$$A_\mu(\omega) = -\lim_{z\to\omega+i0^+}\text{Im}\frac{2z}{\pi}D_\mu(z). \tag{S19}$$

and its normalization properties are independent of the bare phonon frequency

$$\int_{-\infty}^{\infty}d\omega A_\mu(\omega) = 2 \qquad \int_0^\infty d\omega A_\mu(\omega) = 1 \tag{S20}$$

as proved in Section C. Finally, the peak profile for the vibrational Raman contribution is:

$$\mathcal{P}_\mu^{vib}(\omega) = \frac{A_\mu(\omega)}{2\omega}. \tag{S21}$$

Wrapping up the comparison between these two conventions, we see how the linear relation between the two operators

$$\phi_\mu = \sqrt{\frac{M_s 2\omega_\mu^0}{\hbar}}\xi_{\mu,s\alpha}u_{s\alpha} \tag{S22a}$$

propagates to the following quantities:

$$\left\langle|V_{\mathbf{k}l,\mu}|^2\right\rangle_{S_F} = 2\omega_\mu^0\left\langle|g_{\mathbf{k}l,\mu}|^2\right\rangle_{S_F} \tag{S22b}$$

$$\Sigma_\mu(z) = 2\omega_\mu^0\tilde{\Sigma}_\mu(z) \tag{S22c}$$

$$D_\mu(z) = \frac{1}{2\omega_\mu^0}\tilde{D}_\mu(z) \tag{S22d}$$

$$A_\mu(\omega) = \frac{\omega}{2\omega_\mu^0}\tilde{A}_\mu(\omega) \tag{S22e}$$

$$\mathcal{P}_\mu(\omega) = \frac{2\omega_\mu^0}{\omega}\tilde{\mathcal{P}}_\mu(\omega). \tag{S22f}$$

As a consequence of this mapping, the parameters to be calculated in this convention are $\left\langle |V_{\mathbf{k}l,\mu}|^2 \right\rangle_{S_F}$ quantifying the electron-phonon coupling of the Fermi surface; and $\omega_\mu^0$ describing the insulator-like contribution that does not include the Fermi surface contribution. As detailed in the method section, the following identities allows to easily calculated these parameters in a DFPT framework:

$$-\hbar 2N_{\mathrm{F}} \left\langle |V_{\mathbf{k}l\mu}|^2 \right\rangle_{S_F} = (\omega_\mu^{\mathrm{stat}})^2 - (\omega_\mu^{\mathrm{dyn}})^2 \tag{S23}$$

$$\omega_\mu^0 = \omega_\mu^{\mathrm{dyn}}. \tag{S24}$$

## C  Spectral function

A similar arbitrariness is present in the definition of the phonon spectral function. We find useful the following definition for the phonon spectral function due to its normalization properties (S19).

The reasons for adopting this phonon spectral function definition are to be found in the following proof of the normalization properties of Eq. (S19) follows. We can calculate the norm of the spectral function solving the following integral

$$\int_{-\infty}^{\infty} d\omega A_\mu(\omega + i0^+) = \frac{-2}{\pi} \operatorname{Im} \int_{-\infty}^{\infty} d\omega \frac{\omega + i0^+}{(\omega + i0^+)^2 - (\omega_\mu^{\mathrm{dyn}})^2 - \Sigma_\mu(\omega + i0^+)} \tag{S25}$$

in the complex space taking advantage of the of the analytic properties of the phonon self-energy. Since the imaginary part of the phonon self-energy is an odd function of the frequency and has negative values for positive frequencies, the $A_\mu(z)$ has only poles in the lower half of the complex space.

Then, the normalization integral can be calculated considering a closed line-integral in the upper half-plane consisting in a $2R$ length segment of the real axis and an half-circumference of radius $R$. Since there are no poles enclosed in such simple closed curve, the total integral is null thanks to the Cauchy theorem. As a consequence, the integral over the real axis is equal to minus the half-circumference contribution in the limit of infinite radius $R$. We proceed calculating the half-circumference contribution:

$$\frac{-2}{\pi} \operatorname{Im} \lim_{R\to\infty} \int_0^\pi d\theta(iRe^{i\theta}) \frac{Re^{i\theta}}{(Re^{i\theta})^2 - (\omega_\mu^{\mathrm{dyn}})^2 - \Sigma_\mu(Re^{i\theta})}. \tag{S26}$$

Assuming that asymptotically the self-energy becomes a constant due to the absence of high energy interaction, we can simplify the denominator in the asymptotic regime defining the complex constant $\mathcal{Z} = (\omega_\mu^{\mathrm{dyn}})^2 + \Sigma_\mu(Re^{i\theta})$ for $R \gg R_c$ and thus:

$$\int_{-\infty}^{\infty} d\omega A_\mu(\omega + i0^+) = \frac{2}{\pi} \operatorname{Im} \lim_{R\to\infty} \int_0^\pi d\theta \frac{i(Re^{i\theta})^2}{(Re^{i\theta})^2 - \mathcal{Z}} = \frac{2}{\pi} \operatorname{Im} \int_0^\pi d\theta \left[ \lim_{R\to\infty} \frac{i}{1 - \frac{\mathcal{Z}}{(Re^{i\theta})^2}} \right] = \frac{2}{\pi} \operatorname{Im} \int_0^\pi d\theta i = 2. \tag{S27}$$

# III    Pressure estimation from Raman spectrum

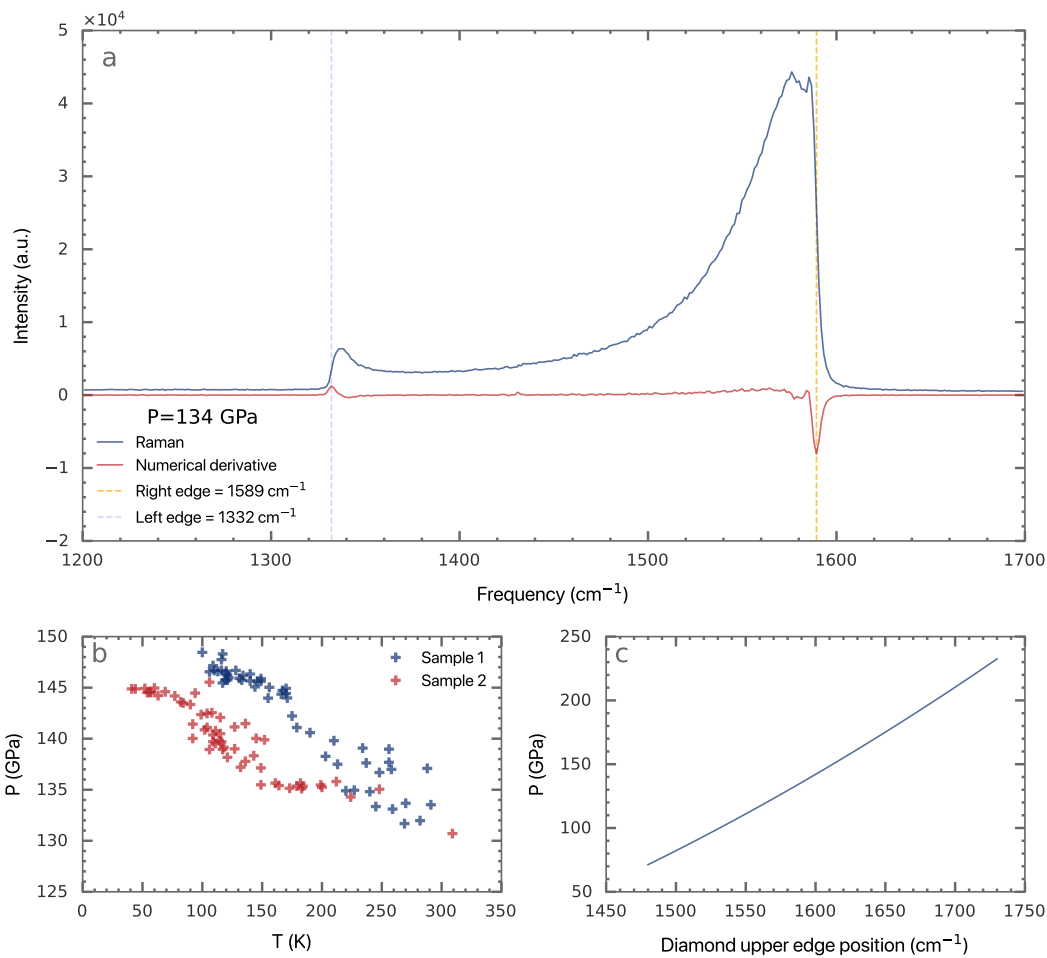

**Supplementary Figure 3**:  **Pressure determination from the stressed diamond Raman phonon. Panel a** shows the Raman from the stressed diamond (blue curve) and its derivative (red curve). The maxima and the minima in the derivative mark the unstressed and stressed edges of the diamond signal, respectively. The two edges are marked by vertical dashed lines and are then used to determine the pressure Akahama and Kawamura 2006. As shown in **panel b**, our datasets demonstrate an increase in pressure when decreasing temperature, typical for high-pressure studies.

# IV    Spectra collection

In this section, we present the complete set of Raman spectra analysed, along with the results of the multi-peak fitting procedure. Not all fitted peaks were used to extract the parameters of the Migdal-Eliashberg model or plotted in the article figures. In the following collection of plots, the peaks included in the ME analysis are shown as shaded regions, while the unselected ones are displayed as hatched areas.

To ensure an unbiased selection of the data, we applied a filtering criterion based on the ratio ($r = \Delta x/|x|$) between the fitting uncertainty ($\Delta x$) and the fitted value ($x$) of each peak parameter –namely amplitude, position, and linewidth ($c_i, \omega_i, \Gamma_i$ in Eq. (2)). For sample 1, we excluded from the analysis all fits with ratios exceeding $r_A > 0.12$ for peak A, $r_D > 0.14$ for peak D, and $r_E > 0.30$ for peak E. An additional criterion was adopted for peak A, due to its sharpness, consisting in studying only spectra obtained with the 1200 gr/mm interferometer grating. For sample 2, we excluded results with $r_A > 0.28$, $r_D > 0.25$, and $r_E > 0.35$.

No threshold criterion was applied to peaks B and C, as their limited signal-to-noise ratio prevents a detailed Migdal-Eliashberg analysis. For these modes, only the averaged peak positions are reported in **Extended Data Table 1**.

Finally, we verified the robustness of our conclusions by testing multiple selection thresholds, finding no significant changes in the resulting analysis.

# A  Spectra sample 1

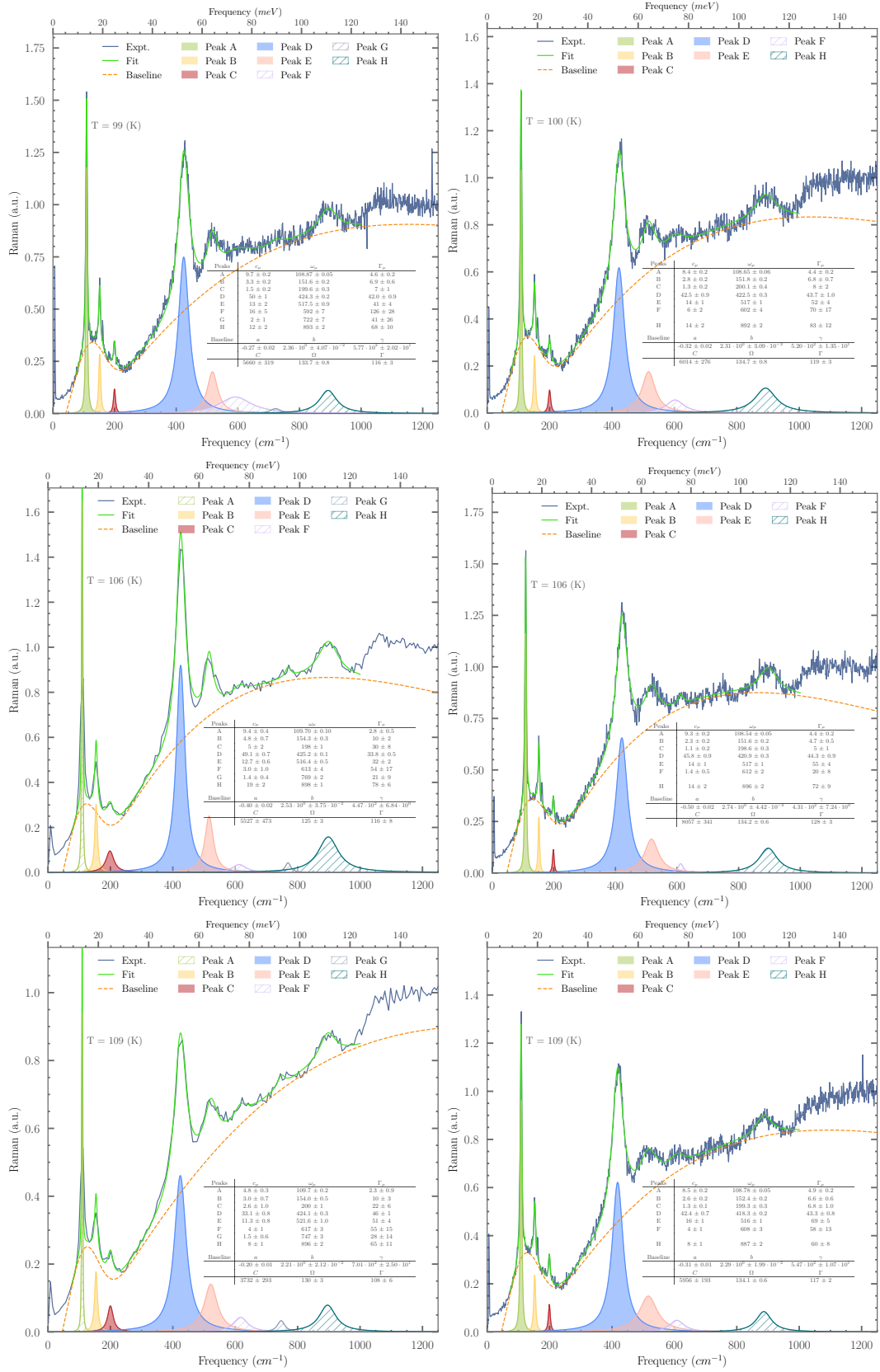

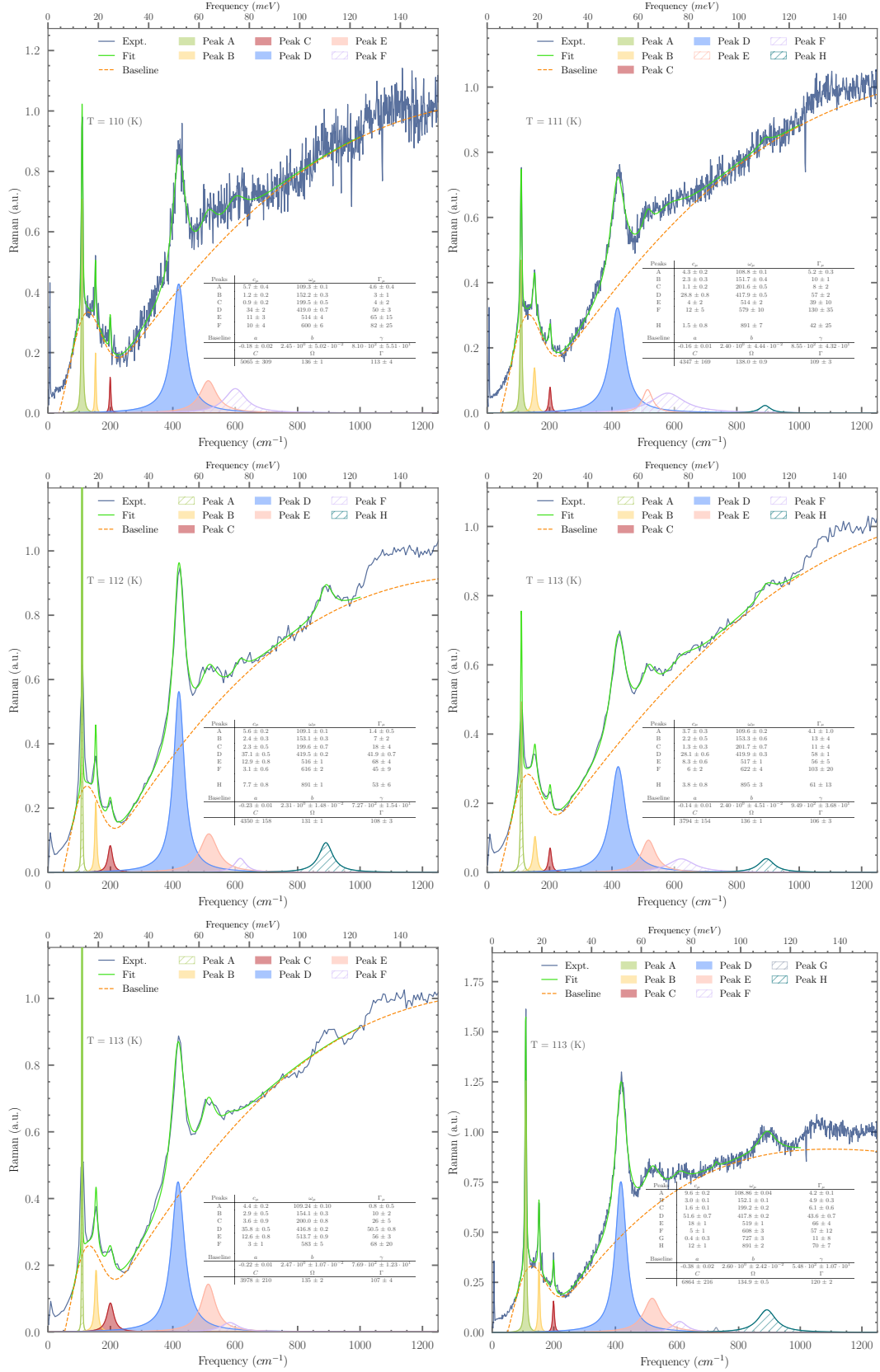

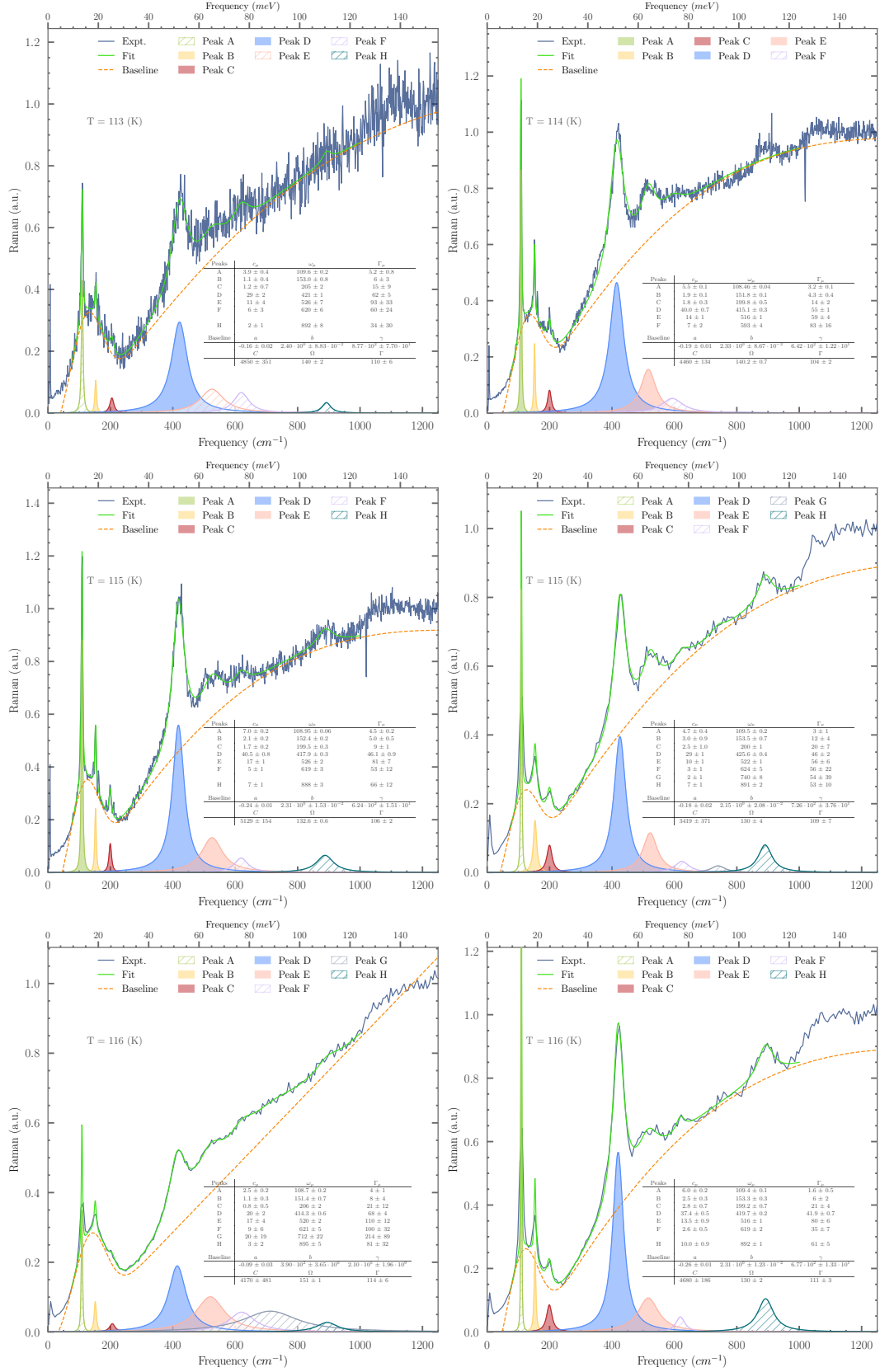

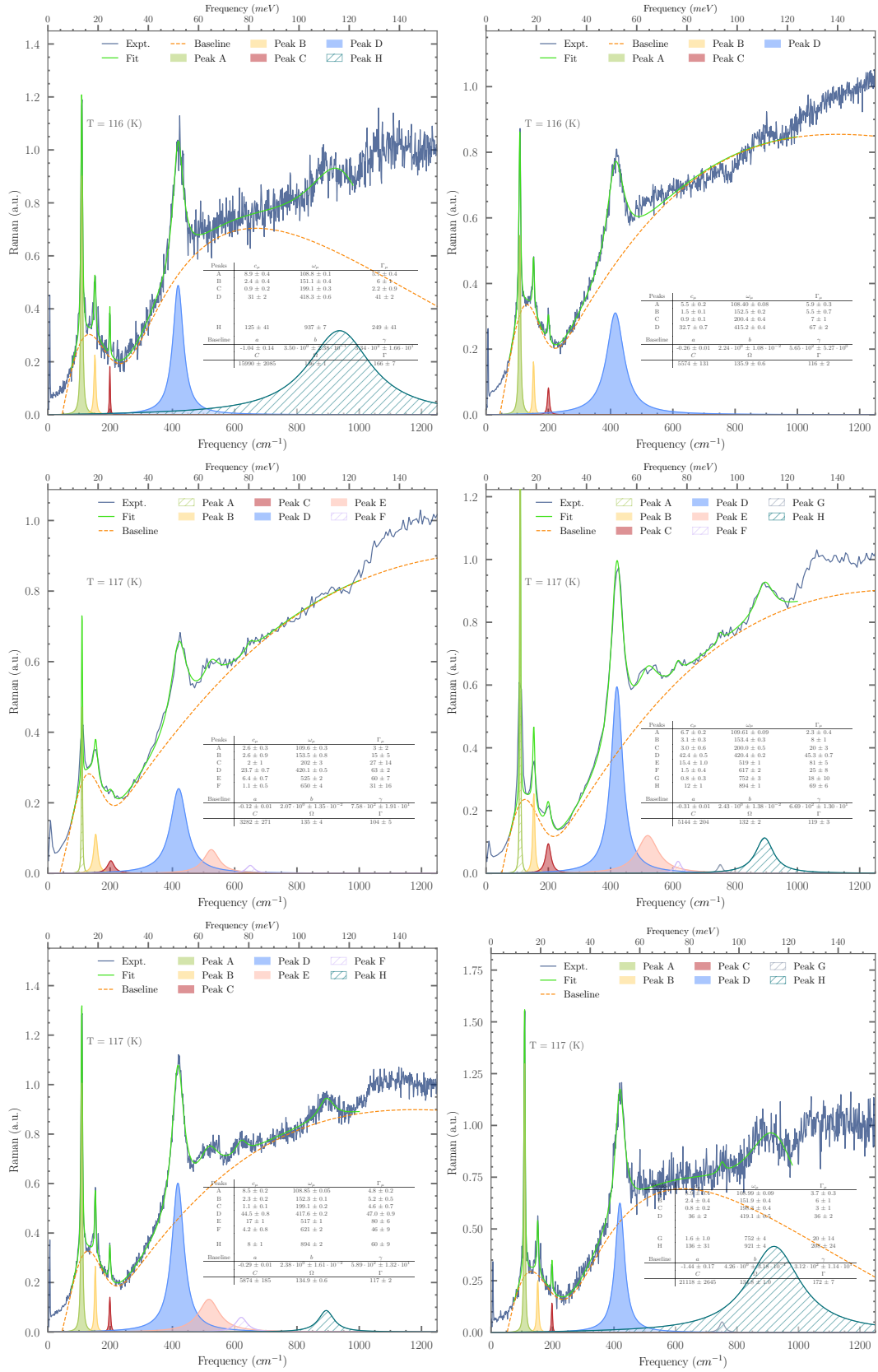

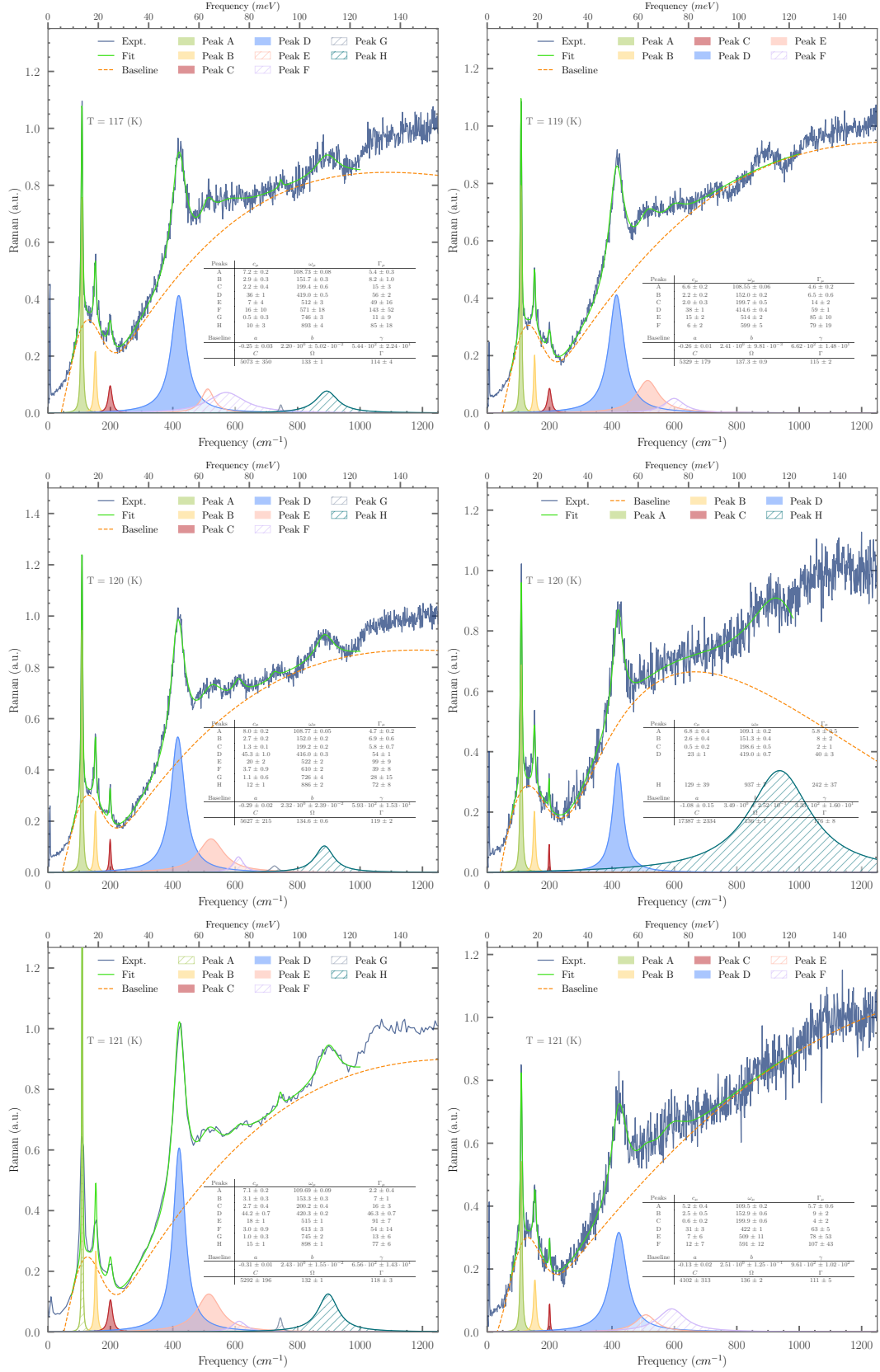

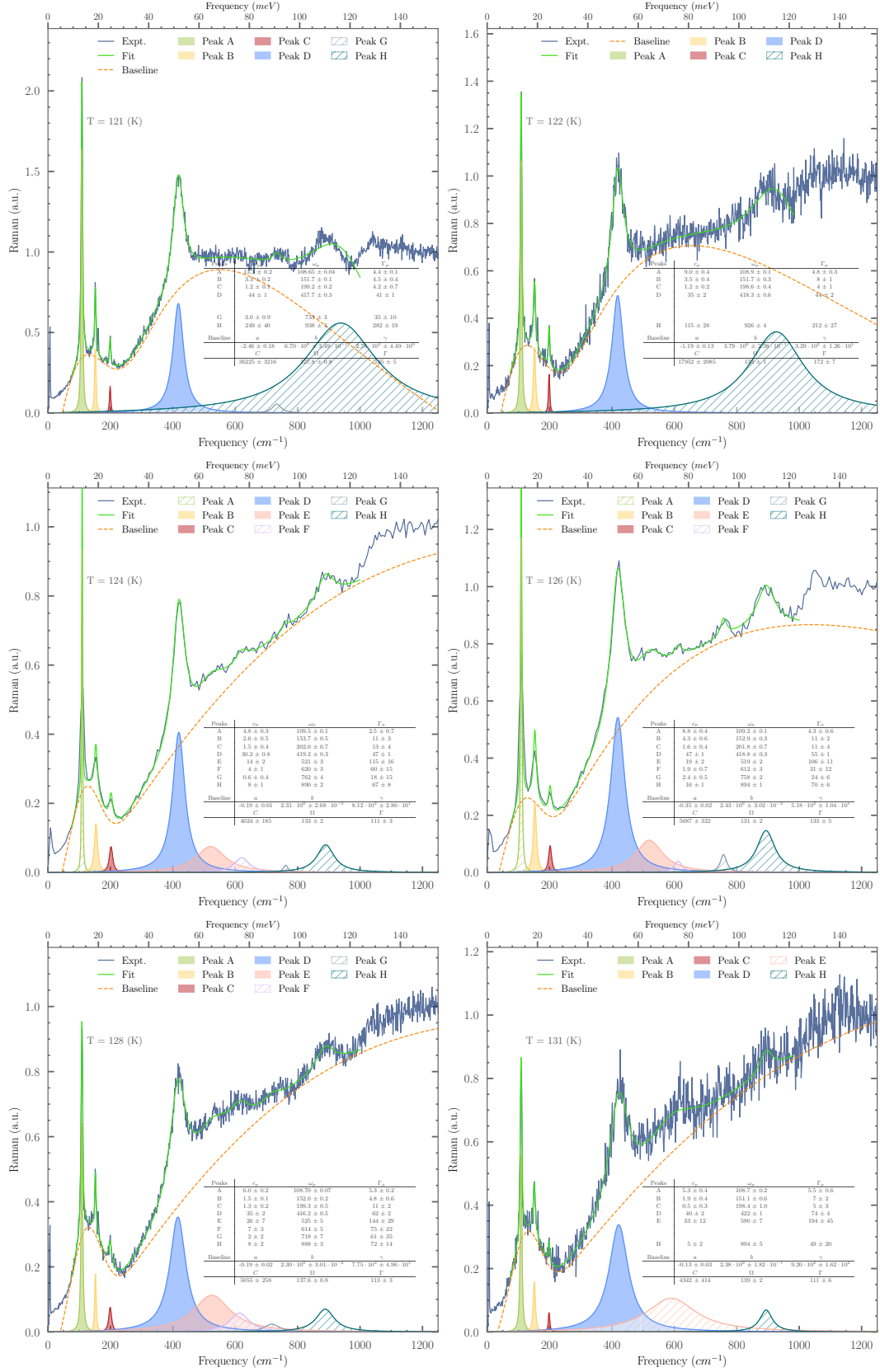

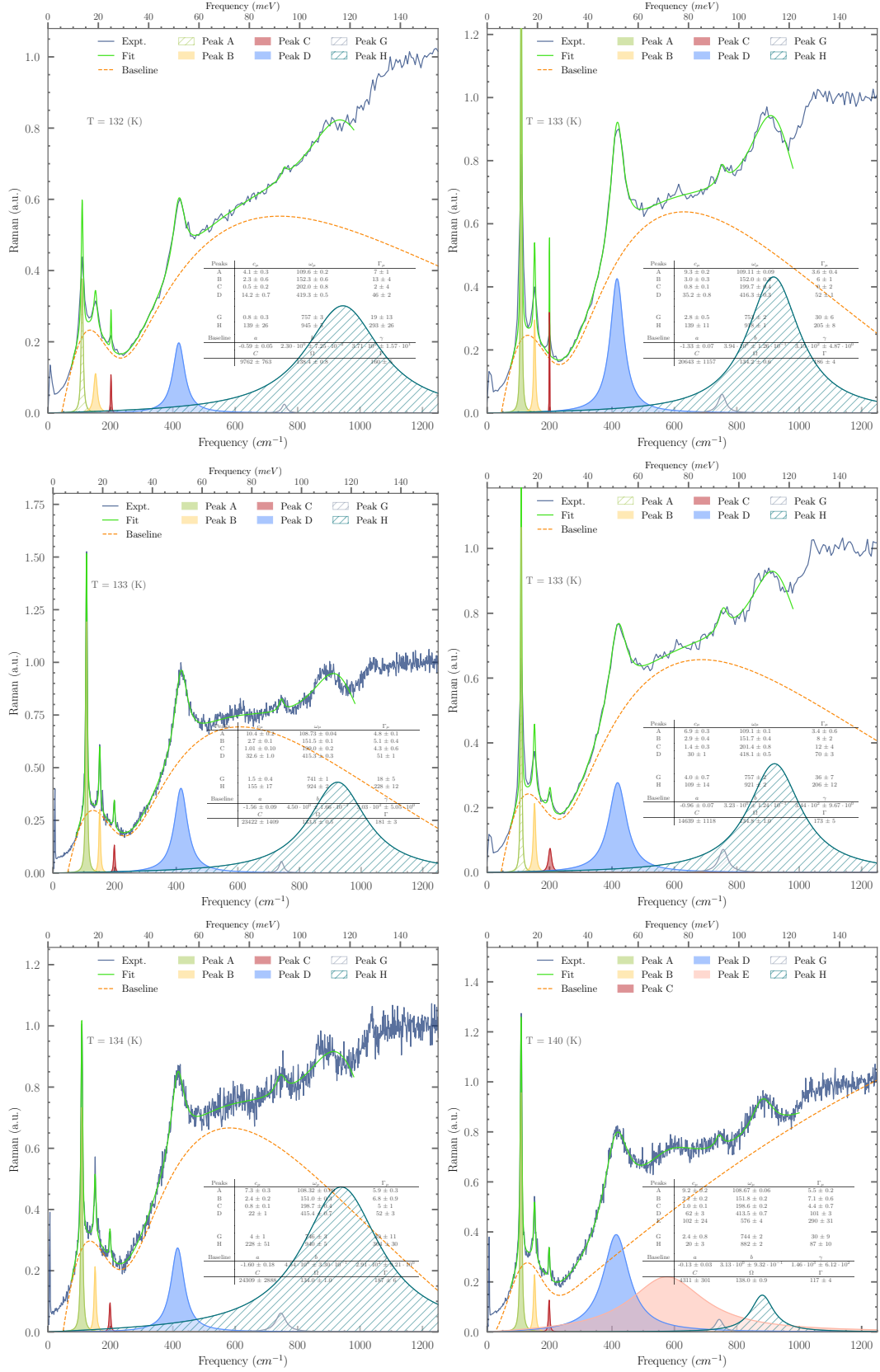

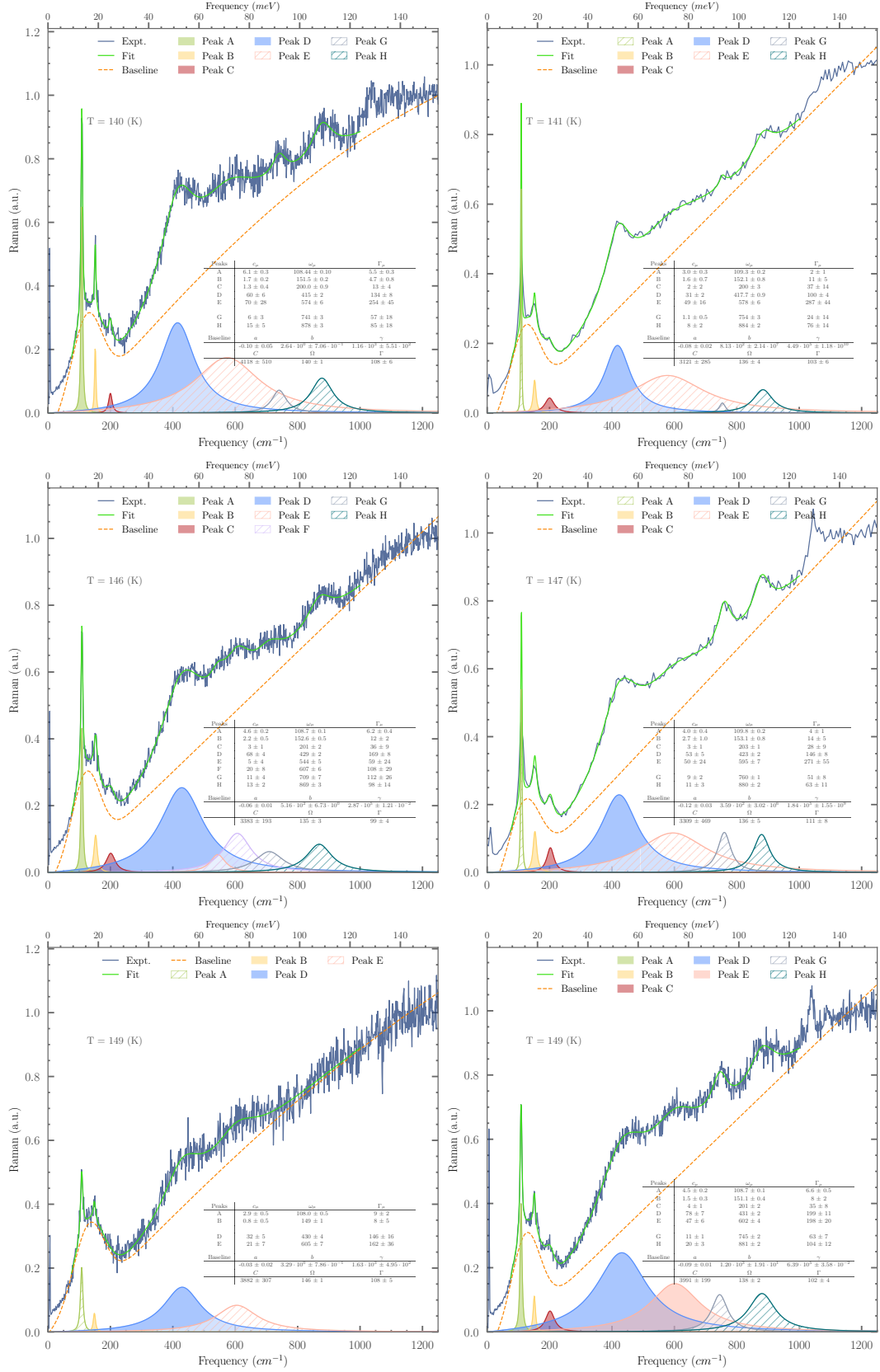

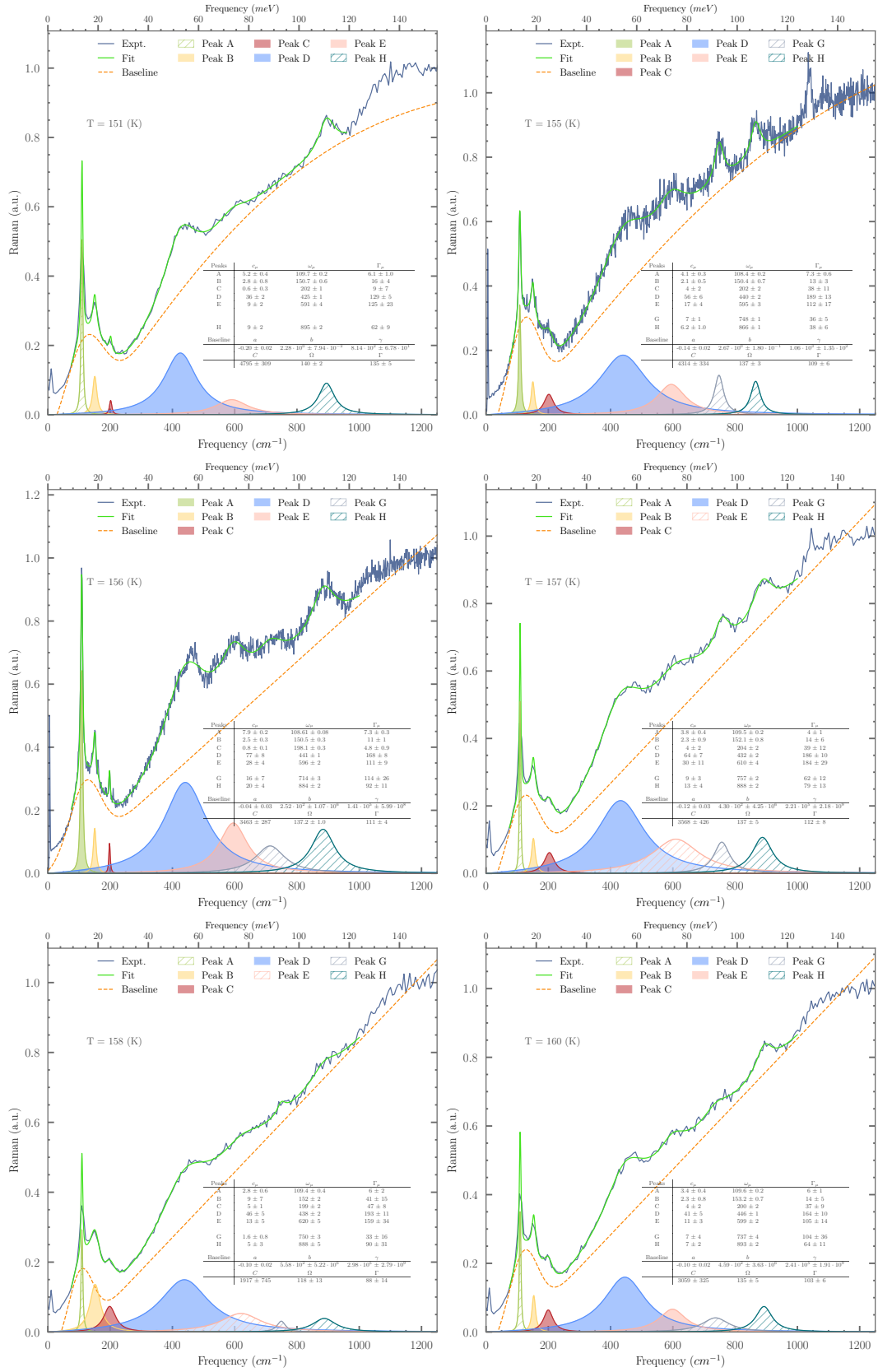

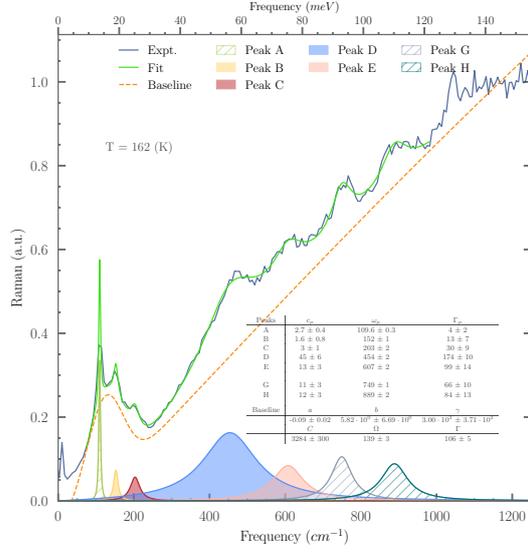

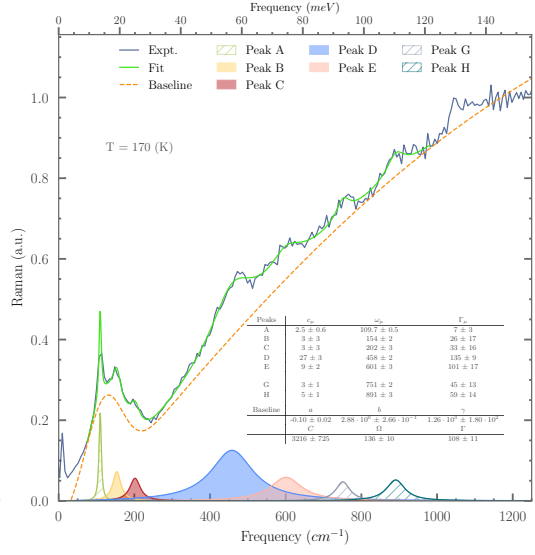

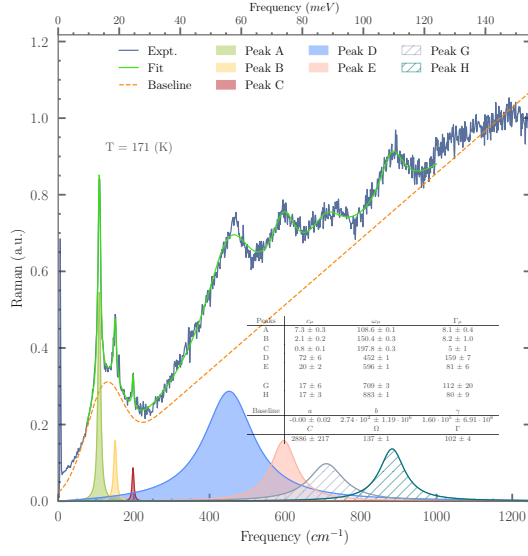

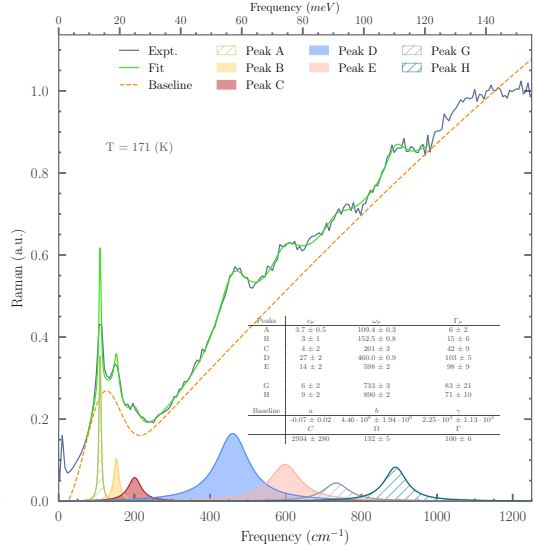

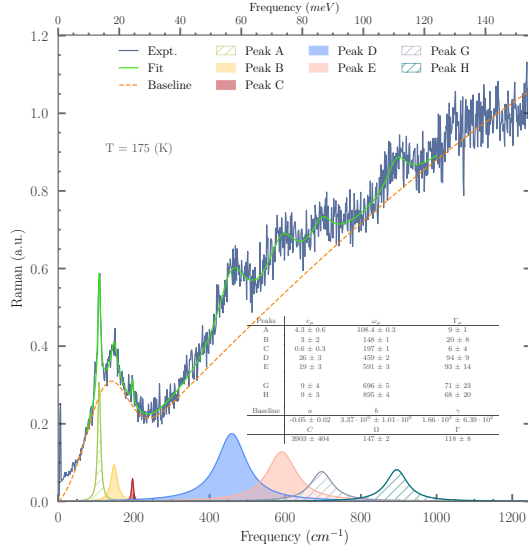

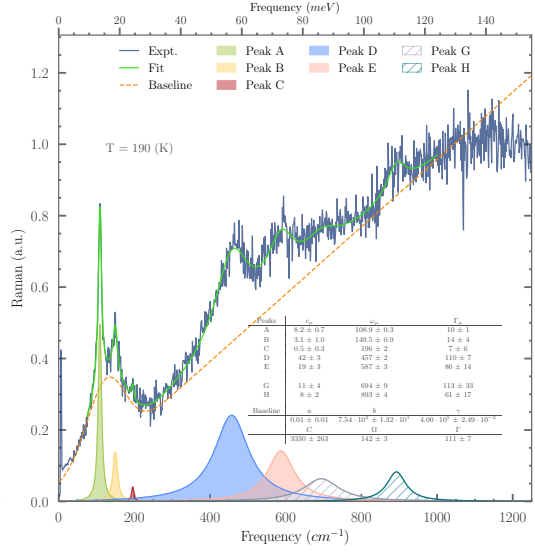

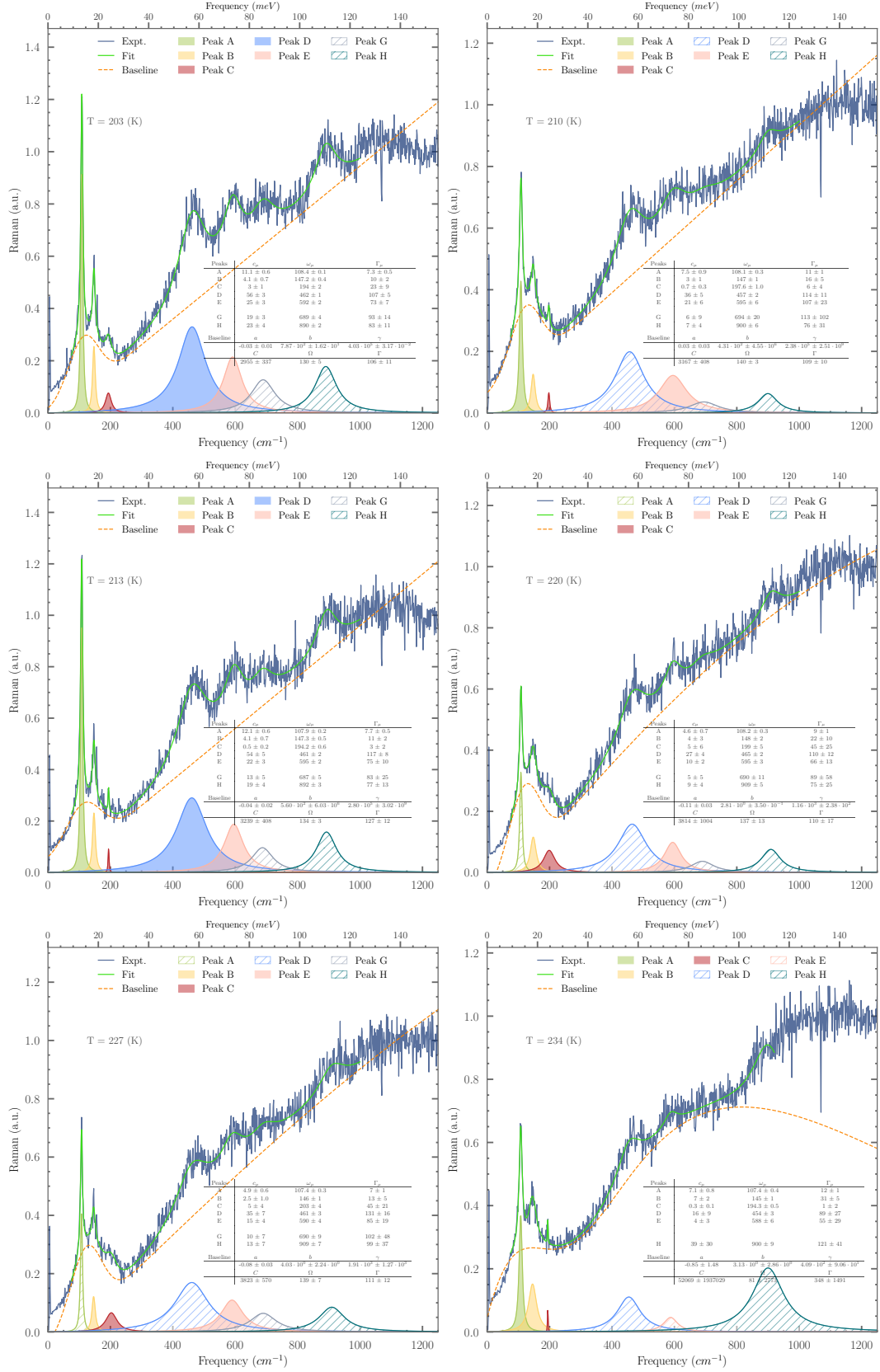

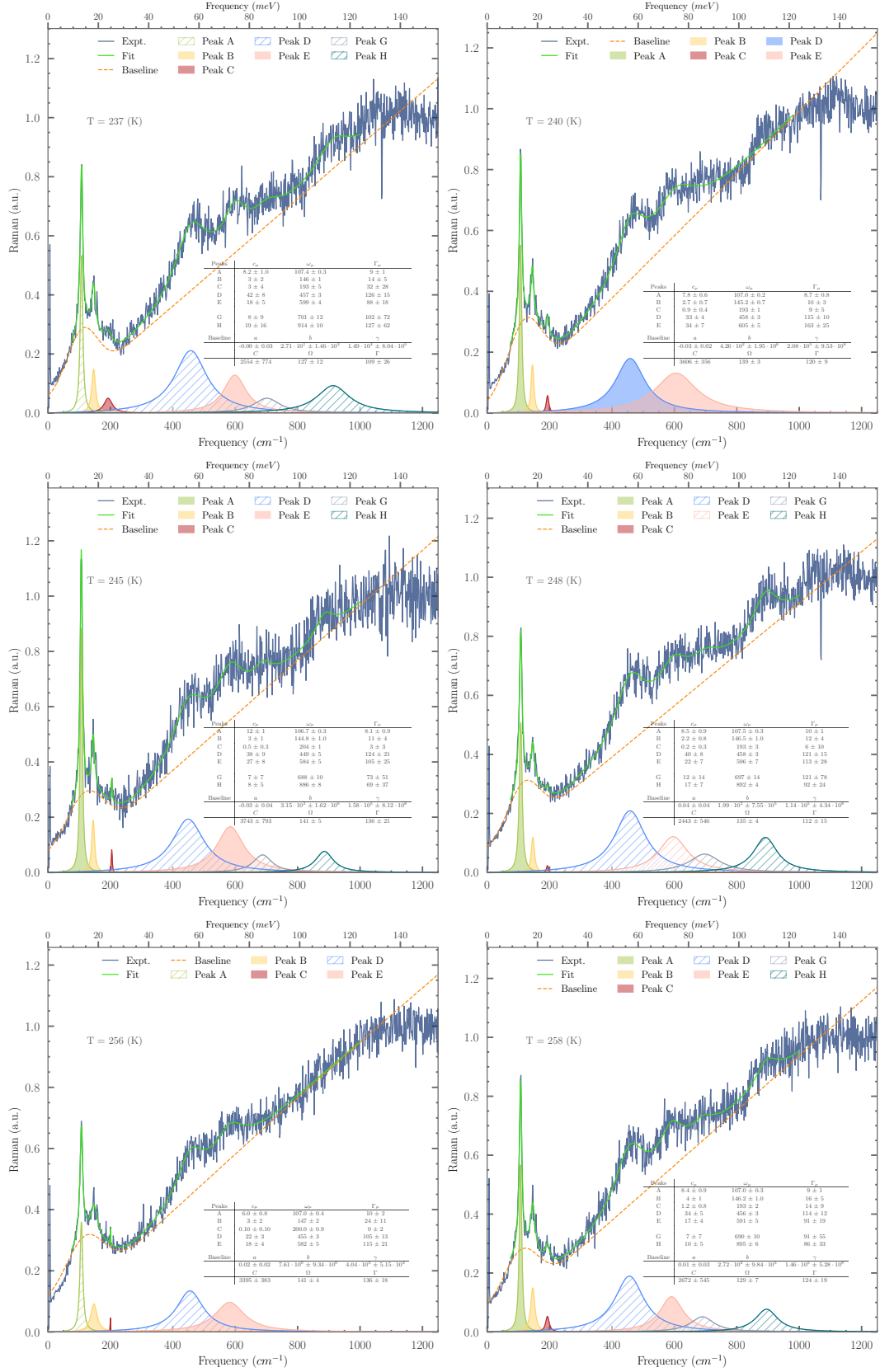

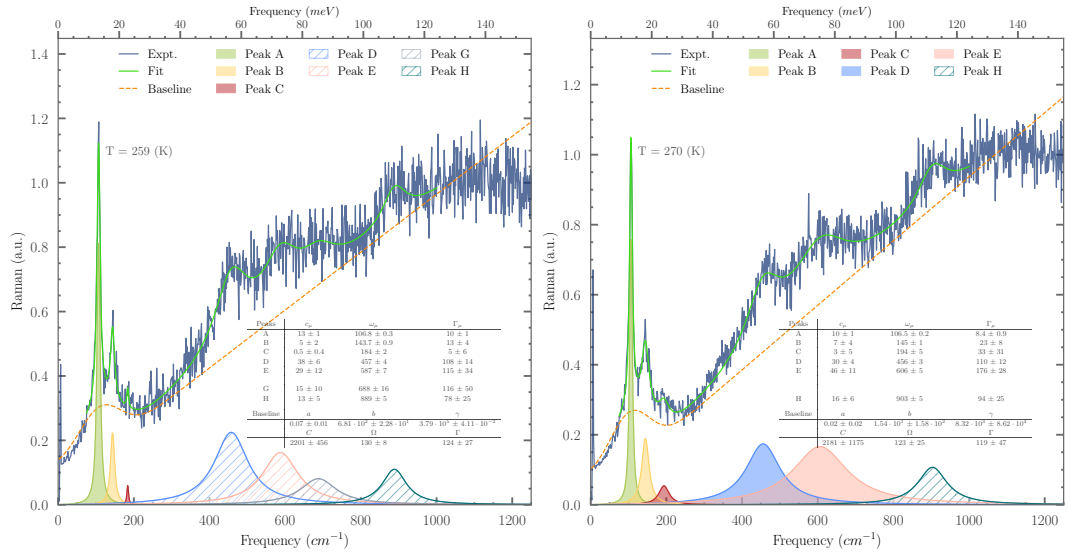

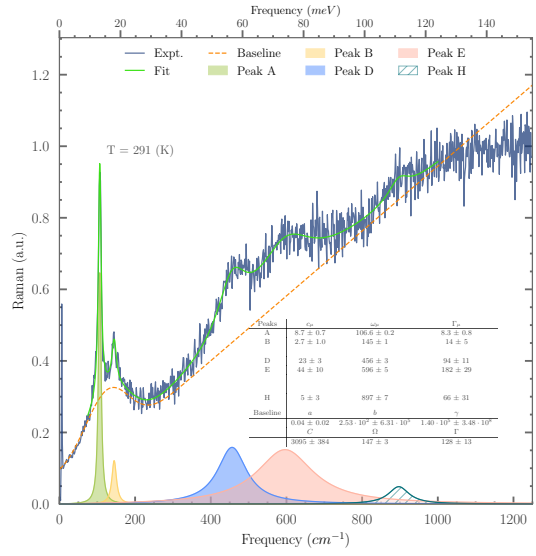

# B  Spectra sample 2

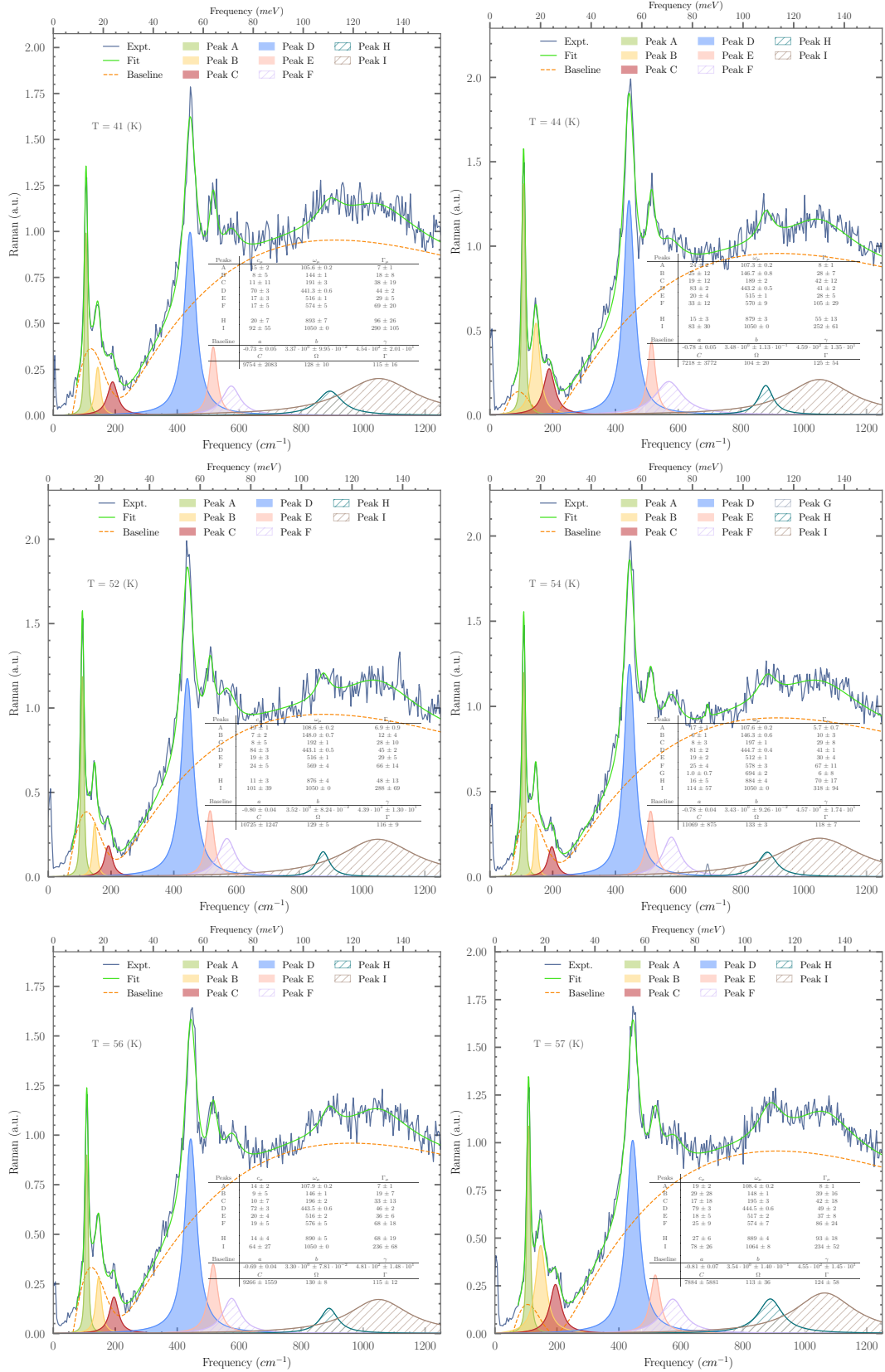

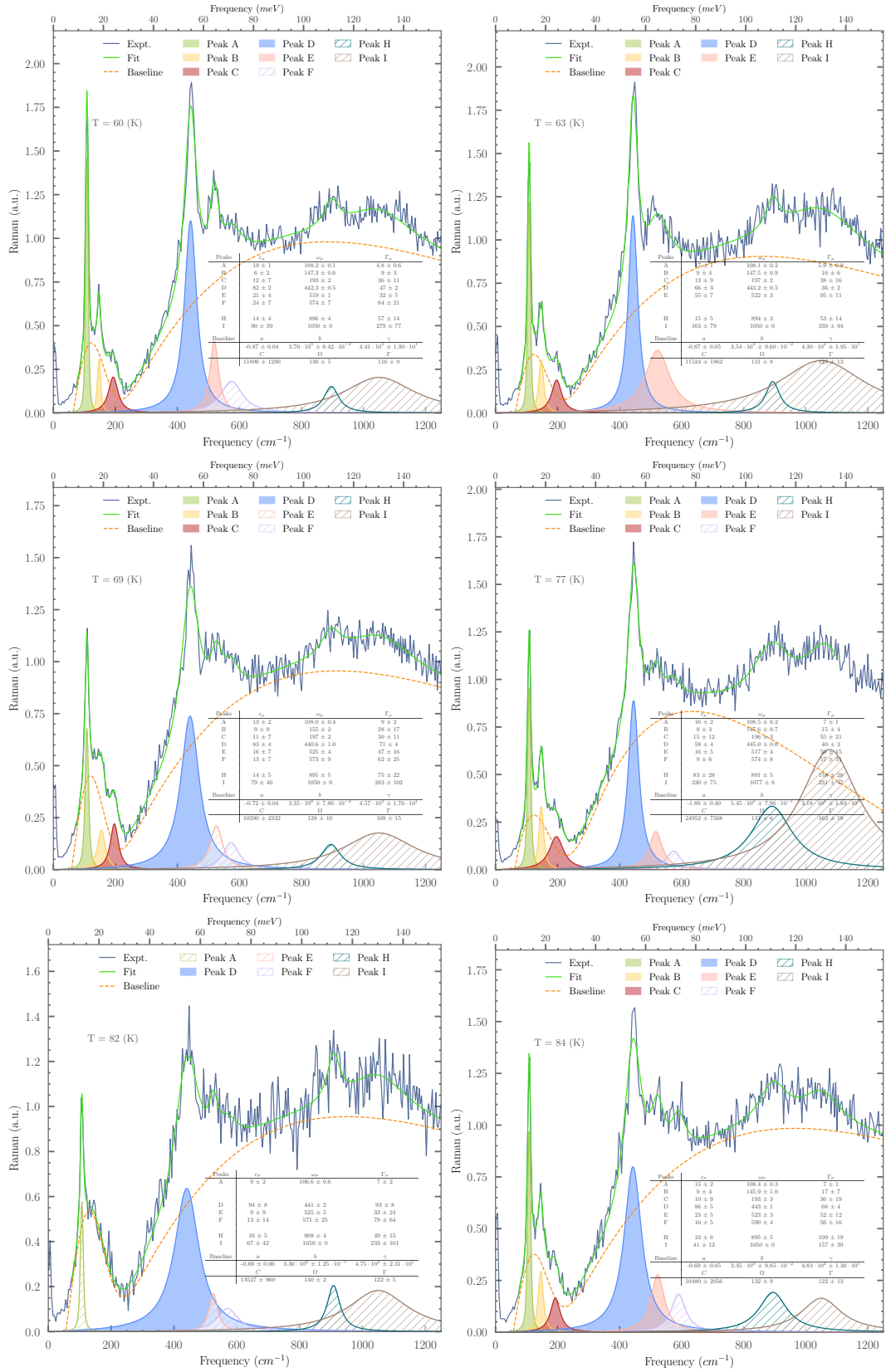

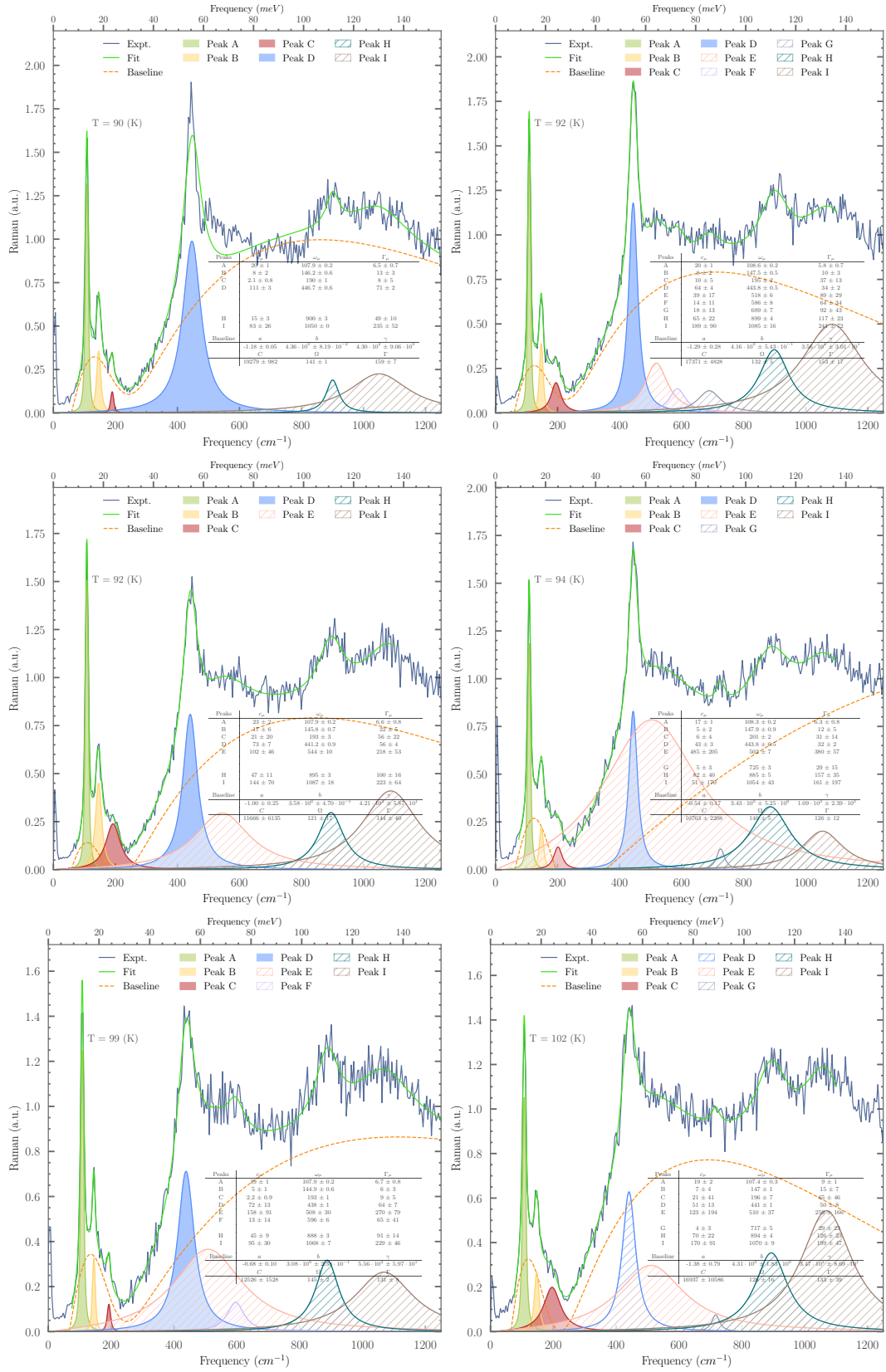

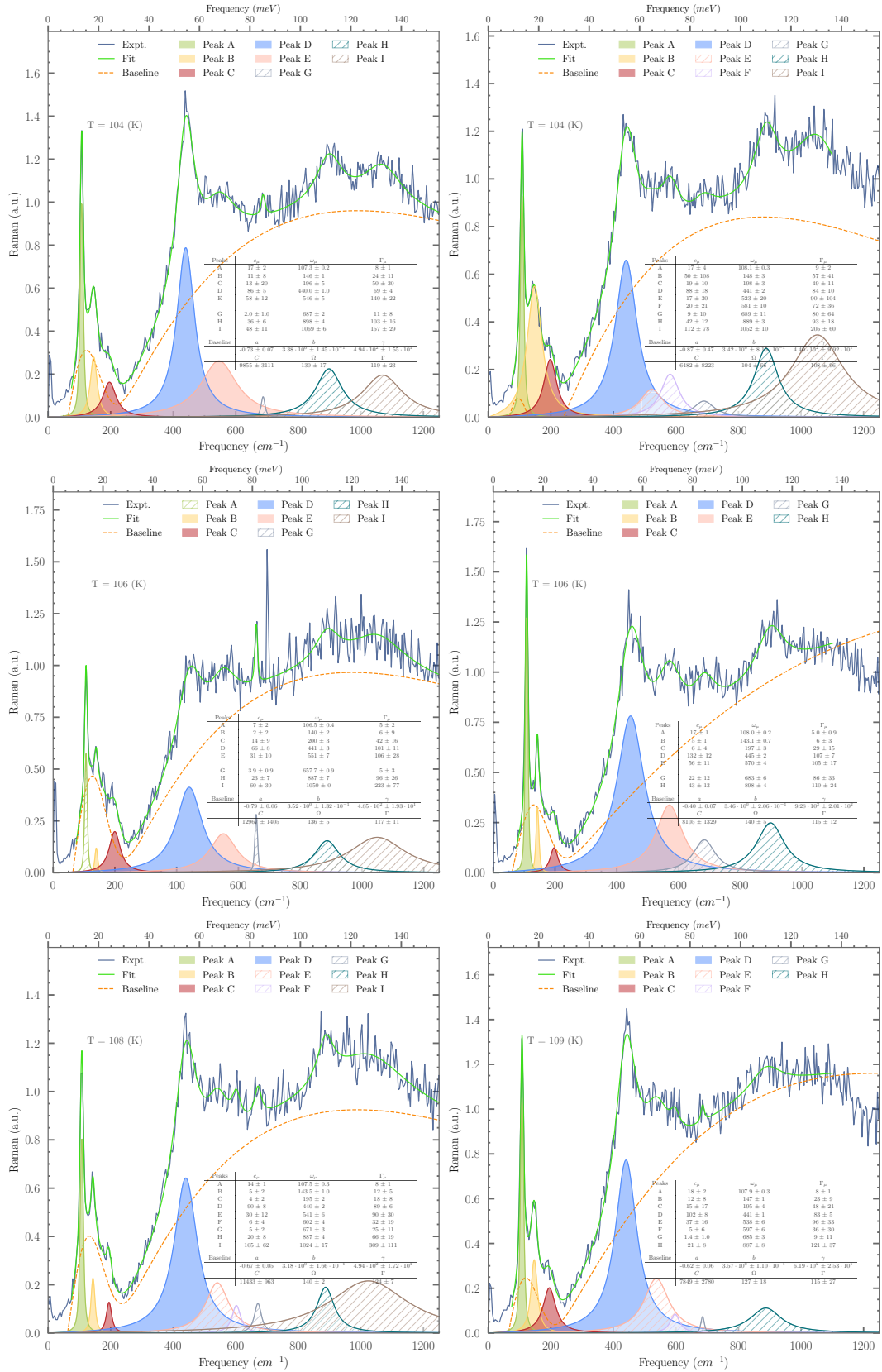

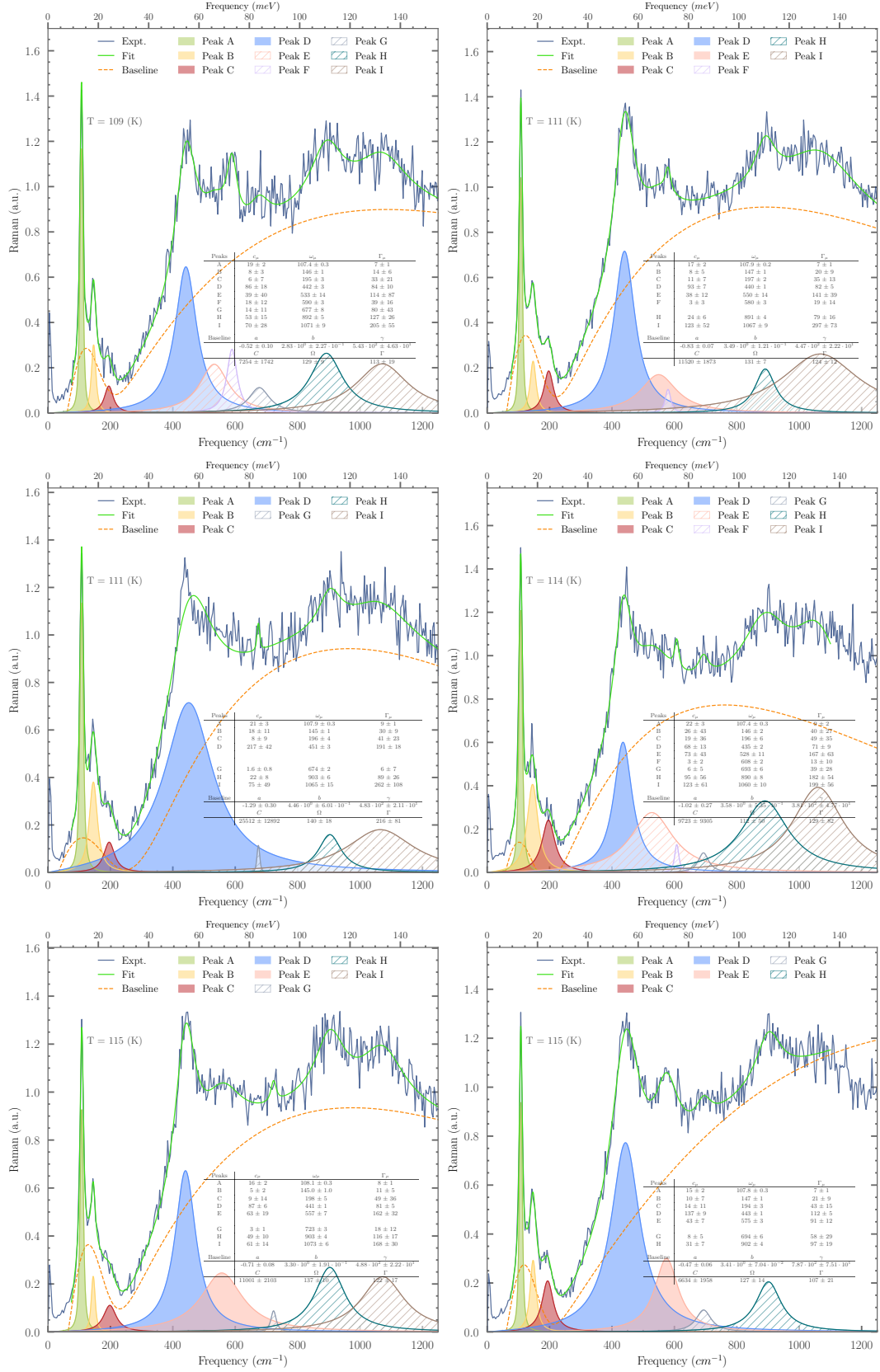

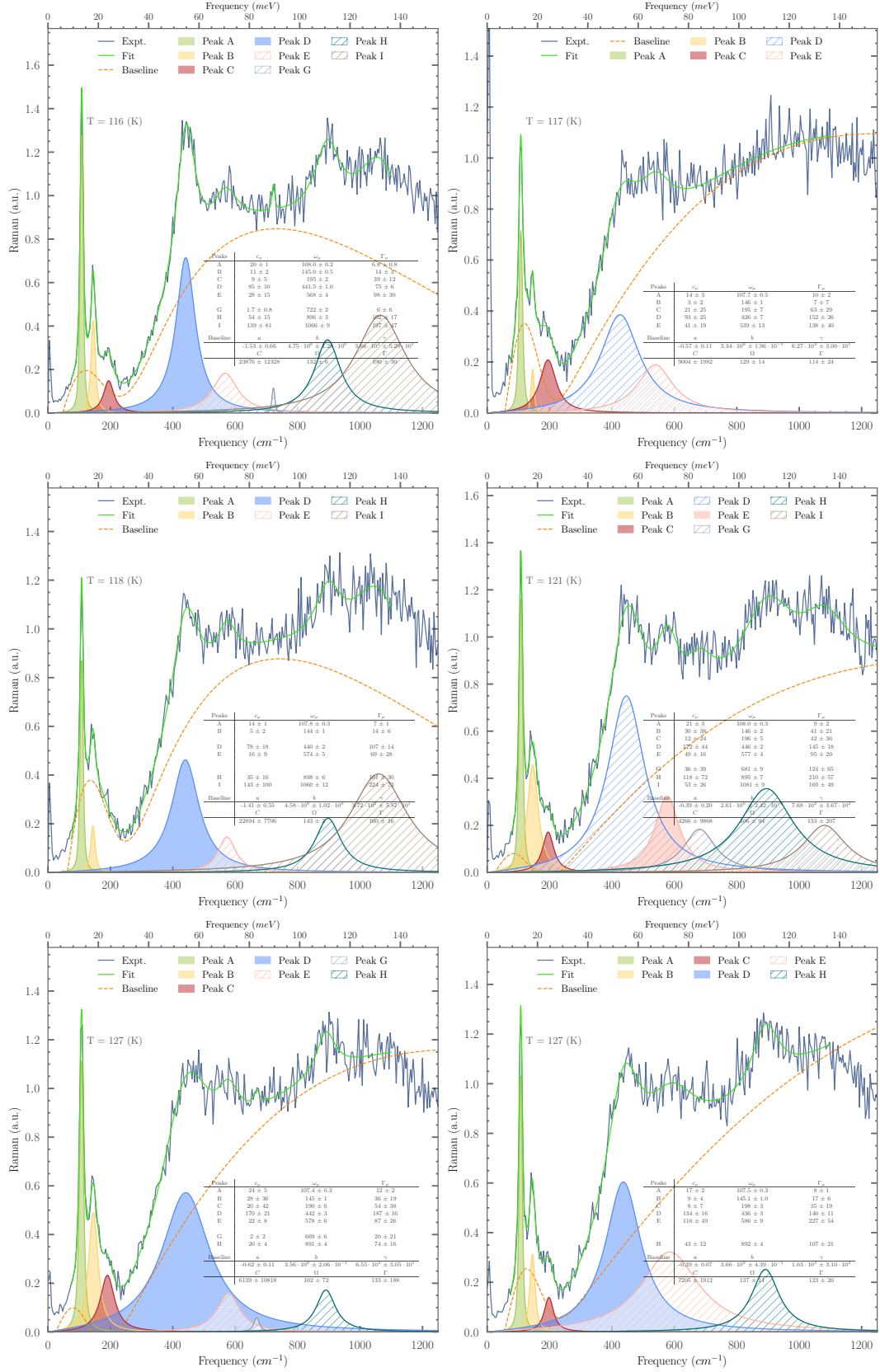

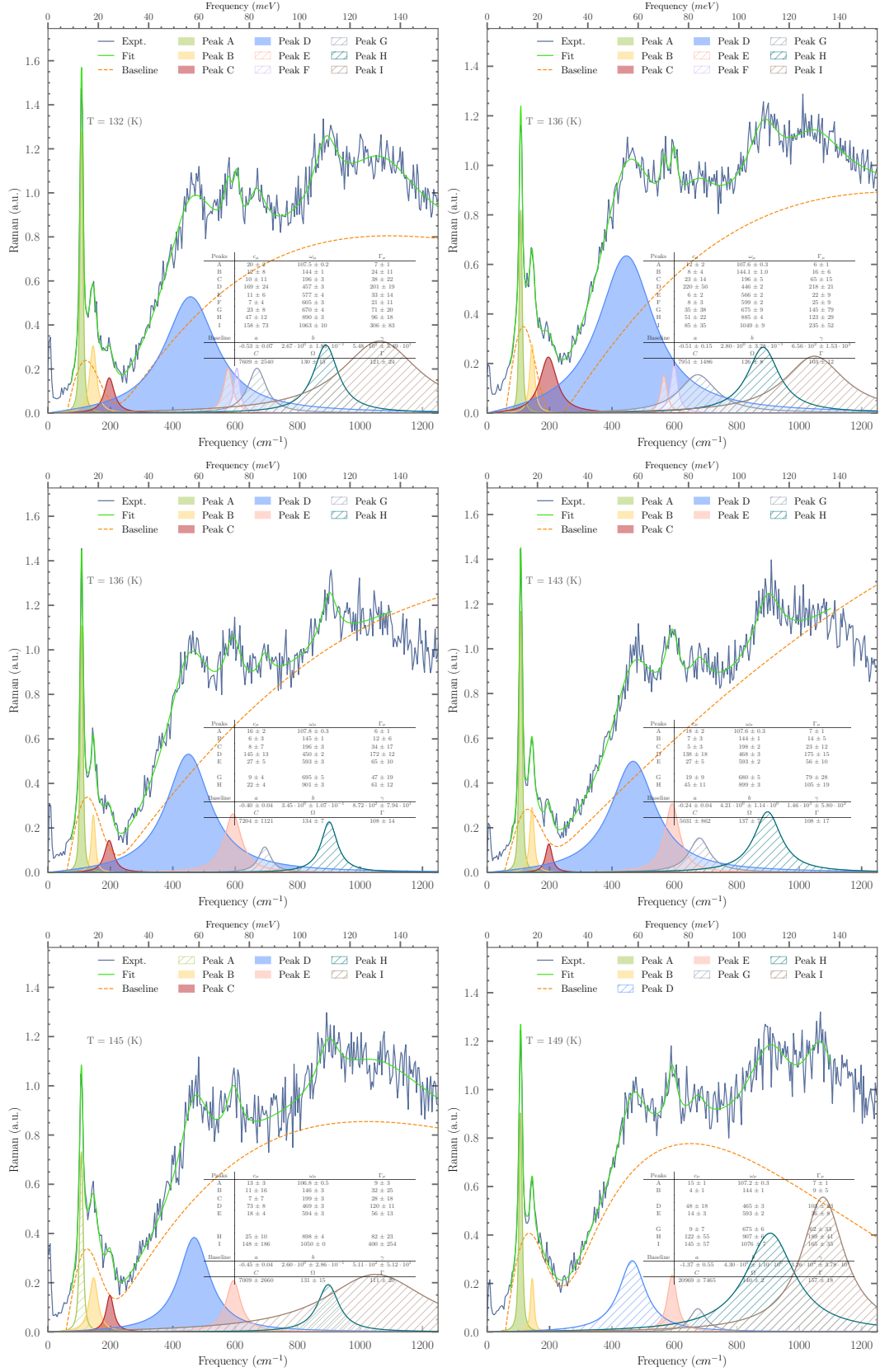

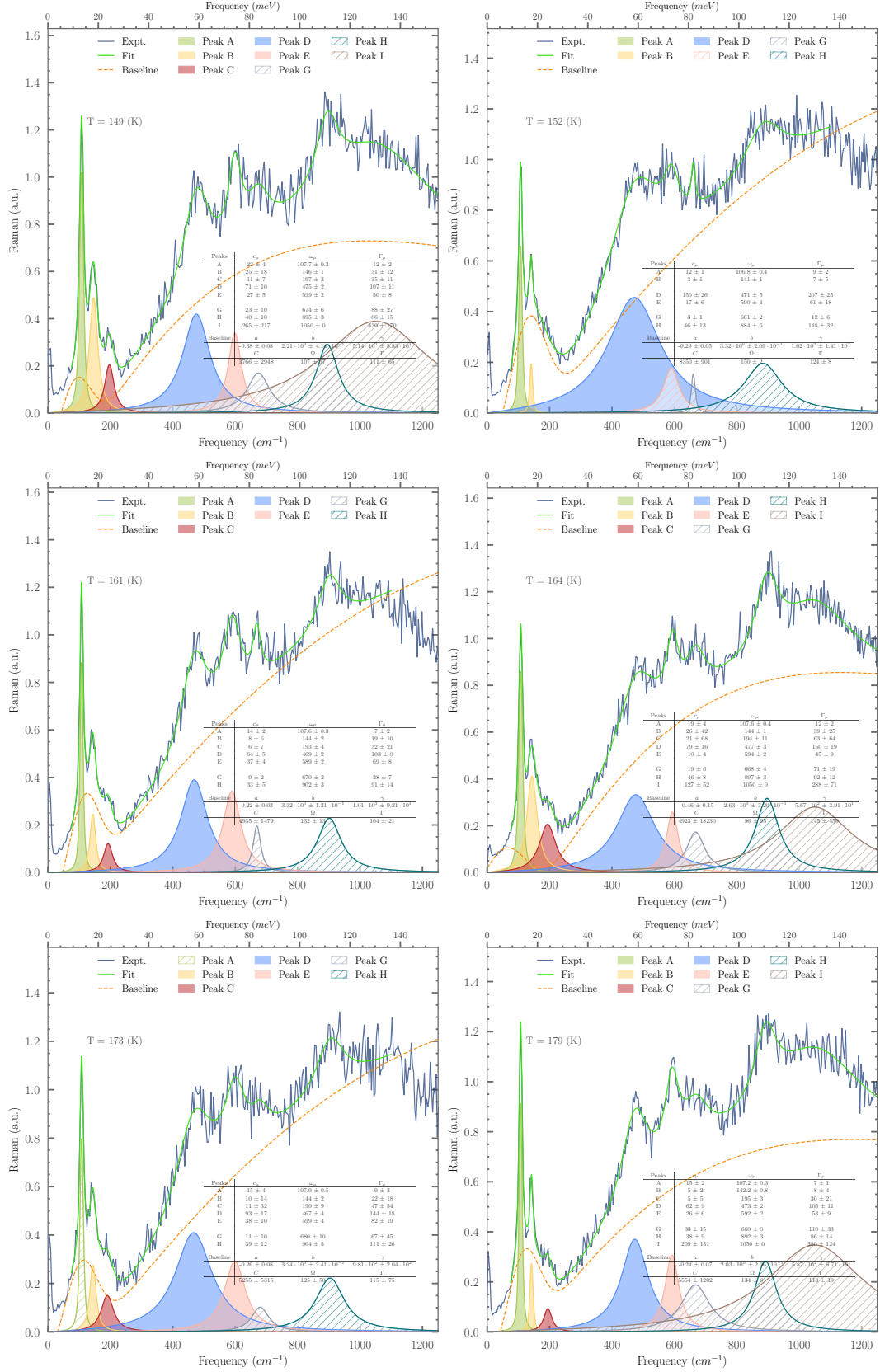

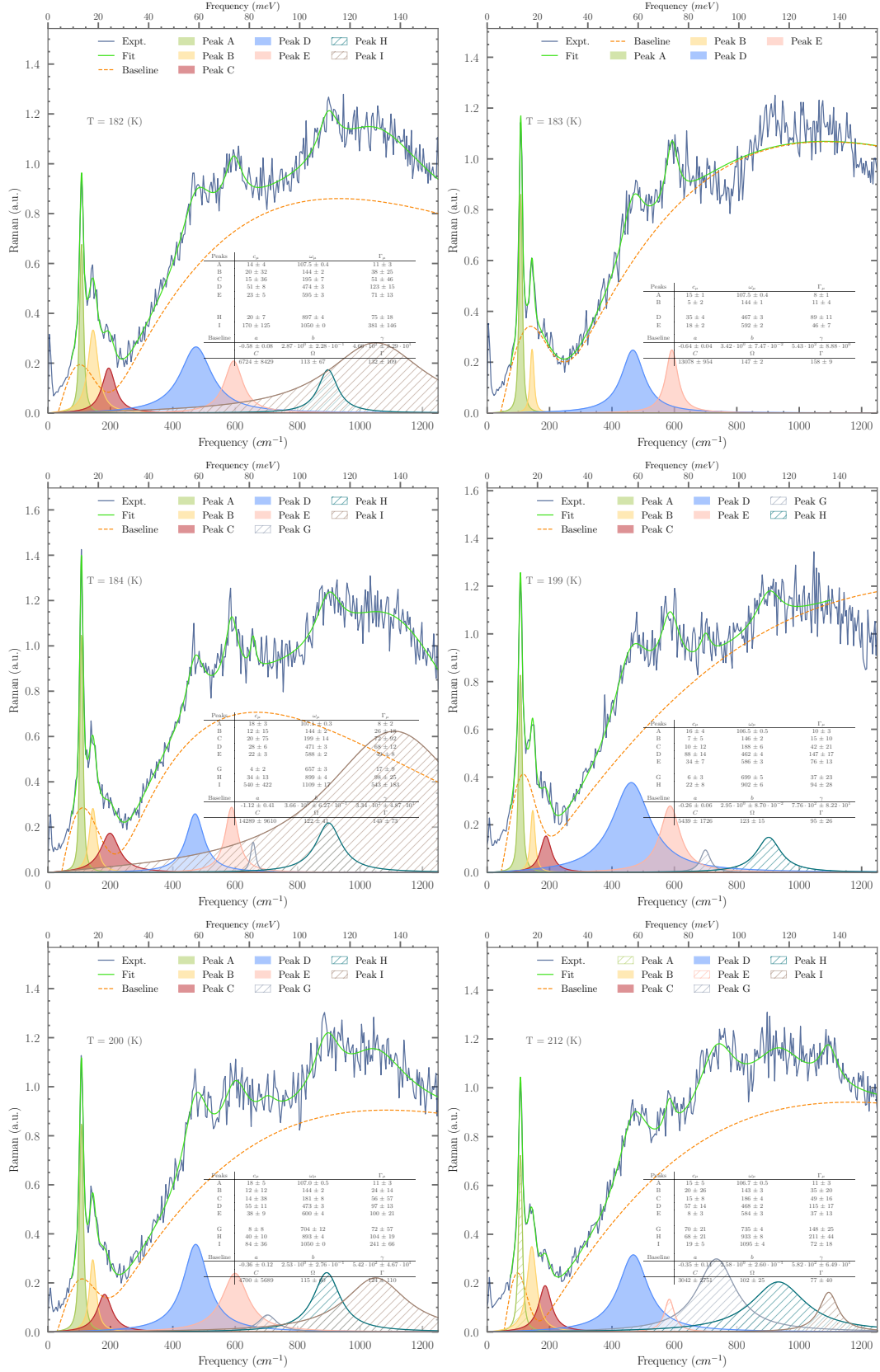

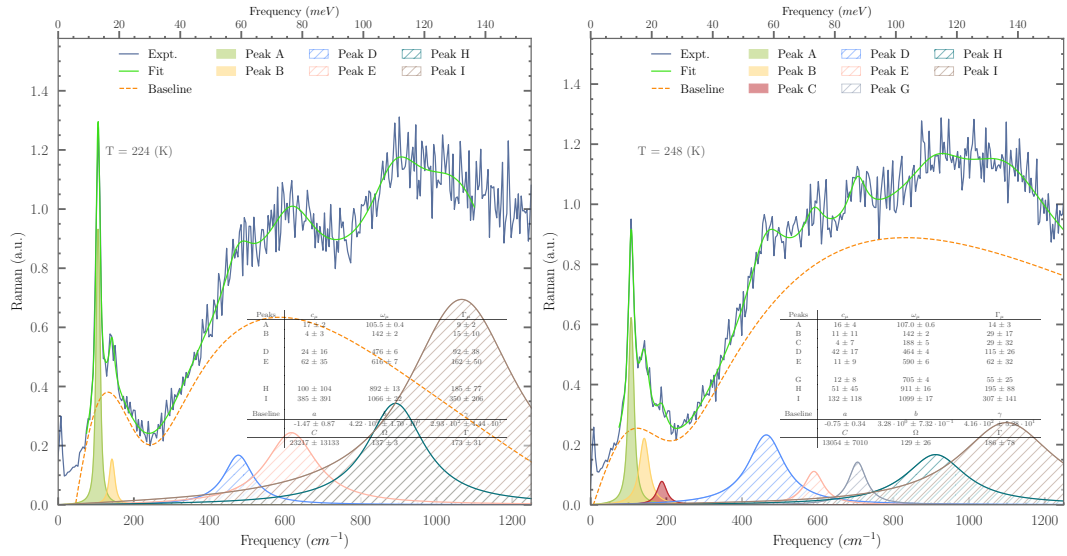

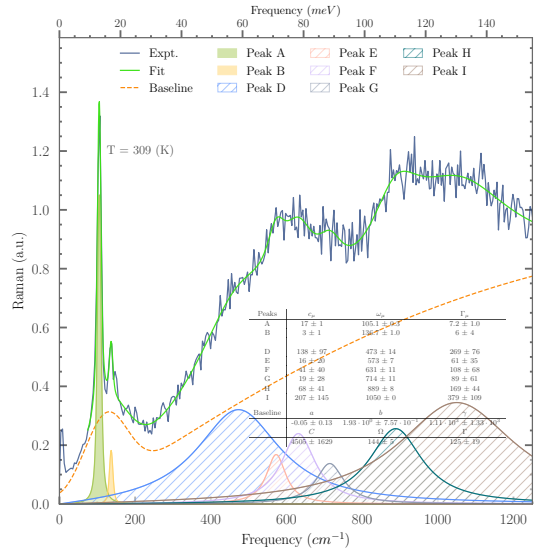



\end{document}